\documentclass[iop,apj,numberedappendix]{emulateapj}

\slugcomment{Received ; accepted ; published }

\shorttitle{Pushing 1D CCSNe to explosions: model and SN~1987A}
\shortauthors{Perego et al.}

\usepackage{natbib}
\usepackage{graphicx}
\usepackage{amsmath}
\usepackage{natbib}
\usepackage{epsfig}
\usepackage{txfonts}
\usepackage{epstopdf}

\newcommand{\push}{\mathrm{push}}

\usepackage{color}
\usepackage[normalem]{ulem}  
\usepackage{xspace}
\definecolor{green2}{rgb}{0,0.5,0}
\definecolor{green2}{rgb}{0.2,0.6,0.1}

\newcommand{\msun}{M$_{\odot}$\xspace}
\newcommand{\kpush}{$k_{\rm push}$\xspace}
\newcommand{\trise}{$t_{\rm rise}$\xspace}

\begin{document}

\title{PUSHing core-collapse supernovae to explosions in spherical symmetry I:\\ The model and the case of SN~1987A}

\author{%
A.\ Perego\altaffilmark{1},
M.\ Hempel\altaffilmark{2},
C.\ Fr\"ohlich\altaffilmark{3}, 
K.\ Ebinger\altaffilmark{2},
M.\ Eichler\altaffilmark{2},
J.\ Casanova\altaffilmark{3},
M.\ Liebend\"orfer\altaffilmark{2},
F.-K.\ Thielemann\altaffilmark{2}%
}
\altaffiltext{1}{Institut f\"ur Kernphysik, Technische Universit\"at Darmstadt, D-64289 Darmstadt, Germany}
\altaffiltext{2}{Departement f\"ur Physik, Universit\"at Basel, CH-4056 Basel, Switzerland}
\altaffiltext{3}{Department of Physics, North Carolina State University, Raleigh NC 27695}
\email{carla\_frohlich@ncsu.edu}

\date{\today}

\begin{abstract}
We report on a method, PUSH, for artificially triggering core-collapse supernova explosions of massive stars in spherical symmetry. We explore basic explosion properties and calibrate PUSH to reproduce SN~1987A observables.
Our simulations are based on the GR hydrodynamics code AGILE combined with the neutrino transport scheme IDSA for electron neutrinos and ASL for the heavy flavor neutrinos. To trigger explosions in the otherwise non-exploding simulations, the PUSH method increases the energy deposition in the gain region proportionally to the heavy flavor neutrino fluxes. We explore the progenitor range 18~--21~M$_{\odot}$.
Our studies reveal a distinction between high compactness (HC) (compactness parameter $\xi_{1.75}>0.45$) and low compactness (LC) ($\xi_{1.75}<0.45$) progenitor models, where LC models tend to explode earlier, with a lower explosion energy, and with a lower remnant mass. HC models are needed to obtain explosion energies around 1~Bethe, as observed for SN~1987A. However, all the models with sufficiently high explosion energy overproduce $^{56}$Ni and fallback is needed to reproduce the observed nucleosynthesis yields. $^{57-58}$Ni yields depend sensitively on the electron fraction and on the location of the mass cut with respect to the shell structure of the progenitor. We identify a progenitor and a suitable set of parameters that fit the explosion properties of SN~1987A assuming 0.1~M$_{\odot}$ of fallback. We predict a neutron star with a gravitational mass of 1.50~M$_{\odot}$. We find correlations between explosion properties and the compactness of the progenitor model in the explored mass range. However, a more complete analysis will require exploring of a larger set of progenitors.
\end{abstract}

\keywords{hydrodynamics, nucleosynthesis, stars: neutron, supernovae: general, supernovae: individual (SN~1987A)}

\section{Introduction}
\label{sec:intro}
Core-collapse supernovae (CCSN) occur at the end of the life of massive stars ($M \gtrsim 8 - 10 \, \rm{M}_{\odot}$). In these violent events, the core of the star gravitationally collapses and triggers a shock wave, leading to the supernova explosion. Despite many decades of theoretical and numerical modeling, the detailed explosion mechanism is not yet fully understood. Simulations in spherical symmetry including detailed neutrino transport and general relativity fail to explode self-consistently, except for the lowest-mass core-collapse progenitors \citep{fischer10,huedepohl10}. There are many ongoing efforts using multi-dimensional fluid dynamics, magnetic fields, and rotation to 
address various remaining open questions in core-collapse supernova theory (see, e.g., \citet{janka12,janka12b,burrows13}). Among those are also technical issues, for example the consequences of neutrino transport approximations, the convergence of simulation results, or the dependence of the simulation outcome on the dimensionality of the model. Awareness of this dependence is especially important because not all investigations can be performed in a computationally very expensive three-dimensional model. While sophisticated multi-dimensional models are needed for an accurate investigation of the explosion mechanism, they are currently too expensive for systematic studies that have to be based on a large number of progenitor models. But such a large number of simulations is required to address the following fundamental questions: What are the conditions for explosive nucleosynthesis as a function of progenitor properties? What is the connection between the progenitor model and the compact remnant? How do 
these aspects relate to the explosion dynamics and energetics? The lack of readily calculable supernova simulations with self-consistent explosions is a problem for many related fields, in particular for predicting nucleosynthetic yields of supernovae.  As we will continue to argue below, spherically symmetric models of the explosion of massive stars are still a pragmatic method to study large numbers of stellar progenitors, from the onset of the explosion up to several seconds after core bounce.

In the past, supernova nucleosynthesis predictions relied on artificially triggered explosions, either using a piston \citep[e.g.][]{ww95,Limongi2006,Chieffi2013} or a thermal energy bomb \citep[e.g.,][]{tnh96,Umeda2008}. For the piston model, the motion of a mass point is specified along a ballistic trajectory. For the thermal energy bomb, explosions are triggered by adding thermal energy to a mass zone. In both cases, additional energy is added to the system to trigger an explosion. In addition, the mass cut (bifurcation between the proto-neutron star (PNS) and the ejecta) and the explosion energy are free parameters which have to be constrained from the mass of the $^{56}$Ni ejecta. While these approaches are appropriate for the outer layers, where the nucleosynthesis mainly depends on the strength of the shock wave, they are clearly incorrect for the innermost layers. There, the conditions and the nucleosynthesis are directly related to the physics of collapse and bounce, and to the details of the 
explosion mechanism.
Besides the piston and thermal bomb methods, another widely used way to artificially trigger explosions is the so-called ``neutrino light-bulb''. In this method, the PNS is excised and replaced with an inner boundary condition which contains an analytical prescription for the neutrino luminosities. The neutrino transport is replaced by neutrino absorption and emission terms in optically thin conditions. Suitable choices of the neutrino luminosities and energies can trigger neutrino-driven explosions \citep[e.g.,][]{Burrows1993,Yamasaki2005,iwakami08,iwakami09,yamamoto2013}.
The light-bulb method has also been used to investigate models with respect to their dimensionality. The transition from spherical symmetry (1D) to axisymmetry (2D) delivers the new degree of freedom to bring cold accreting matter down to the neutrinospheres while matter in other directions can dwell longer in the gain region and efficiently be heated by neutrinos \citep{Herant94}. The standing accretion shock instability \citep[SASI, e.g.,][]{Blondin2003,Blondin2006,Scheck2008,iwakami09,Fernandez2010,Guilet2012} strongly contributes to this effect in 2D light-bulb models and leads to strong polar oscillations of expansion during the unfolding of the explosion \citep{Murphy2008}. It was first expected that the trend toward a smaller critical luminosity for successful explosions will continue as one goes from 2D to three-dimensional (3D) models \citep{Nordhaus2010,handy14}, but other studies pointed toward the contrary \citep{Hanke2012,Couch2013a}. One has to keep in mind, that a light bulb approach might not 
include the full coupling between the accretion rate and the neutrino luminosity. However, recent models that derive the neutrino luminosity from a consistent evolution of the neutron star support the result that 2D models show faster explosions than 3D models \citep{Mueller.Janka.ea:2012, Bruenn.Mezzacappa.ea:2013, Dolence2013, takiwaki2014}. Most important for this work is a finding that is consistent with all above investigations: In 3D there is no preferred axis. The 3D degrees of freedom lead to a more efficient cascade of fluid instabilities to smaller scales. In spite of vivid fluid instabilities, the 3D models show in their overall evolution a more pronounced sphericity than the 2D models. Hence their average conditions resemble more closely the shock expansion that would be obtained by an exploding 1D model.

In a 1D model with detailed Boltzmann neutrino transport two other methods to trigger explosions using neutrinos have been used \citep{cf06a,fischer10}.
These ``absorption methods'' aim at increasing the neutrino energy deposition in the heating region by mimicking the expected net effects of multi-dimensional simulations. In one case, the neutral-current scattering opacities on free nucleons are artificially decreased to values between 0.1 and 0.7 times the original values. This leads to increased diffusive neutrino fluxes in regions of very high density. The net results are a faster deleptonization of the PNS and higher neutrino luminosities in the heating region. In the other case, explosions are enforced by multiplying the reaction rates for neutrino absorption on free nucleons by a constant factor. To preserve detailed balance, the emission rates also have to be multiplied by the same factor. This reduces the timescale for neutrino heating and again results in a more efficient energy deposition in the heating region. However, the energy associated with these explosions were always weak.

Recently, \cite{Ugliano.Janka.ea:2012} have presented a more sophisticated light-bulb method to explode spherically symmetric models using neutrino energy deposition in post-shock layers. They use an approximate, grey neutrino transport and replace the innermost 1.1~\msun of the PNS by an inner boundary. The evolution 
of the neutrino boundary luminosity
is based on an analytic cooling model of the PNS, which depends on a set of free parameters. These parameters are set by fitting observational properties of SN~1987A for progenitor masses around 20~\msun (see also \citet{ertl2015}).

Artificial supernova explosions have been obtained by other authors using a grey leakage scheme that includes neutrino heating via a parametrized charged-current absorption scheme \citep{OConnor2010} in spherically symmetric simulations \citep{OConnor.Ott:2011}.

In this paper, we report on a new approach, PUSH, for artificially triggering explosions of massive stars in spherical symmetry. In PUSH, we deposit a fraction of the luminosity of the heavy flavor neutrinos emitted by the PNS in the gain region to increase the neutrino heating efficiency. We ensure an accurate treatment of the electron fraction of the ejecta through a spectral neutrino transport scheme for $\nu_e$ and $\bar{\nu}_e$ and a detailed evolution of the PNS. We calibrate our new method by comparing the explosion energies and nucleosynthesis yields of different progenitor stars with observations of SN~1987A. This method provides a framework to study many important aspects of core-collapse supernovae for large sets of progenitors: explosive supernova nucleosynthesis, neutron-star remnant masses, explosion energies, and other aspects where full multi-dimensional simulations are still too expensive and traditional piston or thermal bomb models do not capture all the relevant physics. With PUSH we can  investigate general tendencies and perform systematic parameter variations, providing complementary information to ``ab-initio'' multi-dimensional simulations.

The article is organized as follows: 
Section~\ref{sec:method} describes our simulation framework, the stellar progenitor models, the new method PUSH, and our post-processing analysis. 
In Section~\ref{sec:results}, we present a detailed exploration of the PUSH method and the results of fitting it to observables of SN~1987A.
We also analyze aspects of the supernova dynamics and progenitor dependency.
In Section~\ref{sec:discuss}, we discuss further implications of our results and also compare with other works from the literature.
A summary is given and conclusions are drawn in Section~\ref{sec:concl}.

\section{Method and input}
\label{sec:method}

\subsection{Hydrodynamics and neutrino transport}

We make use of the general relativistic hydrodynamics code AGILE in spherical symmetry \citep{Liebendoerfer.Agile}. For the stellar collapse, we apply the deleptonization scheme of \citet{Liebendoerfer.delept:2005}. For the neutrino transport, we employ the Isotropic Diffusion Source Approximation (IDSA) for the electron neutrinos $\nu_e$ and electron anti-neutrinos $\overline{\nu}_{e}$ \citep{Liebendoerfer.IDSA:2009}, and an Advanced Spectral Leakage scheme (ASL) for the heavy-lepton flavor neutrinos $\nu_x = \nu_{\mu}, \overline{\nu}_{\mu}, \nu_{\tau}, \overline{\nu}_{\tau}$ \citep{Perego2014}. We discretize the neutrino energy using 20 geometrically increasing energy bins, in the range $3 \, {\rm MeV} \leq E_{\nu} \leq 300\, {\rm MeV}$. The neutrino reactions included in the IDSA and ASL scheme are summarized in Table~\ref{tab:nu_interact}. They represent the minimal set of the most relevant weak processes in the post-bounce phase, particularly up to the onset of an explosion. Note that electron captures 
on heavy nuclei and neutrino scattering on electrons, which are relevant in the collapse phase \citep[see, for example, ][]{Mezzacappa1993a,Mezzacappa1993c}, are not included explicitly in the form of reaction rates, but as part of the parameterized deleptonization scheme.
In the ASL scheme, we omit nucleon-nucleon bremsstrahlung, $N+N \leftrightarrow N+N +\nu_x+ \bar{\nu}_x$, where $N$ denotes any nucleon \citep[see, for example,][]{Hannestad98,Bartl14}. We have observed that its inclusion would overestimate $\mu$ and $\tau$ neutrino luminosities during the PNS cooling phase, due to the missing neutrino thermalization provided by inelastic scattering on electrons and positrons at the PNS surface. However, we have also tested that the omission of this process does not significantly change the $\mu$ and $\tau$ neutrino luminosities predicted by the ASL scheme before the explosion sets in. Thus, neglecting $N$-$N$ bremsstrahlung is only relevant for the cooling phase, where it improves the overall behavior when compared to simulations obtained with detailed Boltzmann neutrino transport \cite[][e.g.]{fischer10}.

The equation of state (EOS) of \citet{Hempel.SchaffnerBielich:2010} (HS) that we are using includes various light nuclei, such as alphas, deuterons or tritons (details below). However, the inclusion of all neutrino reactions for this detailed nuclear composition would be beyond the standard approach implemented in current supernova simulations, where only scattering on alpha particles is typically included. To not completely neglect the contributions of the other light nuclei, we have added their mass fractions to the unbound nucleons. This is motivated by their very weak binding energies and, therefore, by the idea that they behave similarly as the unbound nucleon component.

In all our models we use 180 radial zones which include the progenitor star up to the helium shell. This corresponds to a radius of $R \approx (1.3-1.5) \times 10^{10}$~cm. With this setup, we model the collapse, bounce, and onset of the explosion. The grid of AGILE is adaptive, with more resolution where the gradients of the thermodynamic variables are steeper. Thus, in the post-bounce and explosion phases, the surface of the PNS and the shock are the better resolved regions. The simulations are run for a total time of 5~s, corresponding to $\gtrsim 4.6$~s after core bounce. At this time, the shock has not yet reached the external edge of our computational domain.

\begin{table}[h,t]
\begin{center}
\caption{Relevant neutrino reactions.\label{tab:nu_interact}}
\begin{tabular}{lll}
    \tableline \tableline
    Reactions  & Treatment & Reference \\
    \tableline
    $e^- + p \leftrightarrow n + \nu_e$ & IDSA & a \\
    $e^+ + n \leftrightarrow p + \overline{\nu}_e$ & IDSA & a \\
    $N + \nu \leftrightarrow N + \nu$ & IDSA \& ASL & a \\
    $(A,Z) + \nu \leftrightarrow (A,Z) + \nu$ & IDSA \& ASL & a \\
    $e^- + e^+ \leftrightarrow \nu_{\mu,\tau} + \overline{\nu}_{\mu,\tau}$ & ASL & a, b \\
    \hline
\end{tabular}
\tablecomments{Nucleons are denoted by $N$. The nucleon charged current rates are based on \citet{Bruenn85}, but effects of mean-field interactions \citep{reddy98,roberts12,martinez12,hempel14} are taken into account.}
\tablerefs{ (a) \citet{Bruenn85}; (b) \citet{Mezzacappa1993b}.}
\end{center}
\end{table}

\subsection{Equation of state and nuclear reactions}
\label{sec_eos}
For the high-density plasma in nuclear statistical equilibrium (NSE) the tabulated microphysical EOS HS(DD2) is used. This supernova EOS is based on the model of \citet{Hempel.SchaffnerBielich:2010}. It uses the DD2 parametrization for the nucleon interactions \citep{typel10}, the nuclear masses from \citet{audi2003}, and the \textit{Finite Range Droplet Model} \citep{moller95}. 8140 nuclei are included in total, up to $Z=136$ and to the neutron drip line. The HS(DD2) was first introduced in \citet{fischer14}, where its characteristic properties were discussed and general EOS effects in core-collapse supernova simulations were investigated. \citet{fischer14} showed that the HS(DD2) EOS gives a better agreement with constraints from nuclear experiments and astrophysical observations than the commonly used EOSs of \citet{lattimer91} and \citet{shen98}. Furthermore, additional degrees of freedom, such as various light nuclei and a statistical ensemble of heavy nuclei, are taken into account. The nucleon mean-
field potentials, which are used in the charged-current rates, have been calculated consistently \citep{hempel14}. The maximum mass of a cold neutron star for the HS(DD2) EOS is 2.42~M$_\odot$ \citep{fischer14}, which is well above the limits from \citet{demorest2010} and \citet{antoniadis2013}. 

The EOS employed in our simulations includes an extension to non-NSE conditions. In the non-NSE regime the nuclear composition is described by 25 representative nuclei from neutrons and protons to iron-group nuclei. The chosen nuclei are the alpha-nuclei $^4$He, $^{12}$C, $^{16}$O, $^{20}$Ne, $^{24}$Mg, $^{28}$Si, $^{32}$S, $^{36}$Ar, $^{40}$Ca, $^{44}$Ti, $^{48}$Cr, $^{52}$Fe, $^{56}$Ni, complemented by $^{14}$N and the following asymmetric isotopes: $^{3}$He, $^{36}$S, $^{50}$Ti, $^{54}$Fe, $^{56}$Fe, $^{58}$Fe, $^{60}$Fe, $^{62}$Fe, and $^{62}$Ni. With these nuclei it is possible to achieve a mapping of the abundances from the progenitor calculations onto our simulations which is consistent with the provided electron fraction, i.e., mantaining charge neutrality. All the nuclear masses $M_i$ are taken from \citet{audi2003}. To advect the nuclear composition inside the adaptive grid, we implement the Consistent Multi-fluid Advection (CMA) method by \citet{plewa98}. For given abundances, the non-NSE EOS is 
calculated based on the same underlying description used in the NSE regime \citep{Hempel.SchaffnerBielich:2010}, but with the following modifications:  excited states of nuclei are neglected, excluded volume effects are not taken into account, and the nucleons are treated as non-interacting Maxwell-Boltzmann gases. Such a consistent description of the non-NSE and NSE phases prevents spurious effects at the transitions between the two regimes.

Outside of NSE, an approximate $\alpha$-network is used to follow the changes in composition. Explosive Helium-, Carbon-, Neon-, and Oxygen-burning are currently implemented in the simulation. Note that the thermal energy generation by the nuclear reactions is fully incorporated via the detailed non-NSE treatment. We do not have to calculate explicitly any energy liberation, but just the changes in the abundances. Within our relativistic treatment of the EOS (applied both in the non-NSE and NSE regime) energy conservation means that the specific internal energy $e_{\rm int}$ is not changed by nuclear reactions. This is due to fact that $e_{\rm int}$ includes the specific rest mass energy $e_{\rm mass}$, where $e_{\rm mass}$ is given by the sum over the masses of all nuclei weighted with their yield $Y_i=X_i/A_i$,
\begin{eqnarray}
e_{\rm mass}=\sum_i Y_i M_i \; .
\end{eqnarray}
However, if we define the specific thermal energy $e_{\rm th}$ as
\begin{equation}
 e_{\rm th} = e_{\rm int} - e_{\rm mass} , 
 \label{eqn:e_th}
\end{equation}
the nuclear reactions will decrease the rest mass energy (i.e., increase the binding) and consequently increase the thermal energy. This treatment of the non-NSE EOS is consistent with the convention used in all high-density NSE EOSs.

Due to limitations of our approximate $\alpha$-network, and because we do not include any quasi-statistical equilibrium description, we apply some parameterized burning for temperatures between 0.3 and 0.4~MeV. A temperature dependent burning timescale is introduced, which gradually transforms the initial non-NSE composition towards NSE. For temperatures of 0.4~MeV and above, the non-NSE phase always reaches a composition close to NSE and its thermodynamic properties become very similar to the NSE phase of the HS(DD2) EOS. Even though the two phases are based on the same input physics, small, but unavoidable differences can remain, due to the limited set of nuclei considered in non-NSE. To assure a smooth transition for all conditions, we have introduced a transition region as an additional means of thermodynamic stability. We have chosen a parameterization in terms of temperature and implement a linear transition in the temperature interval from 0.40~MeV to 0.44~MeV. We check that the basic thermodynamic 
stability condition $ds/dT>0$ is always fulfilled.

\subsection{Initial models}
\begin{figure}[htp!]
   \includegraphics[width=0.5\textwidth]{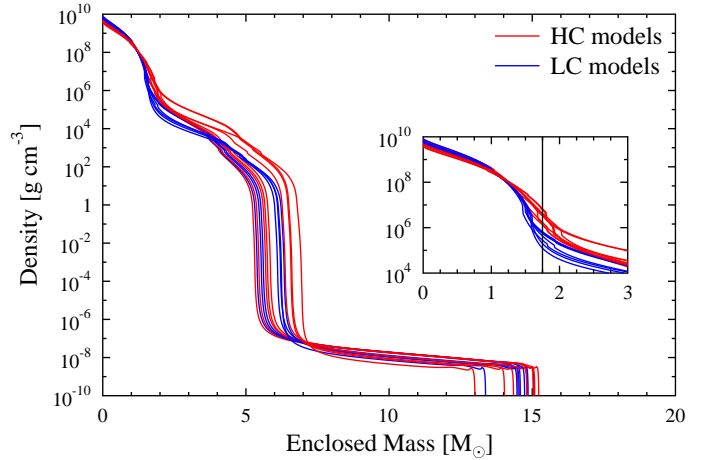}
   \caption{Density profiles as function of ZAMS mass for the progenitor models included in this study (18.0~\msun to 21.0~\msun).  HC models are shown in red, LC models are shown in blue.  The vertical line in the inset is located at 1.75~\msun and indicates that mass at which the compactness parameter $\xi_{1.75}$ is determined (see Equation~\ref{eq:compactness}).
   \label{fig:prog_dens}}
\end{figure}

\begin{figure}[htp!]
   \includegraphics[width=0.5\textwidth]{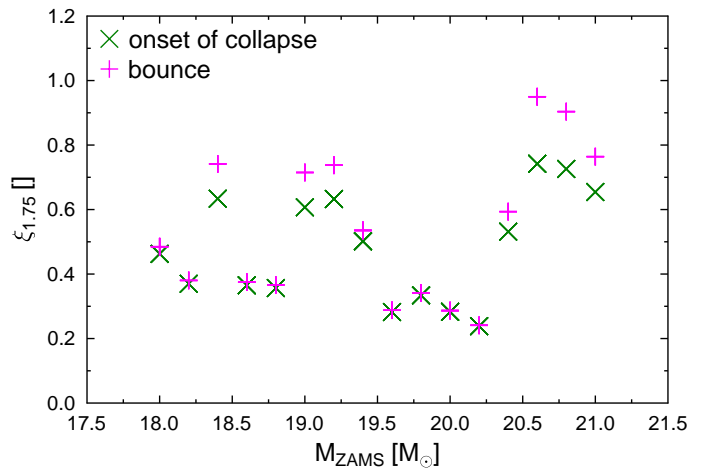}
   \caption{Compactness $\xi_{1.75}$ as function of ZAMS mass for our pre-explosion models at the onset of collapse (green crosses) and at bounce (magenta pluses).
   \label{fig:prog_compact}
}
\end{figure}

For this study, we use solar-metallicity, non-rotating stellar models from the stellar evolution code {\tt KEPLER} \citep{Woosley.Heger:2002}. Our set includes 16 pre-explosion models with zero-age main sequence (ZAMS) mass between 18.0~M$_{\odot}$ and 21.0~M$_{\odot}$ in increments of 0.2~M$_{\odot}$. These models have been selected to have ZAMS mass around 20~M$_{\odot}$, similar to the progenitor of SN~1987A \citep[e.g.,][]{PP.sn1987a:2007}. We label the models by their ZAMS mass. In Figure \ref{fig:prog_dens}, the density profiles of the progenitor models are shown. For each of them the compactness parameter $\xi_{M}$ is defined following \citet{OConnor.Ott:2011} by the ratio of a given mass $M$ and the radius $R(M)$ which encloses this mass:
\begin{equation}
  \xi_{M} \equiv \frac{M/M_{\odot}}{R(M)/1000\mathrm{km}}.
  \label{eq:compactness}
\end{equation}
Typically, either $\xi_{1.75}$ or $\xi_{2.5}$ are used. The compactness can be computed at the onset of collapse or at bounce, as suggested by \citet{OConnor.Ott:2011}. For our progenitors, the difference in the compactness parameter between these two moments is not significant for our discussions. Thus, for simplicity, in the following we will use $\xi_{1.75}$ computed at the onset of the collapse. The progenitor models considered here fall into two distinct families of compactness: low compactness ($\xi_{1.75} < 0.45$; LC models) and high compactness ($\xi_{1.75}> 0.45$; HC models), see Table \ref{tab:prog_compact}. Figure \ref{fig:prog_compact} shows the compactness as function of ZAMS mass for the progenitors of this study. The non-monotonous behavior is a result of the evolution before collapse.  
The mass range between 19 and 21~\msun is particularly prone to variations of the compactness. 
For a detailed discussion of the behavior of the compactness as function of ZAMS mass see \citet{sukhbold.woosley:2014}.

\begin{table*}[h,t]
\begin{center}
\caption{Progenitor properties \label{tab:prog_compact}}
\begin{tabular}{llllllllll}
  \tableline \tableline
  M$_{\rm ZAMS}$& \multicolumn{2}{c}{$\xi_{1.75}$} & \multicolumn{2}{c}{$\xi_{2.5}$} & M$_{\rm prog}$ & M$_{\rm Fe}$ & M$_{\rm CO}$ & M$_{\rm He}$ & M$_{\rm env}$\\
  (M$_{\odot}$) & at collapse  & at bounce    & at collapse & at bounce   & (M$_{\odot}$)  & (M$_{\odot}$)     & (M$_{\odot}$)     & (M$_{\odot}$)     & (M$_{\odot}$)\\
  \tableline
  18.2 & 0.37  & 0.380 & 0.173 & 0.173 & 14.58 & 1.399 & 4.174 & 5.395 & 9.186\\
  18.6 & 0.365 & 0.375 & 0.170 & 0.170 & 14.85 & 1.407 & 4.317 & 5.540 & 9.313\\
  18.8 & 0.357 & 0.366 & 0.166 & 0.166 & 15.05 & 1.399 & 4.390 & 5.613 & 9.435\\
  19.6 & 0.282 & 0.288 & 0.118 & 0.117 & 13.37 & 1.461 & 4.959 & 6.243 & 7.125\\
  19.8 & 0.334 & 0.341 & 0.135 & 0.135 & 14.54 & 1.438 & 4.867 & 6.112 & 8.428\\
  20.0 & 0.283 & 0.287 & 0.125 & 0.125 & 14.73 & 1.456 & 4.960 & 6.215 & 8.517\\
  20.2 & 0.238 & 0.241 & 0.104 & 0.104 & 14.47 & 1.458 & 5.069 & 6.342 & 8.125\\
  \tableline
  18.0 & 0.463 & 0.485 & 0.199 & 0.199 & 14.50 & 1.384 & 4.104 & 5.314 & 9.187\\
  18.4 & 0.634 & 0.741 & 0.185 & 0.185 & 14.82 & 1.490 & 4.238 & 5.459 & 9.366\\
  19.0 & 0.607 & 0.715 & 0.191 & 0.191 & 15.03 & 1.580 & 4.461 & 5.693 & 9.341\\
  19.2 & 0.633 & 0.737 & 0.191 & 0.192 & 15.08 & 1.481 & 4.545 & 5.760 & 9.325\\
  19.4 & 0.501 & 0.535 & 0.185 & 0.185 & 15.22 & 1.367 & 4.626 & 5.860 & 9.365\\
  20.4 & 0.532 & 0.594 & 0.192 & 0.192 & 14.81 & 1.500 & 5.106 & 6.376 & 8.433\\
  20.6 & 0.742 & 0.95  & 0.278 & 0.279 & 14.03 & 1.540 & 5.260 & 6.579 & 7.450\\
  20.8 & 0.726 & 0.904 & 0.271 & 0.272 & 14.34 & 1.528 & 5.296 & 6.609 & 7.735\\
  21.0 & 0.654 & 0.764 & 0.211 & 0.212 & 13.00 & 1.454 & 5.571 & 6.969 & 6.026\\
  \tableline
  \end{tabular}
\tablecomments{ZAMS mass, compactness $\xi_{1.75}$ and $\xi_{2.5}$ at the onset of collapse and at bounce, total progenitor mass at collapse (M$_{\rm prog}$), mass of the iron core (M$_{\rm Fe}$), carbon-oxygen core (M$_{\rm CO}$), and helium core (M$_{\rm He}$), and mass of the hydrogen-rich envelope (M$_{\rm env}$) at collapse,  for all the progenitor models included in this study.  The top part of the table includes the low-compactness progenitors (LC; $\xi_{1.75}<0.4$ at collapse), the bottom part includes the high-compactness progenitors (HC; $\xi_{1.75}>0.45$ at collapse).
}
\end{center}
\end{table*}

\subsection{The PUSH method}
\label{subsec:push}

\subsubsection{Rationale}

The goal of PUSH is to provide a computationally efficient framework to explode massive stars in spherical symmetry to study multiple aspects of core-collapse supernovae. The usage of a spectral transport scheme to compute the $\nu_e$ and $\bar{\nu}_e$ luminosities provides a more accurate evolution of $Y_e$ of the innermost ejecta, which is a crucial aspect for nucleosynthesis. The neutrino luminosities include the accretion contribution, as well as the luminosity coming from PNS mantle and core. The accretion luminosity depends not only on the accretion rate but also on the evolution of the mass and radius of the PNS, which is treated accurately and self-consistently in our models.

In order to trigger explosions in the otherwise non-exploding spherically symmetric simulations, we rely on the delayed neutrino-driven mechanism, which was first proposed by \citet{bethe_85}. Despite the lack of consensus and convergence of numerical results between different groups, recent multi-dimensional simulations of CCSNe have shown that convection, turbulence and SASI in the shocked layers increase the efficiency at which $\nu_e$ and $\bar{\nu}_e$ are absorbed inside the gain region, compared with spherically symmetric models \citep[see, for example,][]{janka96,Nordhaus2010,Hanke2012,Hanke2013,Dolence2013, Couch2013a,Melson2015}. This effect, together with the simultaneous increase in time that a fluid particle spends inside the gain region \citep[e.g.,][]{Murphy2008, handy14}, provides more favorable conditions for the development of an explosion. Moreover, according to multi-dimensional explosion models, the shock revival is followed by a phase where continued accretion and shock expansion coexist 
over a time scale of $\gtrsim 1 \, {\rm s}$ \citep[e.g.,][]{Scheck2006,Marek2009,Bruenn2014,Melson2015}. During this phase, matter accreted through low-entropy downflows onto the PNS continues to power an accretion luminosity. The re-ejection of a fraction of this matter by neutrino heating accelerates the shock and increases the explosion energy. The length of this phase, the exact amount of injected energy, and its deposition rate are still uncertain.

Inspired by the increase of the net neutrino heating that a fluid element experiences due to the above mentioned multi-dimensional effects, PUSH provides a more efficient neutrino energy deposition inside the gain region in spherically symmetric models. However, unlike other methods that use electron flavor neutrinos to trigger artificial 1D explosions (see Section \ref{sec:intro}), in PUSH we deposit a fraction of the luminosity of the heavy flavor neutrinos ($\nu_x$'s) behind the shock to ultimately provide successful explosion conditions. This additional energy deposition is calibrated by comparing the explosion energies and nucleosynthesis yields obtained from our progenitor sample with observations of SN~1987A. This ensures that our artificially increased heating efficiency has an empirical foundation. Thus, we can make predictions in the sense of an effective model.

Despite the fact that $\nu_x$'s contribute only marginally to the energy deposition inside the gain region in self-consistent models \citep[see, for example,][]{bethe_85} and that they only show a weak dependence on the temporal variation of the accretion rate \citep[see, for example,][]{Liebendoerfer2004}, their usage presents a number of advantages for our purposes. They represent one of the largest energy reservoirs available, but they do not directly change the electron fraction $Y_e$ (unlike electron flavor neutrinos). This allows us to trigger an explosion in 1D simulations without modifying $\nu_e$ and $\bar{\nu}_e$ luminosities nor changing charged current reactions. The $\nu_x$ luminosities are calculated consistently within our model. They include dynamical feedback from the accretion history, progenitor properties of each individual model, and the cooling of the forming compact object. As shown by \citet{OConnor.Ott:2013} in broad progenitor studies, during the accretion phase that precedes the 
shock revival, the properties of the $\nu_x$ spectral fluxes correlate significantly with the properties of $\nu_e$'s and $\bar{\nu}_e$'s. Unlike the electron (anti-)neutrino luminosities, that in spherically symmetric models decrease suddenly once the shock has been revived, $\nu_x$ luminosities are only marginally affected by the development of an explosion. This allows PUSH to continue injecting energy inside the expanding shock for a few hundreds of milliseconds after the explosion has set in. Moreover, since this energy injection is provided by the $\nu_x$ fluxes, it changes significantly between different progenitors and correlates with the $\nu_e$ and $\bar{\nu}_e$ accretion luminosities (at least, during the accretion phase).

\subsubsection{Implementation}

The additional energy deposition, that represents the main feature of PUSH, is achieved by introducing a local heating term, $Q^+_{\push} (t,R)$ (energy per unit mass and time), given by
\begin{equation}
   Q^+_{\push} (t,r) = 4 \, \mathcal{G}(t) \int_0^{\infty} q^+_{\push}(r,E) \, dE ,
   \label{eq:push_integral}
\end{equation}
where
\begin{equation}
   q^+_{\push}(r,E) \equiv 
   \sigma_0 \;
   \frac{1}{4 \, m_b} 
   \left( \frac{E}{m_e c^2} \right)^2 
   \frac{1}{4 \pi r^2} 
   \left( \frac{dL_{\nu_x}}{dE} \right)
   \mathcal{F}(r,E) ,
   \label{eq:push_qdot}
\end{equation}
with 
\begin{equation}
  \sigma_0 = \frac{4 G_F^2 \left(m_e c^2 \right)^2}{\pi \left( \hbar c \right)^4 } \approx 1.759 \times 10^{-44} {\rm cm^2} 
\end{equation}
being the typical neutrino cross-section, $m_b \approx 1.674 \times 10^{-24}{\rm g}$ an average baryon mass, 
and $(dL_{\nu_x}/dE)/(4 \pi r^2)$ the spectral energy flux for any single $\nu_x$ neutrino species with energy $E$. 
Note that all four heavy neutrino flavors are treated identically by the ASL scheme, and contribute equally to $Q^+_{\push}$ 
(see the factor 4 appearing in Equation (\ref{eq:push_integral})).

The term $\mathcal{F}(r,E)$ in Equation (\ref{eq:push_qdot}) defines the spatial location where $Q^+_{\push} (t,r)$ is active:
\begin{equation}
  \mathcal{F}(r,E) =
  \left\{
    \begin{array}{ll}
      0 & \mbox{if} \quad ds/dr > 0 \quad \mbox{or} \quad \dot{e}_{\nu_e,\overline{\nu}_e} < 0 \\
      \exp (- \tau_{\nu_e}(r,E)) & \mbox{otherwise} \\
      \end{array}
  \right. ,
  \label{eq:push_F}
\end{equation}
where $\tau_{\nu_e}$ denotes the (radial) optical depth of the electron neutrinos, $s$ is the matter entropy and $\dot{e}_{\nu_e,\bar{\nu}_e}$ the net specific energy rate due to electron neutrinos and anti-neutrinos. 
The two criteria above are a crucial ingredient in our description of triggering CCSN explosions: 
PUSH is only active where electron-neutrinos are heating ($\dot{e}_{\nu_e,\overline{\nu}_e} > 0$) and where neutrino-driven convection can occur ($ds/dr < 0$).

\begin{figure}[htp!]
   \includegraphics[width=0.5\textwidth]{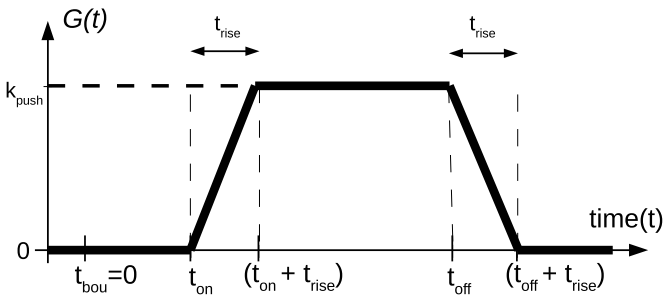}
   \caption{The function $\mathcal{G}(t)$ determines the temporal behavior of the heating due to PUSH.
   The quantity $t_{\rm on}$ is robustly set by multi-dimensional models. 
We consider a value of 80~ms in our calculations and a value of 120 ms for testing. 
$t_{\rm rise}$ and $k_{\rm push}$ are set by our calibration procedure, 
spanning a range from 50~ms to 250~ms, and from 0 (PUSH off) to $\sim$4, respectively.
Since we assume that the explosion takes place within the first second after core bounce, 
we use $t_{\rm off} = 1 {\rm s}$.
   \label{fig:g_factor}}
\end{figure}

The term $\mathcal{G}(t)$ in Equation~(\ref{eq:push_integral}) determines the temporal behaviour of $Q^+_{\push} (t,r)$.
Its expression reads
\begin{equation}
  \mathcal{G} (t) = k_{{\rm push}} \times
  \left\{
    \begin{array}{ll}
      0                                                           & t \leq t_{\rm{on}} \\
      \left( t-t_{\rm on} \right)/t_{\rm rise}                    & t_{\rm on} < t \leq t_{\rm on} + t_{\rm rise}  \\
      1                                                           & t_{\rm on} + t_{\rm rise}< t \leq t_{\rm off}  \\
      \left( t_{\rm off} + t_{\rm rise} - t \right)/t_{\rm rise}  & t_{\rm off} < t \leq t_{\rm off} + t_{\rm rise} \\
      0                                                           & t > t_{\rm off} + t_{\rm rise}
      \end{array}
  \right. ,
  \label{eq:push_G}
\end{equation}
and it is sketched in Figure \ref{fig:g_factor}. 
Note that throughout the article we always measure the time relative to bounce, if not noted otherwise.

The {\em cumulative energy} deposited by PUSH, $E_{\rm push}$, can be calculated from the {\em energy deposition rate} 
${\rm d}E_{\rm push}/{\rm d}t$ as
\begin{equation}
 E_{\rm push}(t) = \int_{t_{\rm on}}^t \: \left( \frac{{\rm d}E_{\rm push}}{{\rm d}t} \right) \, {\rm d}t' =
 \int_{t_{\rm on}}^t \: \left( \int_{V_{\rm gain}} \: Q_{\rm push}^+ \, \rho \, {\rm d}V \right) {\rm d}t' . \label{eq_epush}
\end{equation}
where $V_{\rm gain}$ is the volume of the gain region. 
Both these quantities have to be distinguished from the corresponding energy and energy rate obtained by IDSA:
\begin{equation}
 E_{\rm idsa}(t) = \int_{0}^{t} \: \left( \frac{{\rm d}E_{\rm idsa}}{{\rm d}t} \right) {\rm d}t' = 
 \int_{0}^{t} \: \left( \int_{V_{\rm gain}} \: \dot{e}_{\nu_e,\bar{\nu}_e} \, \rho \, {\rm d}V \right) {\rm d}t . \label{eq_eidsa}
\
\end{equation}

The definition of $\mathcal{G}(t)$ introduces a set of (potentially) free parameters:
\begin{itemize}
\item 
$k_{\rm push}$ is a global multiplication factor that controls directly the amount of extra heating provided by PUSH. The choices of $\sigma_0$ as reference cross-section and of the $\mu$ and $\tau$ neutrino luminosity as energy reservoir suggest $k_{\rm push} \gtrsim 1$.
\item 
$t_{\rm on}$ sets the time at which PUSH starts to act. We relate $t_{\rm on}$ to the time when deviations from spherically symmetric behavior appear in multi-dimensional models. Matter convection in the gain region sets in once the advection time scale $\tau_{\rm adv}$ and the convective growth time scale $\tau_{\rm conv}$ satisfy $\tau_{\rm adv} / \tau_{\rm conv} \gtrsim 3$ \citep{Foglizzo2006}. For all the models we have explored, this happens around $t = 0.06-0.08 \, {\rm s}$.  In the above estimates, $\tau_{\rm adv} = \dot{M}_{\rm shock}/M_{\rm gain}$, where $\dot{M}_{\rm shock}$ is the accretion rate at the shock and $M_{\rm gain}$ the mass in the gain region, and $\tau_{\rm conv} = f_{B-V}^{-1}$, where $f_{B-V}$ is the Brunt-V\"{a}is\"{a}la frequency.  Considering that $\tau_{\rm conv} \sim 4-5 \, {\rm ms}$, we expect $t_{\rm on} \sim 0.08-0.10 \, {\rm s}$, in
agreement with recent multi-dimensional simulations \citep[see, for example, the 1D-2D comparison of the 
shock position in][]{Bruenn.Mezzacappa.ea:2013}.
\item 
$t_{\rm rise}$ defines the time scale over which $\mathcal{G}(t)$ increases from zero to \kpush. We connect $t_{\rm rise}$ with the time scale that characterizes the growth of the largest multi-dimensional perturbations between the shock radius ($R_{\rm shock}$) and the gain radius ($R_{\rm gain}$) \citep[e.g.][]{janka96}. \cite{Foglizzo2006} showed that convection in the gain region can be significantly stabilized by advection, especially if $\tau_{\rm adv} / \tau_{\rm conv}$ only marginally exceeds 3, and that the growth rate of the fastest growing mode is diminished. Thus, $t_{\rm rise} \gg \tau_{\rm conv}$. On the other hand, a lower limit to $t_{\rm rise}$ is represented by the overturn time scale, $\tau_{\rm overturn}$, defined as
\begin{equation}
 \tau_{\rm overturn} \sim \frac{\pi (R_{\rm shock}-R_{\rm gain})}{\langle v \rangle_{\rm gain}} ,
\end{equation}
where $ \langle v \rangle_{\rm gain} $ is the average fluid velocity inside the gain region. In our simulations, we have found $\tau_{\rm overturn} \approx 0.05 \, {\rm s}$ around and after $t_{\rm on}$. In case of a contracting shock, SASIs
are also expected to develop around $0.2-0.3 \, {\rm s}$ after bounce \citep{Hanke2013}. Hence, we assume $0.05 \, {\rm s} \lesssim t_{\rm rise} \lesssim \left( 0.30 \, {\rm s} - t_{\rm on} \right) $.
\item
$t_{\rm off}$ sets the time after which PUSH starts to be switched off. We expect neutrino driven explosions to develop for $t \lesssim 1$~s due to the fast decrease of the luminosities during the first seconds after core bounce. Hence, we fix $t_{\rm off} = 1$~s. PUSH is not switched off suddenly at the onset of the explosion, but rather starts decreasing naturally even before 1~s after core bounce due to the decreasing neutrino luminosities 
and due to the rarefaction of the gain region above the PNS.
The subsequent injection of energy by neutrinos in the accelerating shock is qualitatively
consistent with multi-dimensional simulations, where accretion and explosion can coexist during the early stages of the shock expansion. The decrease of ${\rm d}E_{\rm push}/{\rm d}t$ 
on a time scale of a few hundreds of milliseconds after 
the explosion has been launched makes our results largely independent of the choice of $t_{\rm off}$ for explosions happening not too close to it.
\end{itemize}
While $t_{\rm on}$ is relatively well constrained and $t_{\rm off}$ is robustly set, $t_{\rm rise}$ and especially $k_{\rm push}$ are still undefined. We will discuss their impact on the model and on the explosion properties in detail in Sections \ref{sec:method_kpush}-\ref{sec:method_kpush_trise}. Ultimately, we will fix them using a calibration procure detailed in Section \ref{sec:method_sn1987a}.

\subsection{Post-processing analysis}
For the analysis of our results we determine several key quantities for each simulation. These quantities are obtained from a post-processing approach. We distinguish between the {\em explosion properties}, such as the explosion time, the mass cut,or the explosion energy, and the {\em nucleosynthesis yields}. The former are calculated from the hydrodynamics profiles. The latter are obtained from detailed nuclear network calculations for extrapolated trajectories.

\subsubsection{Accretion rates and explosion properties}
\label{sec:def_eexpl}
For the accretion process, we distinguish between the accretion rate at the shock front, $\dot{M}_{\rm shock} = {\rm d} M(R_{\rm shock}) / {\rm d} t$, and the accretion rate on the PNS, $\dot{M}_{\rm PNS} = {\rm d} M(R_{\rm PNS}) / {\rm d} t$. In these expressions, $M(R)$ is the baryonic mass enclosed in a radius $R$, $R_{\rm shock}$ is the shock radius, and $R_{\rm PNS}$ is the PNS radius that satisfies the condition $\rho(R_{\rm PNS})=10^{11}{\rm g \, cm^{-3}}$.

We consider the explosion time $t_{\rm expl}$ as the time when the shock reaches $500 \, {\rm km}$, measured with respect to core bounce (cf. \citet{Ugliano.Janka.ea:2012}). In all our models, the velocity of matter at the shock front has turned positive at that radius and the explosion has been irreversibly launched. There is no unique definition of $t_{\rm expl}$ in the literature and some other studies (cf. \citet{janka96,handy14}) use different definitions, e.g., the time when the explosion energy increases above $10^{48}$~erg. However, we do not expect that the different definitions give qualitatively different explosion times.

For the following discussion, we will use the total energy of the matter between a given mass shell $m_0$ up to the stellar surface:
\begin{equation}
  E_{\mathrm{total}}(m_0,t)=-\int_M^{m_0} \: e_{\rm total}(m,t) \, {\rm d}m \;.
\label{eq:total energy}
\end{equation}
$M$ is the enclosed baryonic mass at the surface of the star and $m_0$ is a baryonic mass coordinate ($ 0 \leq m_0 \leq M$). $e_{\rm total}$ is the specific total energy, given by 
\begin{equation}
 e_{\rm total}=e_{\rm int}+e_{\rm kin}+e_{\rm grav} \; ,
\end{equation}
i.e., the sum of the (relativistic) internal, kinetic, and gravitational specific energies. For all these quantities we make use of the general-relativistic expressions in the laboratory frame \citep{fischer10}. The integral in Equation~(\ref{eq:total energy}) includes both the portion of the star evolved in the hydrodynamical simulation and the outer layers, which are considered as stationary profiles from the progenitor structure.

The explosion energy emerges from different physical contributions (see, for example, the appendix of \cite{Scheck2006} and the discussion in \cite{Ugliano.Janka.ea:2012}). In our model, we are taking into account:
(i) the total energy of the neutrino-heated matter that causes the shock revival;
(ii) the nuclear energy released by the recombination of nucleons and alpha particles into heavy nuclei at the transition to non-NSE conditions;
(iii) the total energy associated with the neutrino-driven wind developing after the explosion up to the end of the simulation;
(iv) the energy released by the explosive nuclear burning in the shock-heated ejecta; and
(v) the total (negative) energy of the outer stellar layers (also called the ``overburden'').
We are presently not taking into accout the variation of the ejecta energy due to the appearance of late-time fallback. This is justified as long as the fallback represents only a small fraction of the total ejected mass.

To compute the explosion energy, we assume that the total energy of the ejecta with rest-masses subtracted eventually converts into kinetic energy of the expanding supernova remnant at $t \gg t_{\rm expl}$. The quantity $e_{\rm total}$ includes the rest mass contribution via $e_{\rm int}$, see Equation~(\ref{eqn:e_th}). Instead, if we want to calculate the explosion energy, we have to consider the thermal energy $e_{\rm th}$.  Therefore, we define the specific explosion energy as
\begin{equation}
 e_{\rm expl}=e_{\rm th}+e_{\rm kin}+e_{\rm grav} \; ,
\end{equation}
and the time- and mass-dependent explosion energy for the fixed mass domain between $m_0$ and $M$ as
\begin{equation}
  H_{\mathrm{expl}}(m_0,t)=-\int_M^{m_0} \: e_{\rm expl}(m,t) \, {\rm d}m \;.
\label{eq:expl energy mt}
\end{equation}
This can be interpreted as the total energy of this region in a non-relativistic EOS approach, where rest masses are not included. 

The actual explosion energy (still time-dependent) is given by
\begin{equation}
  E_{\rm{expl}}(t) = H_{\rm{expl}}(m_{\rm cut}(t),t) \; .
\label{eq:expl energy t}
\end{equation}
i.e., for the matter above the mass cut $m_{\rm cut}(t)$.

To identify the mass cut, we consider the expression suggested by Bruenn in \citet{fischer10}:
\begin{equation}
  m_{\rm{cut}}(t) = m\left(\max(H_{\rm{expl}}(m,t))\right) \, ,
\label{eq:mass cut}
\end{equation}
where the maximum is evaluated outside the homologous core ($m \gtrsim 0.6$~\msun), which has large positive values of the specific explosion energy $e_{\rm expl}$ once the PNS has formed due to the high compression. In the outer stellar envelope, before the passage of the shock wave, $e_{\rm expl}$ is dominated by the negative gravitational contribution.  However, it is positive in the neutrino-heated region and in the shocked region above it.  Hence, the above definition of $m_{\rm cut}$ locates essentially the transition from gravitationally unbound to bound layers. The final mass cut is obtained for $t=t_{\rm final}$.

Our final simulation time $t_{\rm final} \gtrsim 4.6$~s is always much larger than the explosion time and, as we will show later, it allows $E_{\rm expl}(t)$ to saturate. Thus, we consider $E_{\rm{expl}}(t=t_{\rm final})$ as the ultimate explosion energy of our models. In the following, if we use $E_{\rm expl}$ without the time as argument, we mean this final explosion energy.

\subsubsection{Nucleosynthesis yields}
\label{sec:winnet}
To predict the composition of the ejecta, we perform nucleosynthesis calculations using the full nuclear network {\sc Winnet} \citep{Winteler.ea:2012}. We include isotopes up to $^{211}$Eu covering the neutron-deficient as well as the neutron-rich side of the valley of $\beta$-stability. The reaction rates are the same as in \citet{Winteler.ea:2012}. They are based on experimentally known rates where available and predictions otherwise. The n-, p-, and alpha-captures are taken from \citet{Rauscher.FKT:2000}, who used known nuclear masses where available and the \textit{Finite Range Droplet Model} \citep{Moller.ea:1995} for unstable nuclei far from stability. The $\beta$-decay rates are from the nuclear database \textit{NuDat2}\footnote{http://www.nndc.bnl.gov/nudat2/}.

We divide the ejecta into different mass elements of $10^{-3}$~\msun each and follow the trajectory of each individual mass element.  As we are mainly interested in the amounts of $^{56}$Ni, $^{57}$Ni, $^{58}$Ni, and $^{44}$Ti, we only consider the 340 innermost mass elements above the mass cut, corresponding to a total mass of $0.34$~M$_{\odot}$. The contribution of the outer mass elements to the production of those nuclei is negligible. 

For $t<t_{\rm final}$, we use the temperature and density evolution from the hydrodynamical simulations as inputs for our network. For each mass element we start the nucleosynthesis post-processing when the temperature drops below 10~GK, using the NSE abundances (determined by the current electron fraction $Y_e$) as the initial composition.  For mass elements that never reach 10~GK we start at the moment of bounce and use the abundances from the approximate $\alpha$-network at this point as the initial composition. Note that for all tracers the further evolution of $Y_e$ in the nucleosynthesis post-processing is determined inside the {\sc Winnet} network.

At the end of the simulations, i.e.\ $t=t_{\rm final}$, the temperature and density of the inner zones are still sufficiently high for nuclear reactions to occur ($T \approx 1$~GK and $\rho \approx 2.5 \times 10^3$~g~cm$^{-3}$). Therefore, we extrapolate the radius, density and temperature up to $t_{\rm end} = 100$~s using:
\begin{align}
  r(t) &= r_{\rm final}  + t v_{\rm final} \label{eq:extrapol_rad} \\
  \rho(t) &= \rho_{\rm final} \left( \frac{t}{t_{\rm final}} \right)^{-3} \\
  T(t) &= T[s_{\rm final},\rho(t),Y_e(t)] \label{eq:extrapol_t9},
\end{align}
where $r$ is the radial position, $v$ the radial velocity, $\rho$ the density, $T$ the temperature, $s$ the entropy per baryon, and $Y_e$ the electron fraction of the mass zone.  The temperature is calculated at each timestep using the equation of state of \citet{Timmes.Swesty:2000}. The prescription in Equations (\ref{eq:extrapol_rad})--(\ref{eq:extrapol_t9}) corresponds to a free expansion for the density and an adiabatic expansion for the temperature (see, for example, \citet{korobkin2012}).

\section{Fitting and Results}
\label{sec:results}

To test the PUSH method, we perform a large number of runs where we vary the free parameters and explore their impact on the explosion properties.  We also analyze in detail the basic features of the simulations and of the explosions in connection with the properties of the progenitor star.  Finally, we fit the free parameters in the PUSH method to reproduce observed properties of SN~1987A for a progenitor star in the range 18-21~\msun.

\subsection{General effects of free parameter variations}
\subsubsection{$k_{\rm push}$}
\label{sec:method_kpush}

The parameter with the most intuitive and strongest impact on the explosion is \kpush. Its value directly affects the amount of extra heating which is provided by PUSH.  As expected, larger values of \kpush (assuming all other parameters to be fixed) result in the explosion being more energetic and occurring earlier.  In addition, a faster explosion implies a lower remnant mass, as there is less time for the accretion to add mass to the forming PNS.

\begin{figure}[htp!]
   \includegraphics[width=0.49\textwidth]{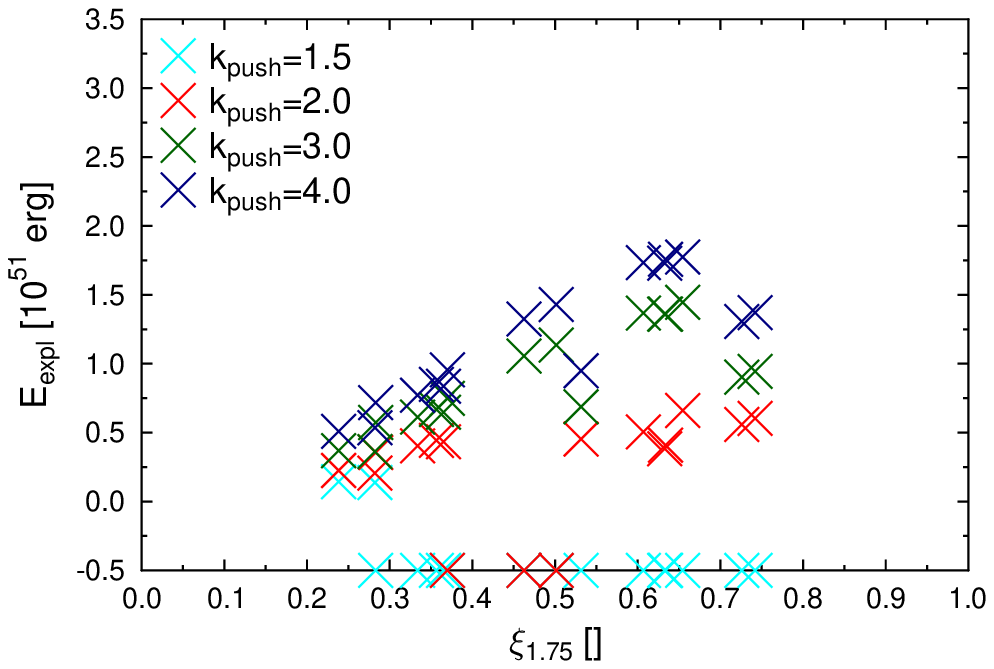} \\
   \includegraphics[width=0.49 \textwidth]{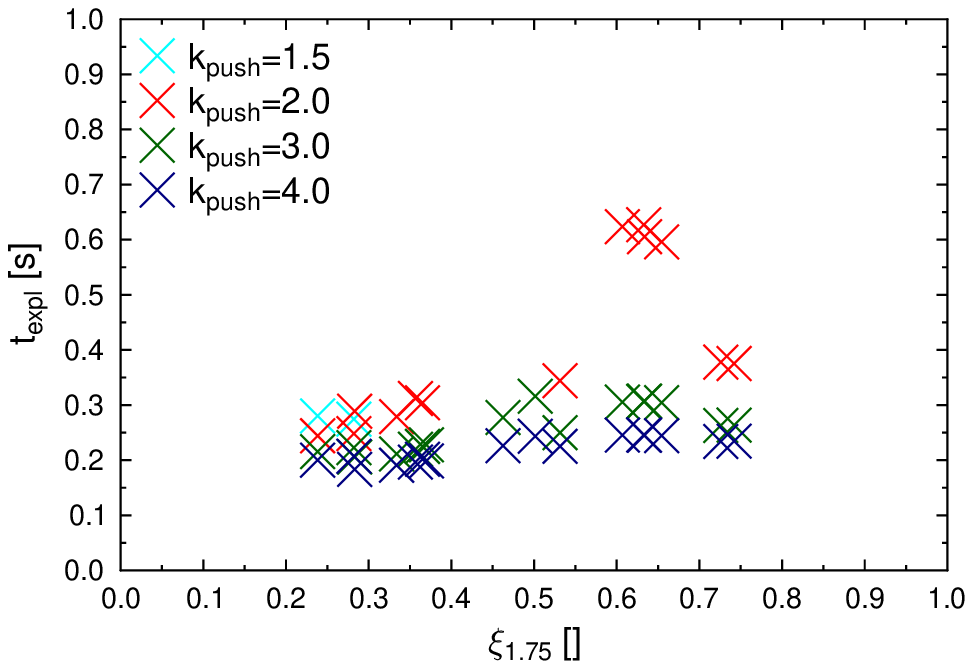} \\
   \includegraphics[width=0.49\textwidth]{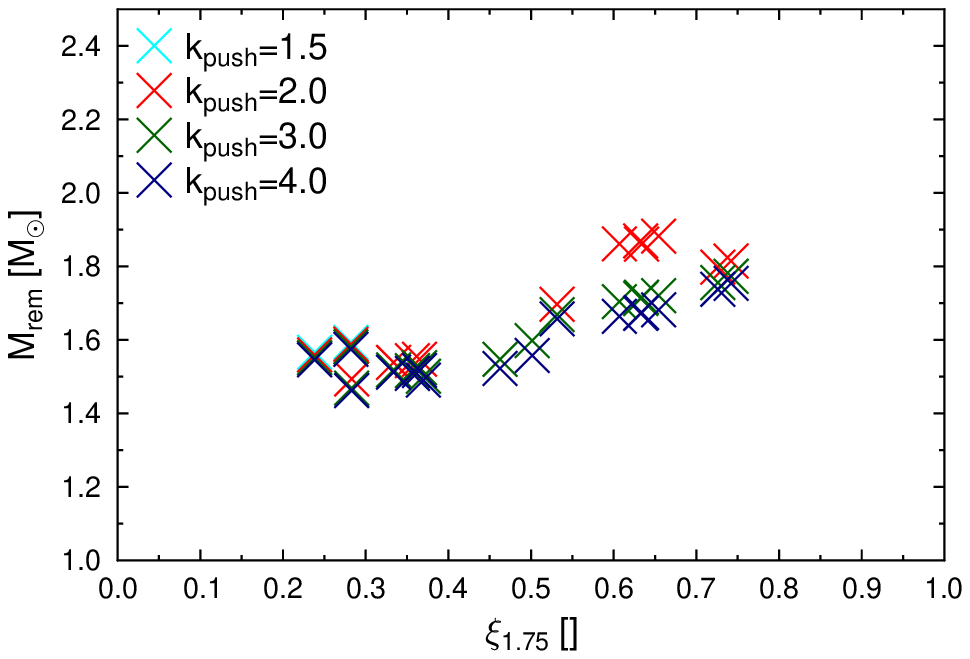}
   \caption{Explosion energies (top), explosion times (middle), and (baryonic) remnant mass (bottom) as function of compactness for \kpush $1.5$, $2.0$, $3.0$, and $4.0$, and fixed $t_{\rm rise}=0.15$~s for all progenitor models included in this studay (ZAMS mass between 18.0 and 21.0~\msun).  Non-exploding models are indicated with $E_{\rm expl}=-0.5$~B in the top panel and are omitted in the other panels.
   \label{fig:kpush} }
\end{figure}

Beyond these general trends with \kpush, the detailed behavior depends also on the compactness of the progenitor.  For all 16~progenitor models in the 18-21~\msun ZAMS mass range, we have explored several PUSH models, varying $k_{\rm push}$ between 0.0 and 4.0 in increments of 0.5 but fixing $t_{\rm on} = 80$~ms and $t_{\rm rise} = 150$~ms.  For $k_{\rm push}\leqslant 1$, none of the models explode and for $k_{\rm push} = 1.5$ only the lowest compactness models explode.  Figure \ref{fig:kpush} shows the explosion energy, the explosion time and the (baryonic) remnant mass as function of the progenitor compactness for $k_{\rm push} = 1.5,2.0,3.0,4.0$.
A distinct behavior between low and high compactness models is seen. The LC models ($\xi_{1.75}<0.4$) result in slightly weaker and faster explosions, with less variability in the explosion energy and in the explosion time for different values of \kpush.  Even for relatively large values of \kpush, the explosion energies remain below 1~Bethe (1~Bethe, abbreviated as 1~B, is equivalent to $10^{51}$~erg). On the other hand, the HC models ($\xi_{1.75}>0.45$) explode stronger and later, with a larger variation in the explosion properties.  In this case, for high enough values of \kpush ($\gtrsim 3.0$), explosion energies of $\gtrsim 1$~Bethe can be obtained.
The HC models also lead to a larger variability of the remnant masses, even though this effect is less pronounced than for the explosion time or energy. For the values of \kpush used here, we obtain (baryonic) remnant masses from approximately 1.4 to 1.9~\msun. The differences of LC and HC models will be investigated further in Section \ref{sec:hc and lc}.

There are three models with $0.37 \lesssim \xi_{1.75} \lesssim 0.50 $ (corresponding to ZAMS masses of 18.0 (HC), 18.2 (LC), and 19.4~M$_{\odot}$ (HC)) which do not follow the general trend. In particular, we find the threshold value of $k_{\rm push}$ for successful explosions to be higher for these models.  A common feature of these three models is that they have the lowest Fe-core mass of all the models in our sample and the highest central densities at the onset of collapse.

The choice of \trise does not affect the observed trends with \kpush: similar behaviors are also seen for $50\, {\rm ms} \lesssim t_{\rm rise} \lesssim  250 \, {\rm ms}$.

\subsubsection{$t_{\rm on}$}
\label{sec:method_ton}
To test the sensitivity of our method to the parameter $t_{\rm on}$, we compute models with $k_{\rm push} = 2.0$ and $t_{\rm rise} = 0.15 \, {\rm s}$ for a very large onset parameter, $t_{\rm on} = 120 \, {\rm ms}$.  We compare the corresponding results with the ones obtained for $t_{\rm on} = 80 \, {\rm ms}$.  As expected, the shock revival happens slightly later (with a temporal shift of $\sim 30 \, {\rm ms}$), the explosion energies are smaller (by $\sim 0.05$~B) and the remnant masses are marginally larger (by 0.08~\msun).  However, all the qualitative behaviours described above, as well as the distinction between high and low compactness models, do not show any dependence on $t_{\rm on}$.  In the following, we will always assume $t_{\rm on}=80 \, {\rm ms}$.

\subsubsection{$k_{\rm push}$ \& $t_{\rm rise}$}
\label{sec:method_kpush_trise}

\begin{figure}[htp!]
   \includegraphics[width=0.49\textwidth]{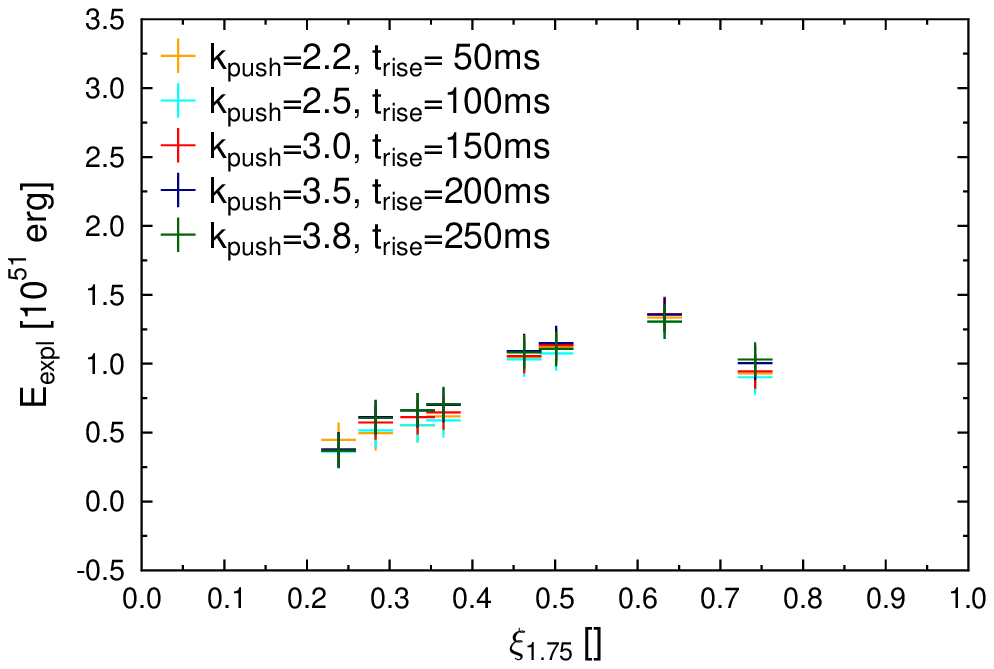} \\
   \includegraphics[width=0.49 \textwidth]{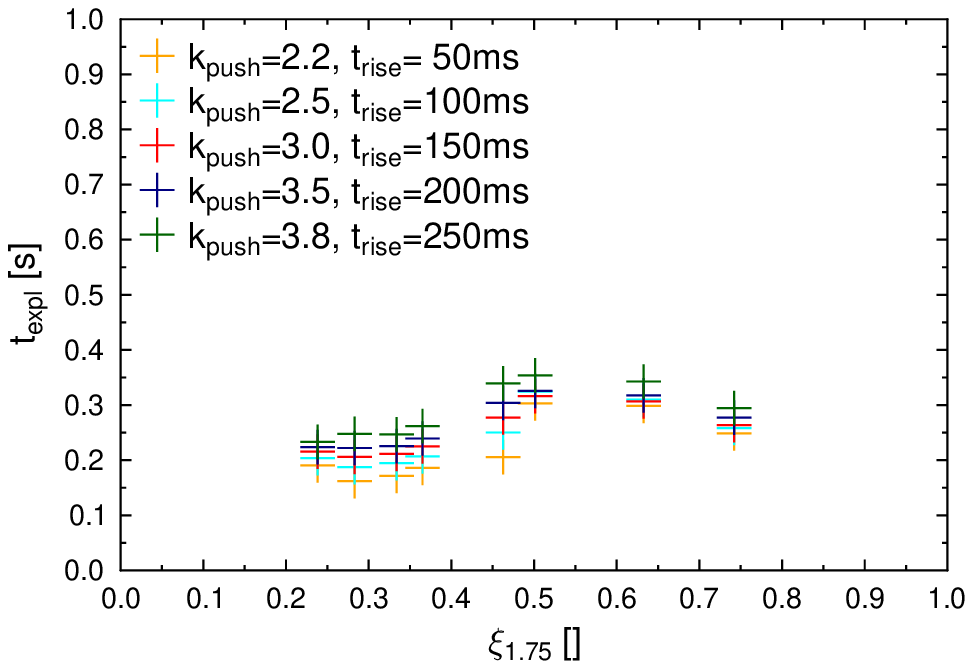} \\
   \includegraphics[width=0.49\textwidth]{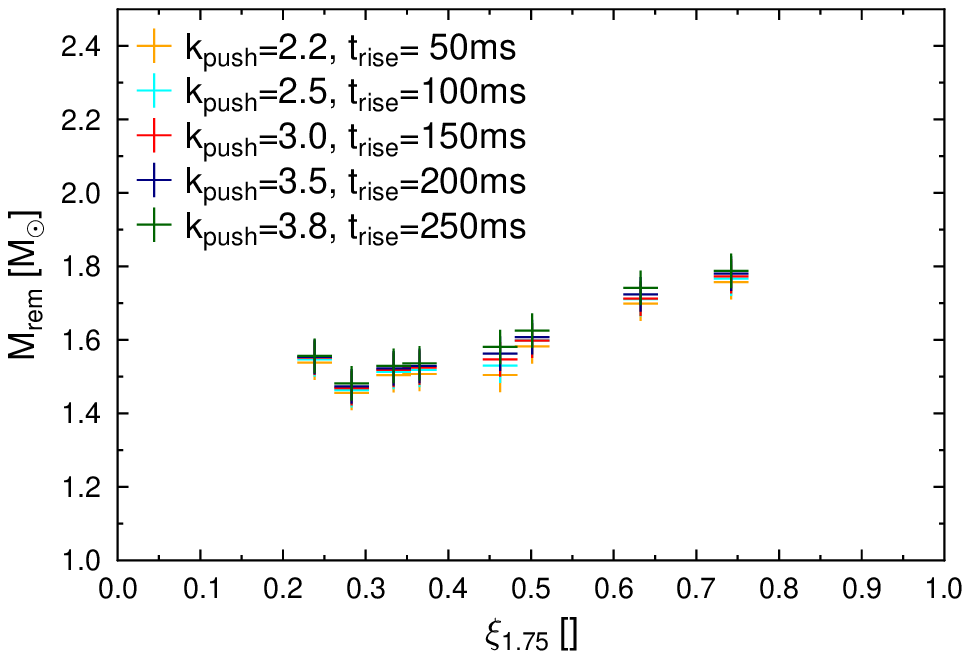}
   \caption{Explosion energies (top), explosion times (middle), and (baryonic) remnant mass (bottom) as function of compactness for pairs of \kpush and \trise, and for the progenitor models with ZAMS mass 18.0, 18.6, 19.2, 19.4, 19.8, 20.0, 20.2, and 20.6~\msun.
   \label{fig:kpush_trise}
}
\end{figure}

\begin{figure}[htp!]
   \includegraphics[width=0.35\textwidth,angle=-90]{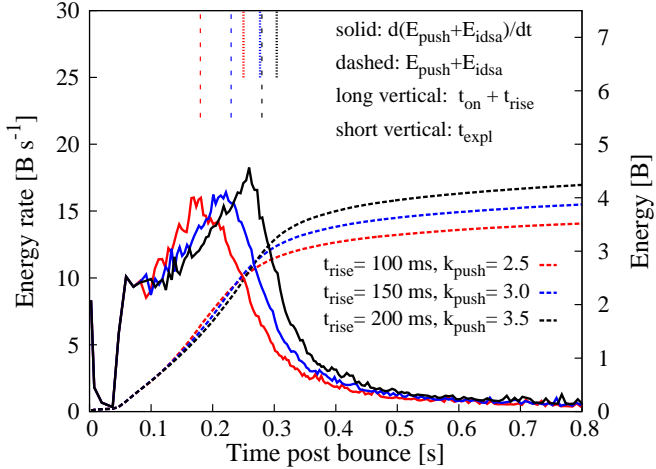}
   \caption{Temporal evolution of the total neutrino energy contribution inside the gain region $( E_{\rm push} + E_{\rm idsa} )$ (solid lines) and of its time derivative (dashed lines), for four runs with the same ZAMS progenitor mass (18.0~\msun), but different combinations of PUSH parameters \trise and \kpush. For each run, the vertical lines correspond to $t = t_{\rm on}+t_{\rm rise}$ (long, dashed) and to $t_{\rm expl}$ (short, dot-dashed).
   \label{fig:energy trise} }
\end{figure}

In Sections \ref{sec:method_kpush} and \ref{sec:method_ton}, we have investigated the dependency of the model on the single parameters \kpush and $t_{\rm on}$. Now, we explore the role of \trise\ in combination with \kpush. For this, we approximately fix the explosion energy to the canonical value of $\sim 1$~B for the high compactness models (corresponding, for example, to the previously examined models with $k_{\rm push} = 3.0$ and $t_{\rm rise} = 150 \, {\rm ms}$), and investigate which other combinations of \kpush and \trise result in the desired explosion energy.  We restrict our explorations to a sub-set of progenitor models (18.0~\msun, 18.6~\msun, 19.2~\msun, 19.4~\msun, 19.8~\msun, 20.0~\msun, 20.2~\msun and 20.6~\msun) that spans the $\xi_{1.75}$-range of all 16 progenitors.  Figure~\ref{fig:kpush_trise} summarizes the explosion energies, explosion times, and remnant masses for various combinations of \kpush and \trise for progenitors of different compactness. The required constraint  can be 
obtained by several combinations of parameters, which lie on a curve in the \kpush-\trise plane.  As a general result, a longer \trise requires a larger \kpush to obtain the same explosion energy. This can be understood from the different roles of the two parameters: while \kpush sets the maximum efficiency at which PUSH deposits energy from the reservoir represented by the $\nu_{\mu,\tau}$ luminosity, \trise sets the time scale over which the mechanism reaches this maximum.  Together, they control the slope of $\mathcal{G}(t)$ in the rising phase (see Figure~\ref{fig:g_factor}).  A model with a longer rise time reaches its maximum efficiency later, at which time the luminosities have already decreased and a part of the absorbed energy has been advected on the PNS or re-emitted in the form of neutrinos. To compensate for these effects, a larger \kpush is required for a longer \trise.  This is seen in Figure~\ref{fig:energy trise}, where we plot the cumulative neutrino contribution $( E_{\rm push}+E_{\rm idsa}
 )$ and its time derivative for four runs of the 18.0~\msun progenitor model, but with different combinations of \trise and \kpush.  Runs with larger parameter values require PUSH to deposit more energy (see $ ( E_{\rm push}+E_{\rm idsa})$ at $t \approx t_{\rm expl}$), and the corresponding deposition rates are shifted towards later times.  Moreover, for increasing values of $t_{\rm rise}$, the explosion time $t_{\rm expl}$ becomes larger, but the interval between $(t_{\rm on}+t_{\rm rise})$ and $t_{\rm expl}$ decreases.  Despite the significant variation of $k_{\rm push}$ between different runs, the peak values of ${\rm d}( E_{\rm push}+E_{\rm idsa} )/{\rm d} t$ at the onset of the shock revival that preceeds the explosion are very similar in all cases.

\subsubsection{$t_{\rm off}$}

Even though PUSH is active up to $t_{\rm off}+t_{\rm rise} \gtrsim 1 \, {\rm s}$, its energy deposition reduces progressively on a timescale  of a few 100~ms after the explosion has set in (see Figure~\ref{fig:energy trise}). This shows explicitly that the value of $t_{\rm off}$ does not have important consequences in our simulations, at least as long as we have typical explosion times well below one second. The observed decrease of the PUSH energy deposition rate after the launch of the explosion will be explained in Section~\ref{sec:hc and lc}.

\subsection{Contributions to the explosion energy}
\label{sec:eexpl_contr}
In the following, we discuss the contributions to and the sources of the explosion energy, i.e., we investigate how the explosion energy is generated. This is done in several steps: first, we have a closer look at the neutrino energy deposition. Then we show how it relates to the increase of the total energy of the ejected layers, and finally how this increase of the total energy transforms into the explosion energy. For this analysis, we have chosen the 19.2 and 20.0~\msun ZAMS mass progenitor models as representatives of the HC and LC samples, respectively. We consider their exploding models obtained with $t_{\rm on}=80$~ms, $t_{\rm rise}=150$~ms, and $k_{\rm push}=3.0$. A summary of the explosion properties can be found in Table \ref{tab:HC vs LC}.

\begin{table}[h,t]  
\begin{center}
\caption{Explosion properties for two reference runs \label{tab:HC vs LC}}
\begin{tabular}{lccc}
  \tableline \tableline
  Quantity                                      &         & HC        & LC          \\
  \tableline
  ZAMS 			                         & (\msun) & 19.2      & 20.0        \\
  $\xi_{1.75}$       		                 & (-)     & 0.637     & 0.283       \\
  $t_{\rm on}$       	                         & (ms)    & \multicolumn{2}{c}{80}  \\
  $t_{\rm rise}$         	                 & (ms)    & \multicolumn{2}{c}{150} \\
  $k_{\rm push}$     	  	                 & (-)     & \multicolumn{2}{c}{3.0} \\
  $t_{\rm expl}$    	                         & (ms)    & 307       & 206         \\ 
  $M_{\rm remn}$                                & (\msun) & 1.713     & 1.469       \\
  $E_{\rm expl}$ ($t_{\rm final}$)              & (B)     & 1.36      & 0.57        \\
  $E_{\rm push}$ ($t_{\rm off} + t_{\rm rise}$) & (B)     & 3.51      & 1.08        \\
  $E_{\rm idsa}$ ($t_{\rm off} + t_{\rm rise}$) & (B)     & 2.76      & 1.01        \\
  $E_{\rm idsa}$ ($t_{\rm final}$)              & (B)     & 4.10      & 2.11        \\
  \tableline
\end{tabular}
\tablecomments{These two runs are used to compare the HC and LC samples.}
\end{center}
\end{table}

The table shows that for both models neutrinos are required to deposit a net cumulative energy $( E_{\rm push}+E_{\rm idsa} )$ much larger than $E_{\rm expl}$ to revive the shock and to lead to an explosion that matches the expected energetics. For the two reference runs, when the PUSH contribution is switched off ($t = t_{\rm off}+t_{\rm rise}$), the cumulative deposited energy is $\sim 4$ times larger than $E_{\rm expl}$. This can also be inferred from Figure~\ref{fig:energy trise} for other runs. That ratio increases further up to $\sim 5.5$ at $t = t_{\rm final}$, due to the neutrino energy deposition happening at the surface of the PNS which generates the $\nu$-driven wind. According to Equations~(\ref{eq_epush}) and (\ref{eq_eidsa}), $E_{\rm push}$ and $E_{\rm idsa}$ are the total energies which are deposited in the (time-dependent) gain region. This neutrino energy deposition increases the internal energy of the matter flowing in that region. However, since the advection timescale is much shorter than 
the explosion timescale, a large fraction of this energy is advected onto the PNS surface by the accreting mass before the explosion sets in, and hence does not contribute to the explosion energy. Only the energy deposited by neutrinos in the region above the final mass cut will eventually contribute to the explosion energy.

\begin{figure*}[htp!]  
\begin{tabular}{cc}
 \includegraphics[width=0.35\textwidth,angle=-90]{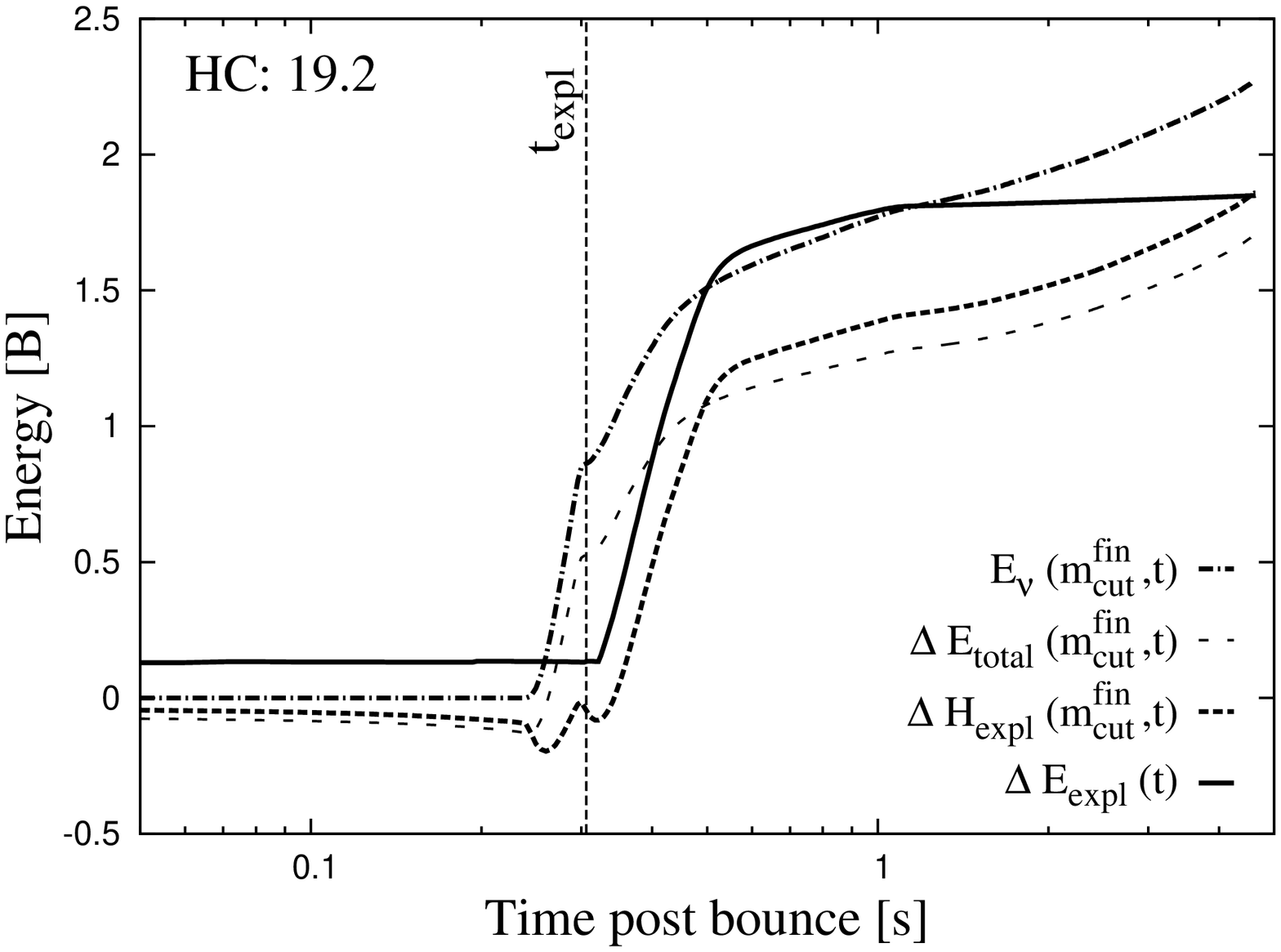}
 \includegraphics[width=0.35\textwidth,angle=-90]{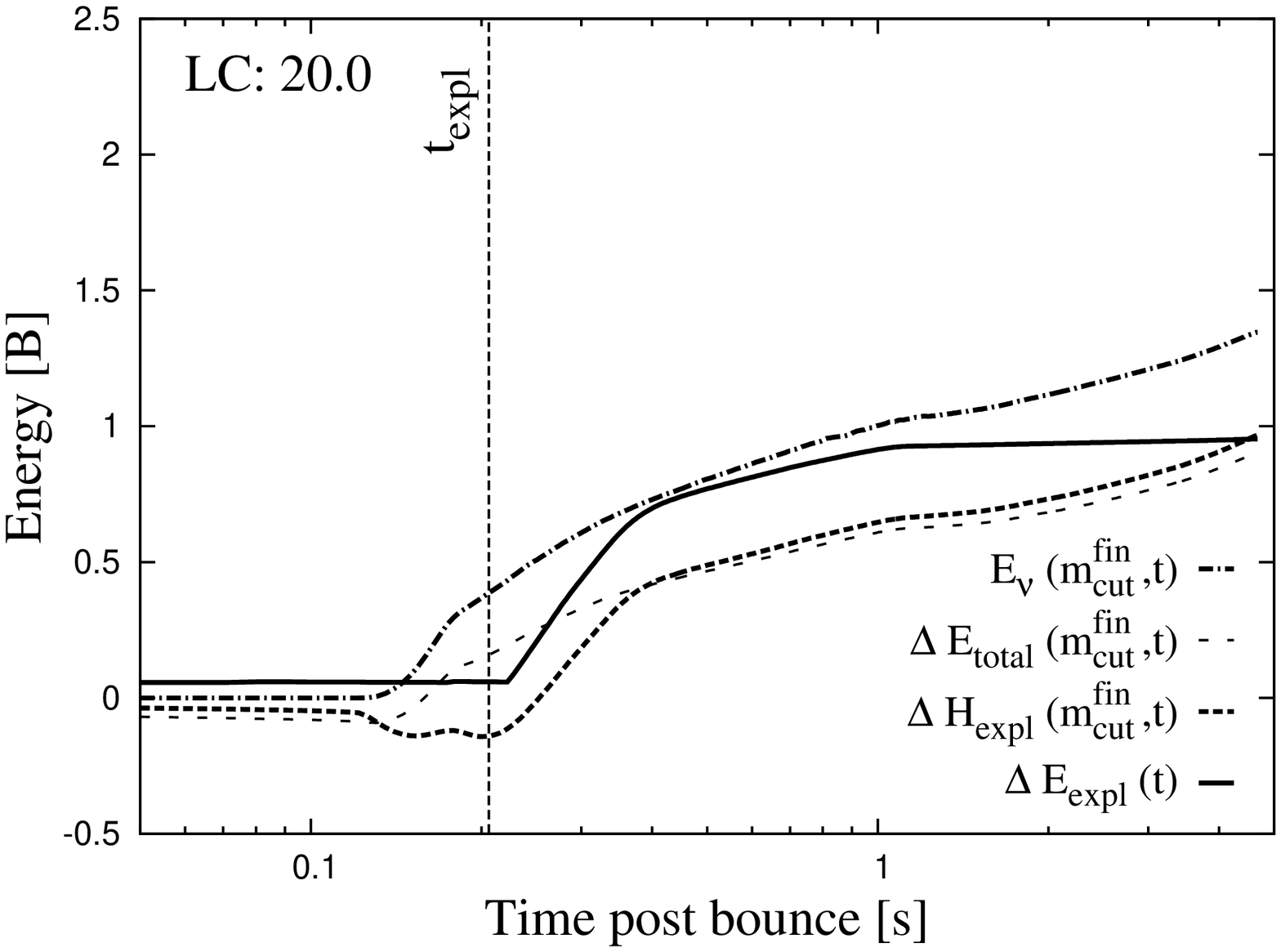}
\end{tabular}
 \caption{Time evolution of the time- and mass-integrated variation of the total energy $\Delta E_{\rm total}$ (thin dashed line), of the neutrino net deposition energy $E_{\nu}$ (dot-dashed line) and of the explosion energy for a fixed domain $\Delta H_{\rm expl}$ (thick dashed line), above $m_{ \rm cut}^{\rm fin} = m_{\rm cut}(t_{\rm final})$, for the HC (left) and for the LC (right) reference runs 
 reported in Table \ref{tab:HC vs LC}. The evolution of the time-dependent explosion energy, $\Delta E_{\rm expl}$, is also shown (solid line). 
 Both $\Delta H_{\rm expl}$ and 
 $\Delta E_{\rm expl}$ are computed with respect to $H_{\rm expl}(m_{\rm cut}^{\rm fin},t_{\rm initial})$. The difference between 
 $\Delta E_{\rm total}(m_{\rm cut}^{\rm fin},t)$ and $E_\nu$ represents the mechanical work, $E_{\rm mech}$; the difference between 
 $\Delta E_{\rm total}(m_{\rm cut}^{\rm fin},t)$ and $\Delta  H_{\rm expl}(m_{\rm cut}^{\rm fin},t)$ represents the released rest-mass energy, 
 $-\Delta E_{\rm mass}$.
\label{fig:total energy contributions} }
\end{figure*}

To identify this \textit{relevant} neutrino contribution, in Figure~\ref{fig:total energy contributions} we show the time evolution of the integrated net neutrino energy deposition $E_{\nu}(m_{ \rm cut}^{\rm fin},t)$ within the domain above the fixed mass $m_{ \rm cut}^{\rm fin} = m_{\rm cut}(t_{\rm final})$. We choose $m_{ \rm cut}^{\rm fin}$ to include all the relevant energy contributions to the explosion energy, up to the end of our simulations. Despite the significant differences in magnitudes, the two models show overall similar evolutions. If we compare $E_{\nu}(m_{ \rm cut}^{\rm fin},t)$ at late times with $( E_{\rm push}(t_{\rm off}+t_{\rm rise})+E_{\rm idsa}(t_{\rm final}) )$ from Table \ref{tab:HC vs LC}, we see that it is significantly smaller. About two thirds of the energy originally deposited in the gain region are advected onto the PNS and hence do not contribute to the explosion energy.

In addition to the neutrino energy deposition, in Figure~\ref{fig:total energy contributions} we also show the variation of the total energy for the domain above $m_{ \rm cut}^{\rm fin}$, i.e., $\Delta E_{\rm total}(m_{ \rm cut}^{\rm fin},t) = E_{\rm total}(m_{ \rm cut}^{\rm fin},t) - E_{\rm total}(m_{ \rm cut}^{\rm fin}, t_{\rm initial})$, where $t_{\rm initial}$ is the time when we start our simulation from the stage of the progenitor star. The variation of the total energy can be separated into the net neutrino contribution and the mechanical work at the inner boundary, $\Delta E_{\rm total}=E_{\nu}+E_{\rm mech}$. We note that in our general relativistic approach the variation of the gravitational mass due to the intense neutrino emission from the PNS is consistently taken into accout. It is visible in Figure~\ref{fig:total energy contributions}, that the net deposition by neutrinos makes up the largest part of the change of the total energy. The transfer of mechanical energy $E_{\rm mech}$ is negative 
because of the expansion work performed by the inner boundary during the collapse and the PNS shrinking. However it is significantly smaller in magnitude than $E_{\nu}$. 

Next, we investigate the connection between the variation of the total energy and the explosion energy. In Figure~\ref{fig:total energy contributions}, we show the variation of the explosion energy above the fixed mass $m_{ \rm cut}^{\rm fin}$, i.e. $\Delta H_{\rm expl}(m_{ \rm cut}^{\rm fin},t)=H_{\rm expl}(m_{ \rm cut}^{\rm fin},t) - H_{\rm expl}(m_{ \rm cut}^{\rm fin}, t_{\rm initial})$, together with the relative variation of the time-dependent explosion energy, $\Delta E_{\rm expl}(t) = E_{\rm expl}(t) - H_{\rm expl}(m_{ \rm cut}^{\rm fin}, t_{\rm initial})$. It is obvious from Equations~(\ref{eqn:e_th}), (\ref{eq:total energy}), and (\ref{eq:expl energy mt}) that the difference between $\Delta H_{\rm expl}(m_{ \rm cut}^{\rm fin},t)$ and $\Delta E_{\rm total}(m_{ \rm cut}^{\rm fin},t)$ is given by the variation of the integrated rest mass energy, $\Delta H_{\rm expl}(m_{ \rm cut}^{\rm fin},t)=\Delta E_{\rm total}(m_{ \rm cut}^{\rm fin},t)-\Delta E_{\rm mass}(m_{ \rm cut}^{\rm fin},t)$. 
In Figure~\ref{fig:total energy contributions}, $-\Delta E_{\rm mass}(m_{ \rm cut}^{\rm fin},t)$ can thus be identified as the difference between the long-thin and the short-thick dashed lines. We find that the overall rest mass contribution to the final explosion energy is positive, but much smaller than the neutrino contribution. Figure~\ref{fig:total energy contributions} also makes evident the conceptual difference between $H_{\rm expl}$ and $E_{\rm expl}$, and, at the same time, shows that $H_{\rm expl}(m_{ \rm cut}^{\rm fin},t) \rightarrow E_{\rm expl}(t)$ for $t \rightarrow t_{\rm final}$, since we have chosen $m_{ \rm cut}^{\rm fin}=m_{\rm cut}(t_{\rm final})$. It also reveals that the explosion energy $E_{\rm expl}$ has practically saturated for $t \gtrsim 1 \, {\rm s}$, while $E_{\nu}$ (and, consequently, $\Delta E_{\rm tot}$ and $\Delta H_{\rm expl}$) increases up to $t_{\rm final}$, when $m_{ \rm cut}^{\rm fin}$ is finally ejected. However, this energy provided by neutrinos is mostly spent to 
unbind matter from 
the PNS surface. Thus, the late $\nu$-driven wind, which occurs for several seconds after 1 s, still increases $E_{\rm expl}$, but at a relative small, decreasing rate.

To summarize, the variation of the explosion energy above $m_{ \rm cut}^{\rm fin}$ can be expressed as
\begin{eqnarray}
 \Delta H_{\rm expl}(m_{ \rm cut}^{\rm fin},t)&=&\Delta E_{\rm total}(m_{ \rm cut}^{\rm fin},t)-\Delta E_{\rm mass}(m_{ \rm cut}^{\rm fin},t) \nonumber \\
&=& E_{\nu}(m_{ \rm cut}^{\rm fin},t)+E_{\rm mech}(m_{ \rm cut}^{\rm fin},t)-\Delta E_{\rm mass}(m_{ \rm cut}^{\rm fin},t) \; . \nonumber \\
&&
\label{eq_deltaeexp}
\end{eqnarray}
The quantity $-\Delta E_{\rm mass}$ is positive, but significantly smaller than $E_{\nu}(m_{ \rm cut}^{\rm fin},t)$. $E_{\rm mech}$ is negative and also smaller than $E_{\nu}(m_{ \rm cut}^{\rm fin},t)$. Therefore, we conclude that in our models the explosion energy is mostly generated by the energy deposition of neutrinos in the eventually ejected layers, especially within the first second after bounce.

\begin{figure*}[htp!]  
\begin{tabular}{cc}
 \includegraphics[width=0.35\textwidth,angle=-90]{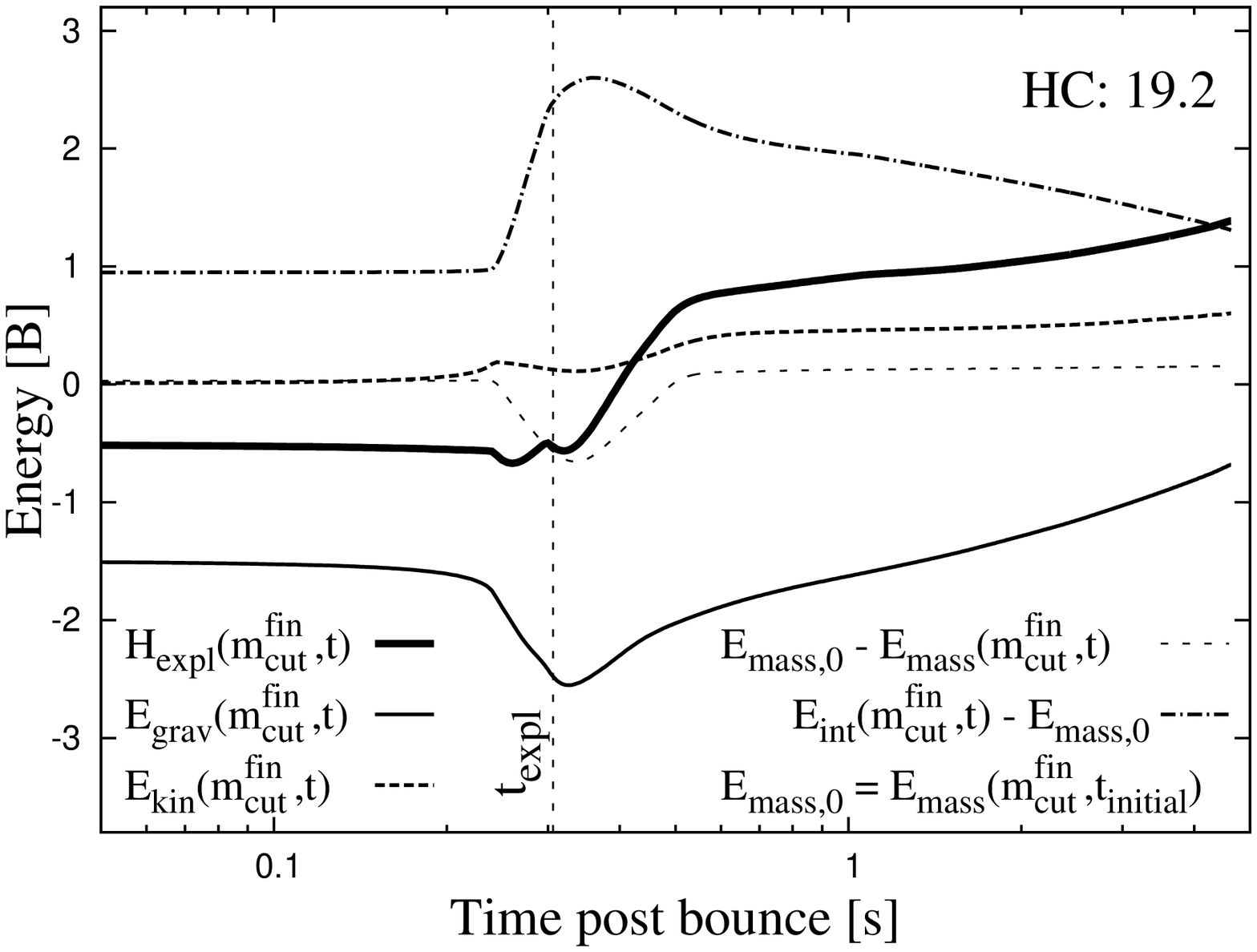}
 \includegraphics[width=0.35\textwidth,angle=-90]{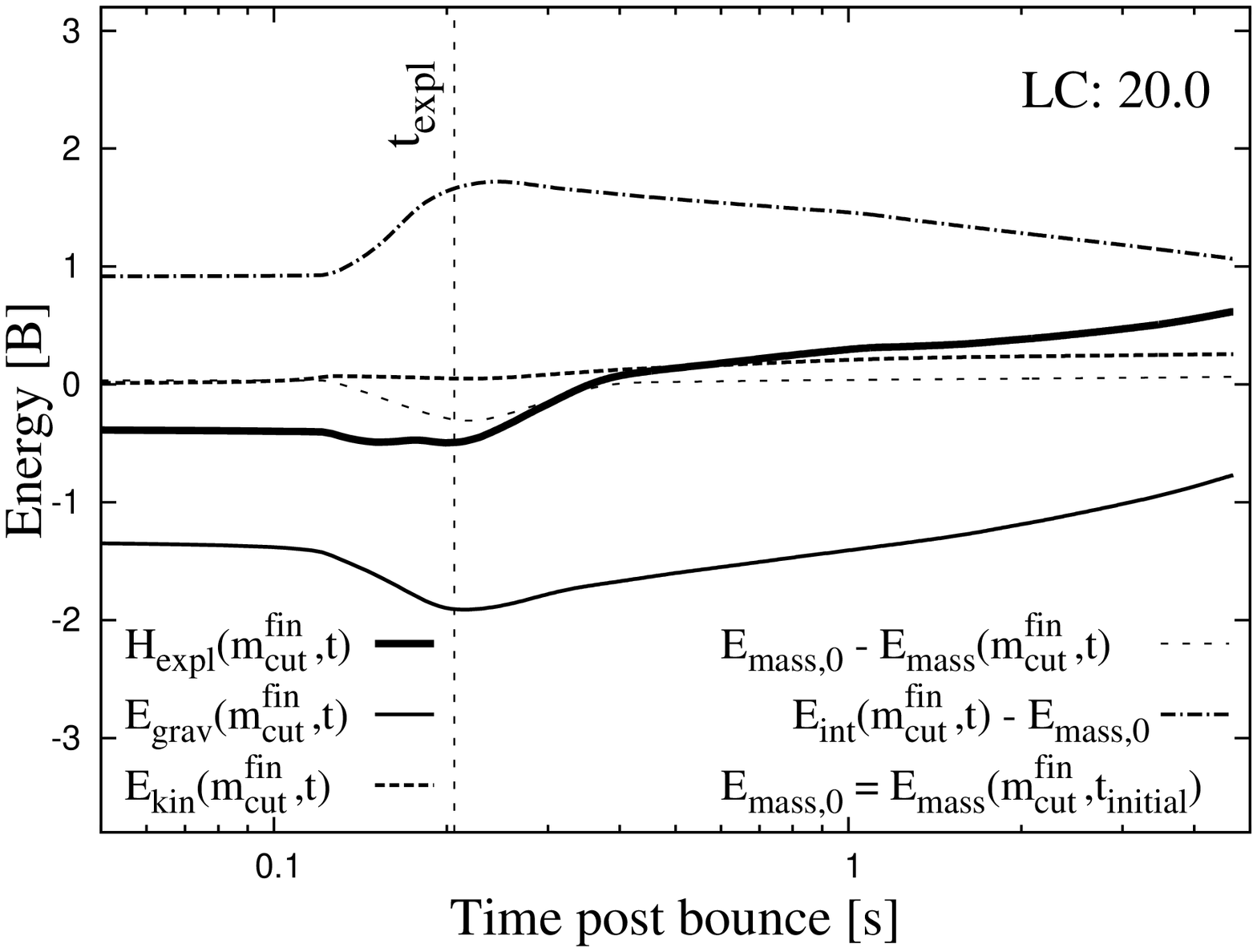}
\end{tabular}  
\caption{Temporal evolution of the gravitational (thin solid), kinetic (long dashed), negative rest mass (short dashed), internal (dot-dashed), and explosion (solid thick) energies above $m_{\rm cut}^{\rm fin} = m_{\rm cut}(t_{\rm final})$, for the HC (left) and for the LC (right) reference runs, see Table \ref{tab:HC vs LC}. The internal and the rest mass energy are given with respect to the initial rest mass, $E_{\rm mass,0}=E_{\rm mass}(m_{\rm cut}^{\rm fin},t_{\rm initial})$. The difference between the internal energy and the rest mass energy represents the thermal energy. 
\label{fig:explosion energy contributions} }
\end{figure*}

To give further insight, in Figure~\ref{fig:explosion energy contributions} we show the time evolution of all energies which contribute to the explosion energy together with the explosion energy itself, for both the HC (left panel) and the LC  model (right panel). We present $E_{\rm int}(m_{ \rm cut}^{\rm fin},t)$, $-E_{\rm mass}(m_{ \rm cut}^{\rm fin},t)$, $E_{\rm grav}(m_{ \rm cut}^{\rm fin},t)$ and $E_{\rm kin}(m_{ \rm cut}^{\rm fin},t)$, which together give a complete decomposition of the explosion energy, i.e., 
\begin{eqnarray}
H_{\rm expl}(m_{ \rm cut}^{\rm fin},t)&=&E_{\rm kin}(m_{ \rm cut}^{\rm fin},t)+E_{\rm grav}(m_{ \rm cut}^{\rm fin},t)\nonumber \\
&&+E_{\rm int}(m_{ \rm cut}^{\rm fin},t)-E_{\rm mass}(m_{ \rm cut}^{\rm fin},t) \;. 
\end{eqnarray}
Compared to Figure~\ref{fig:total energy contributions} we are now not dealing with variations any more but with absolute values. Gravitational energy initially dominates ($H_{\rm expl}(m_{ \rm cut}^{\rm fin},t) < 0$), meaning that the portion of the star above $m_{ \rm cut}^{\rm fin}$ is still gravitationally bound. The HC model is initially more bound than the LC model (for example, $H_{\rm expl}(m_{ \rm cut}^{\rm fin},t=0.1 \, {\rm s}) \approx -0.54 \, {\rm B}$, versus $H_{\rm expl}(m_{ \rm cut}^{\rm fin},t=0.1 \, {\rm s}) \approx -0.40 \, {\rm B}$, respectively). Before providing positive explosion energy, neutrinos have to compensate for this initial negative binding energy as well as for the negative $E_{\rm mech}$. This can be seen explicitly by expressing Equation (\ref{eq_deltaeexp}) as:
\begin{eqnarray}
 H_{\rm expl}(m_{ \rm cut}^{\rm fin},t_{\rm final}) &\sim&  H_{\rm expl}(m_{ \rm cut}^{\rm fin},t_{\rm initial}) + E_\nu(m_{ \rm cut}^{\rm fin}, t_{\rm final})\nonumber \\
&& + E_{\rm mech}(t_{\rm final}) \, , 
\end{eqnarray}
where we have neglected $\Delta E_{\rm mass}$.

In the following, we discuss the evolution of the relevant energies and, in particular, of the rest mass energy (see Section~\ref{sec_eos} for the description of the (non-)NSE EOS and of the related definitions of the internal, thermal and rest mass energies). The innermost part of the ejecta (corresponding to $\sim 0.15$~\msun and $\sim 0.07$~\msun above $m_{ \rm cut}^{\rm fin}$ for the 19.2~\msun and 20.0~\msun model, respectively) is initially composed of intermediate mass nuclei (mainly silicon and magnesium). In the first part of the evolution, during the gravitational collapse, no significant changes of $E_{\rm int}$ and $E_{\rm mass}$ are observed in Figure~\ref{fig:explosion energy contributions}. However, when this matter enters the shock, it is quickly photodissociated into neutrons, protons, and alpha particles. This process increases the rest mass energy, as is visible in Figure~\ref{fig:explosion energy contributions} between roughly 200 and 300~ms for the HC model and between 100 and 200~ms for 
the LC model. At the same time, the release of gravitational energy of the still infalling matter and the dissipation of kinetic energy happening at the shock, together with the large and intense neutrino absorption on free nucleons, increase $E_{\rm int}$. Later, once neutrino heating has halted the collapse and started the explosion, the expanding shock decreases its temperature and free neutrons and protons inside it recombine first into alpha particles and then into iron group nuclei. At the same time, fresh infalling layers are heated by the shock to temperatures above $0.44$~MeV, and silicon and magnesium are converted into heavier nuclei and alpha particles under NSE conditions, leading to an alpha-rich freeze-out from NSE. The production of alpha particles, which are less bound than the heavy nuclei initially present in the same layers, limits the amount of rest mass energy finally released. Thus, these recombination and burning processes liberate in a few hundred milliseconds after $t_{\rm expl}$ an 
amount of rest mass energy larger but comparable to the energy spent by the shock to photodissociate the infalling matter during shock revival and early expansion. We have checked in post-processing that the full nucleosynthesis network {\sc Winnet} confirms these results.

\subsection{Explosion dynamics and the role of compactness} 
\label{sec:hc and lc}

\begin{figure*}[htp!] 
\begin{tabular}{cc}
 \includegraphics[width=0.35\textwidth,angle=-90]{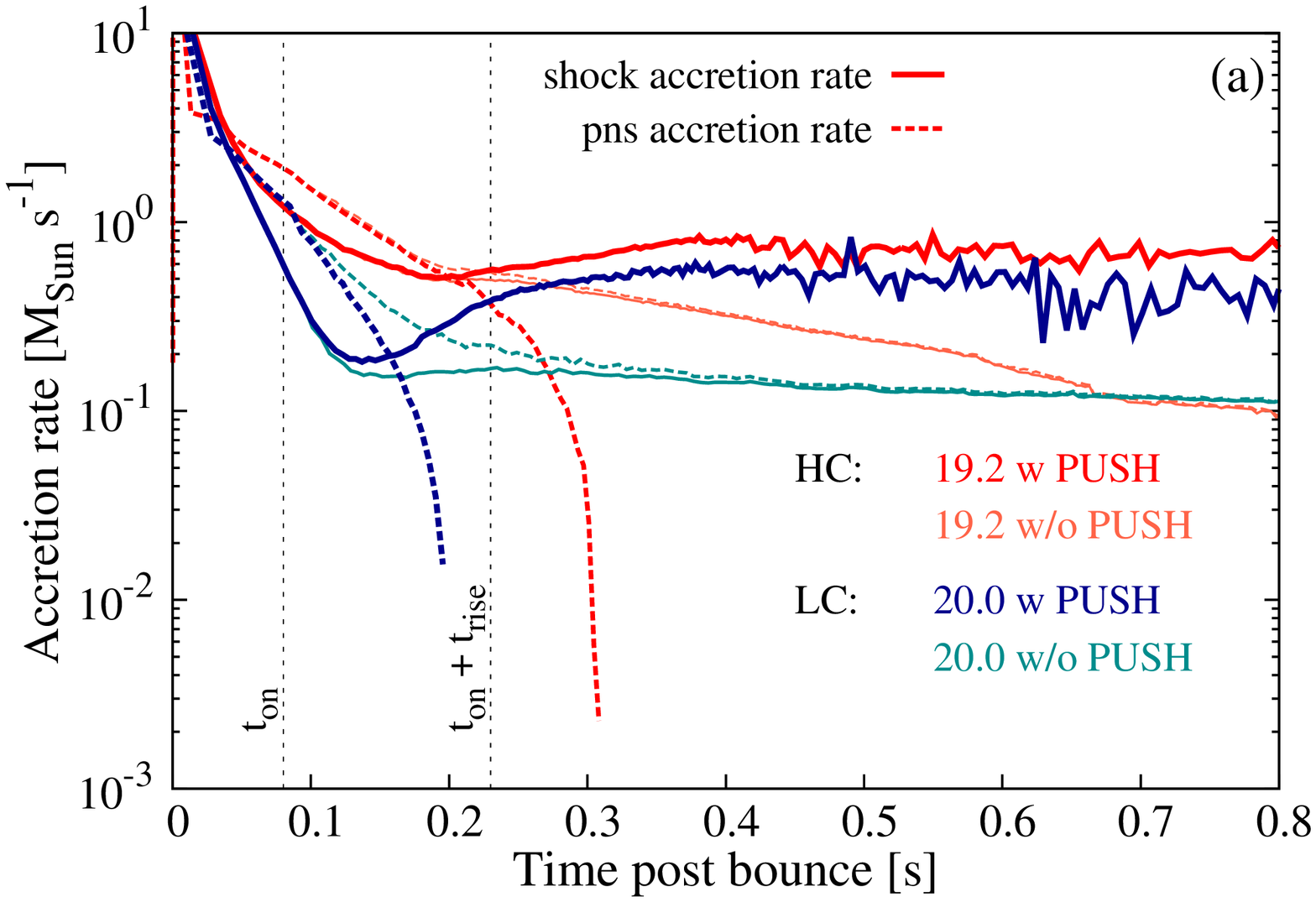} &
 \includegraphics[width=0.35\textwidth,angle=-90]{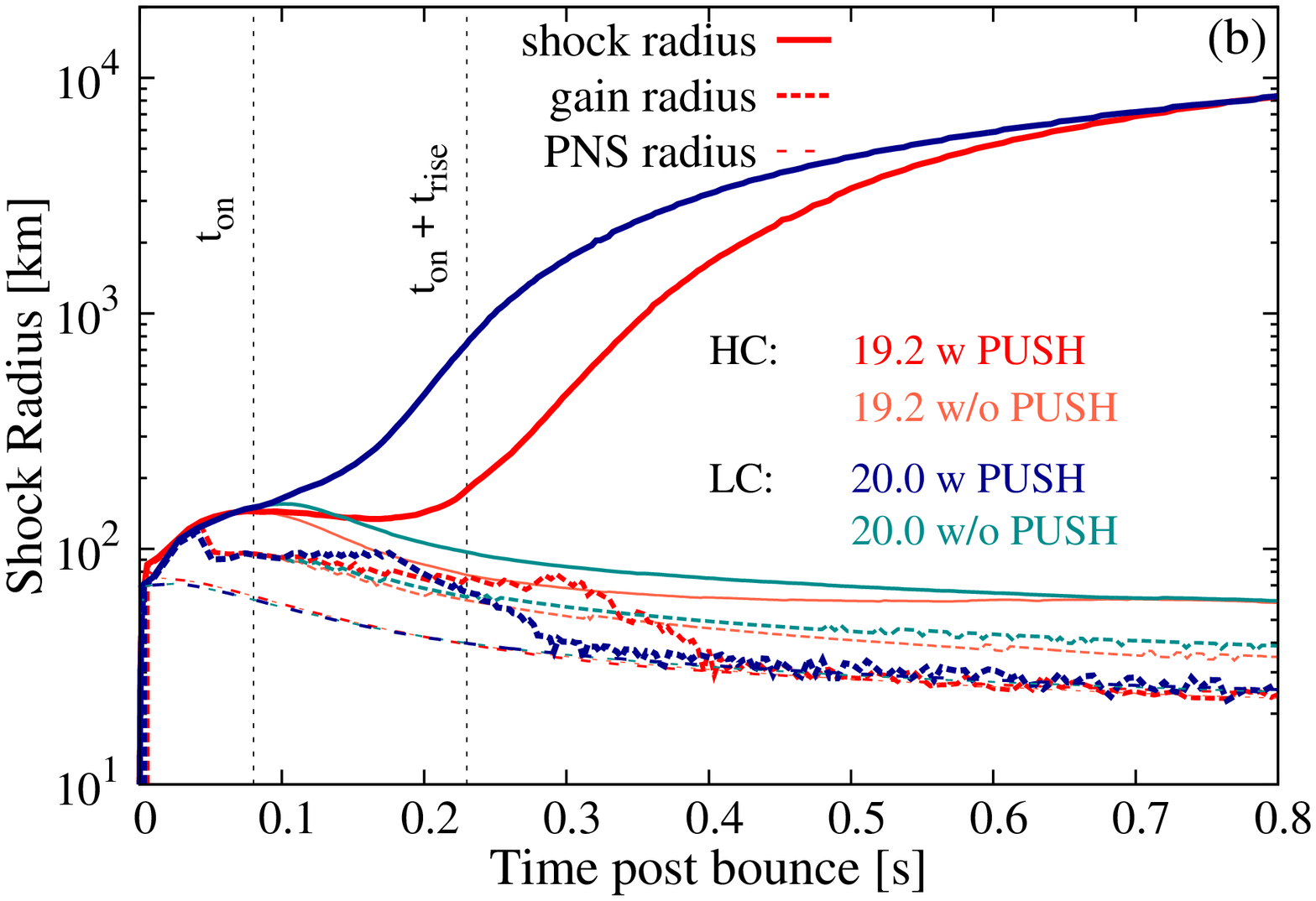}
 \\
 \includegraphics[width=0.35\textwidth,angle=-90]{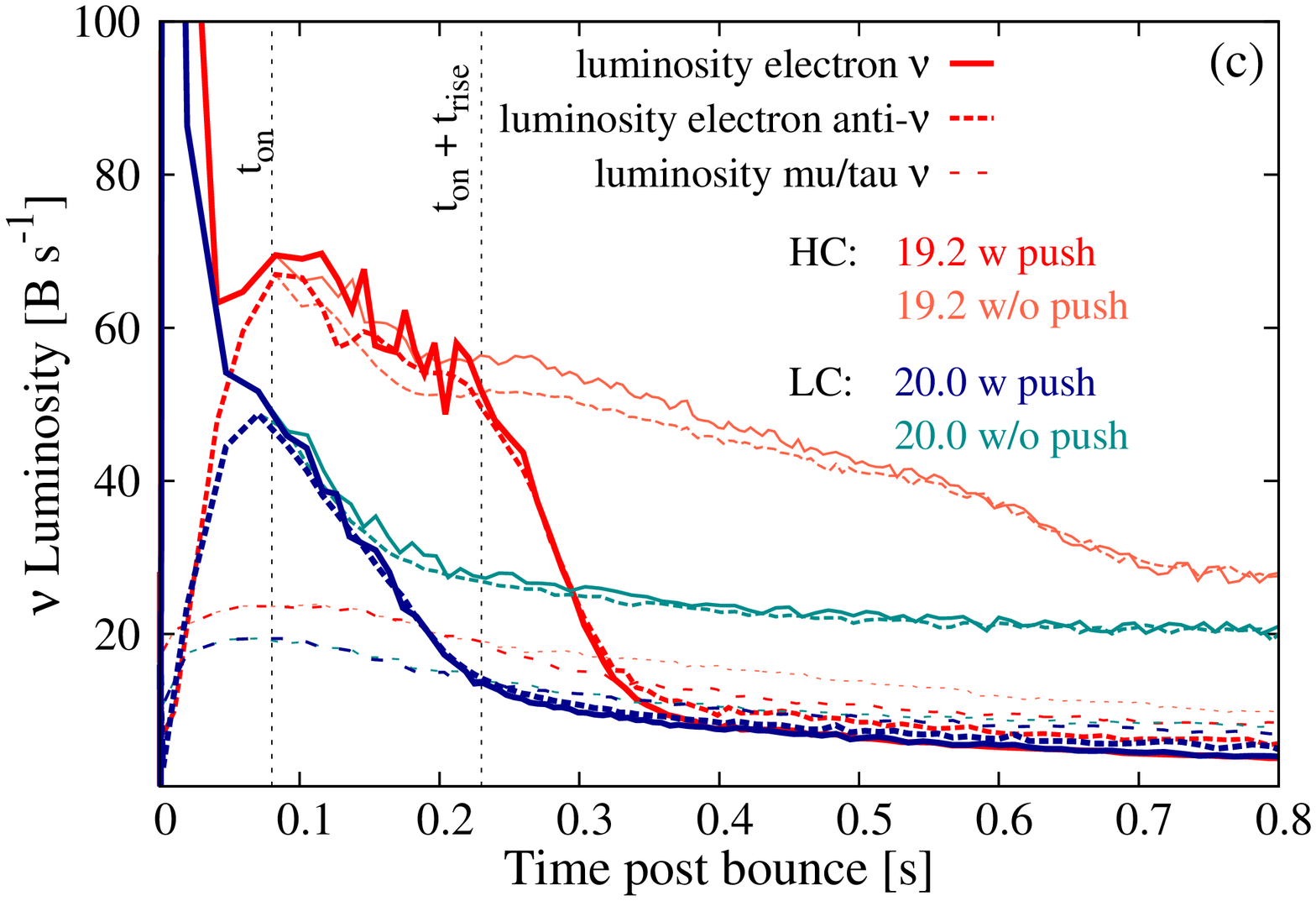} &
 \includegraphics[width=0.35\textwidth,angle=-90]{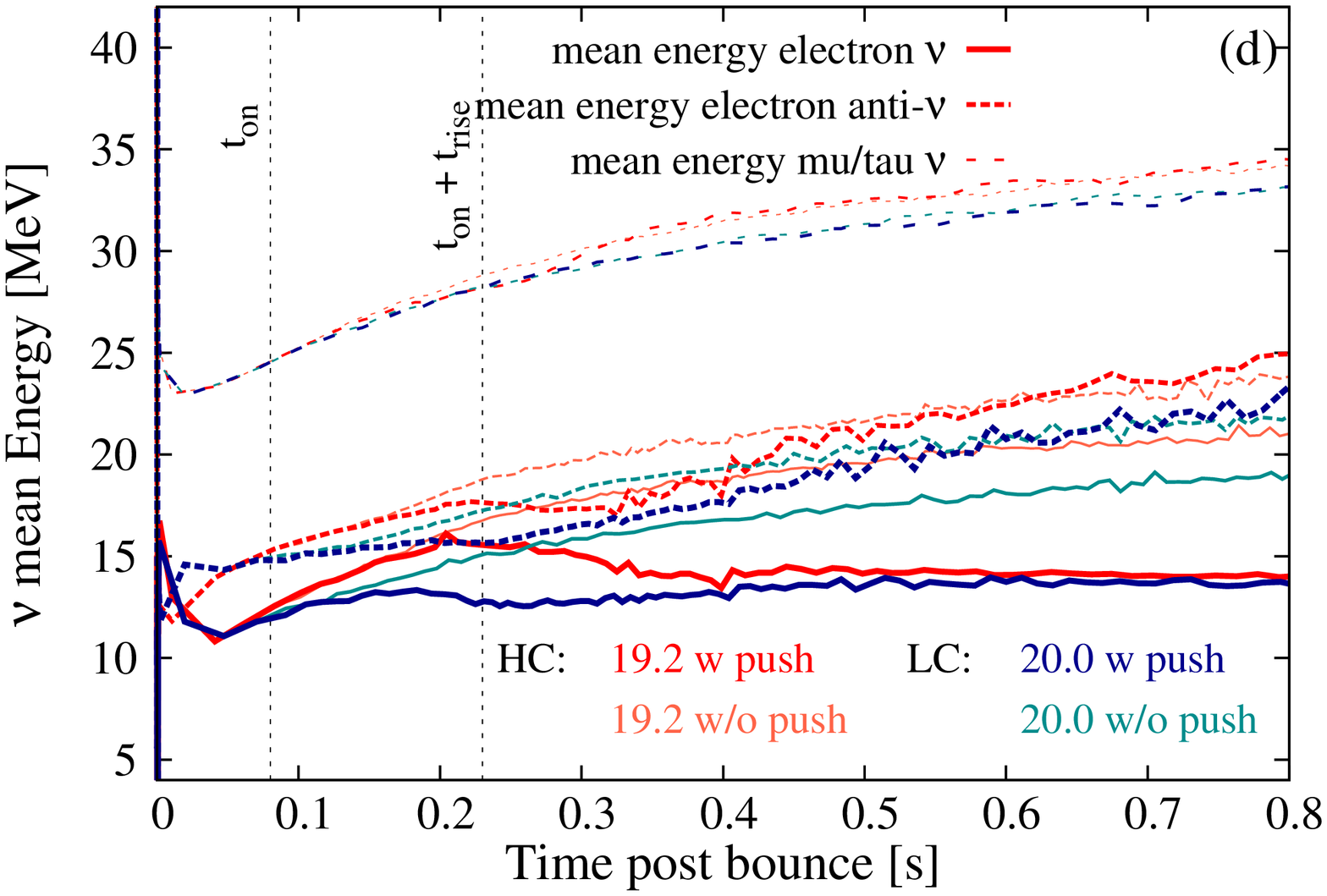}
\end{tabular}  
\caption{Temporal evolution of (a) the accretion rate at the PNS and at the shock, (b) the shock, the gain, and the PNS radii, (c) the neutrino luminosities, and (d) the neutrino mean energies, for all modeled neutrino flavors. In all panels, we present exploding runs for the 19.2~\msun (red lines) and then 20.0~\msun (blue lines) ZAMS mass models obtained with the PUSH parameters reported in Table \ref{tab:HC vs LC}. We also plot the corresponding non-exploding runs obtained by setting $k_{\rm push}=0$ for the 19.2~\msun (light red) and 20.0~\msun (light blue) ZAMS mass progenitor models. 
\label{fig:compactness comparison}}
\end{figure*}

\begin{figure}[ht] 
\begin{center}
 \includegraphics[width=0.35\textwidth,angle=-90]{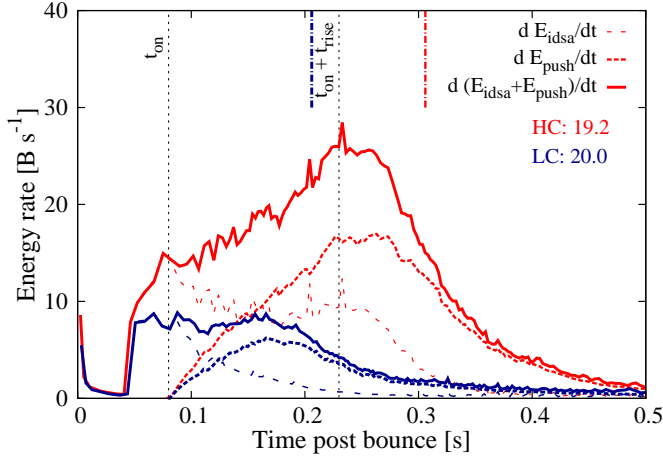}
 \end{center}
 \caption{Temporal evolution of the neutrino energy deposition inside the gain region from $\nu_e$ and $\bar{\nu}_e$  (${\rm d}E_{\rm idsa}/{\rm d} t$, long-thin dashed lines), from PUSH (${\rm d}E_{\rm push}/{\rm d} t$, short-thick dashed lines), and their sum (solid lines).  The 19.2~\msun (HC) ZAMS mass model is represented in red, while the 20.0~\msun (LC) ZAMS mass is in blue, with PUSH parameters reported in Table \ref{tab:HC vs LC}. The short colored vertical lines show the time of explosion.
\label{fig:energy derivaties} }
\end{figure}

\begin{figure}[ht]  
\begin{center}
 \includegraphics[width=0.35\textwidth,angle=-90]{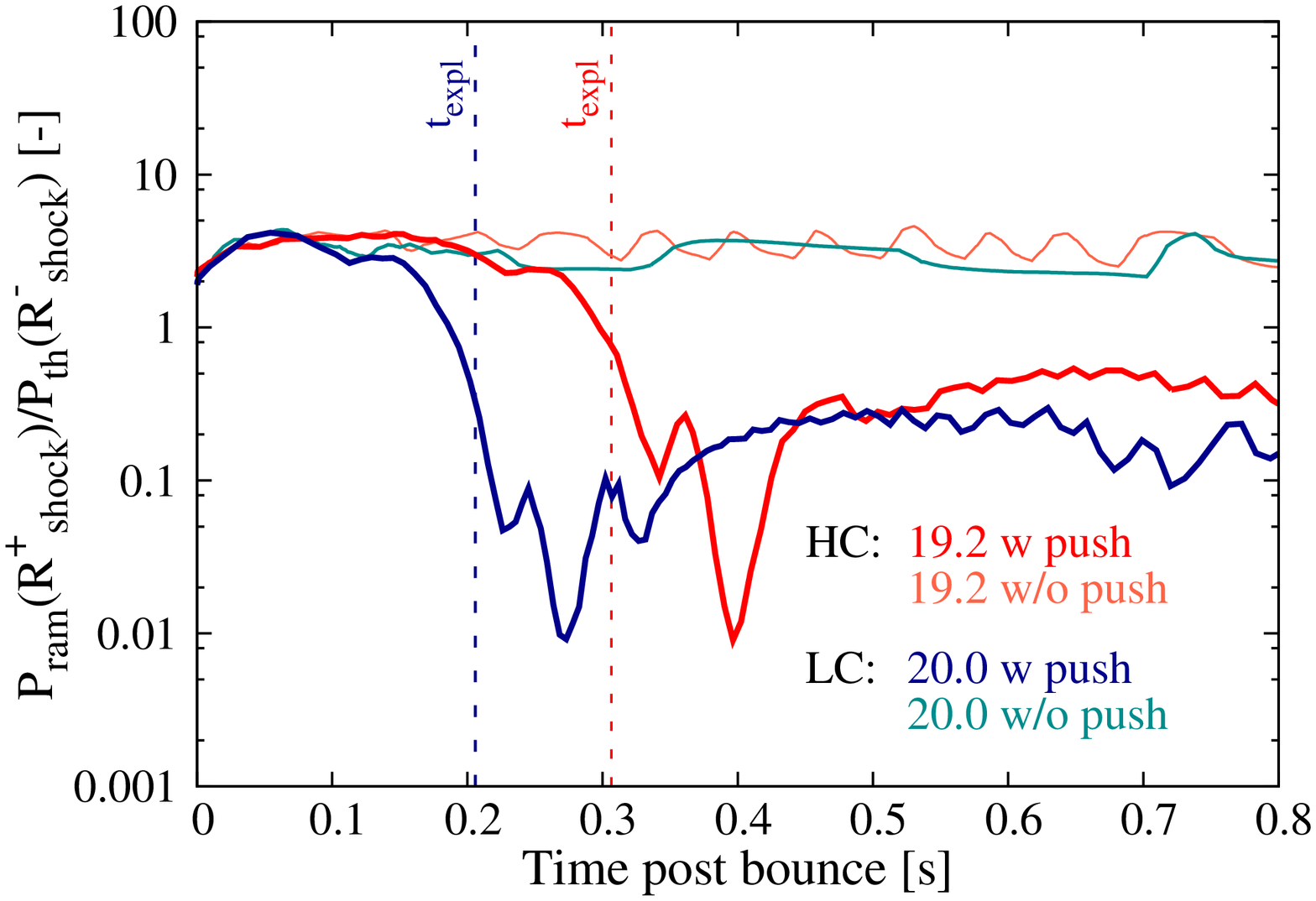}
\end{center}
 \caption{Temporal evolution of the ratio between the ram pressure above the shock and the thermal pressure below the shock. The 19.2~\msun (HC) ZAMS mass model is represented in red, while the 20.0~\msun (LC) ZAMS mass model is in blue. The PUSH parameters are reported in Table \ref{tab:HC vs LC}. Light red and light blue lines represent the corresponding 
 runs without PUSH ($k_{\rm push}=0$). 
\label{fig:pressure ratio} }
\end{figure}

The distributions of the explosion energy and explosion time obtained with PUSH, as well as their variations in response to changes of the model parameters, suggest a possible distinction between high and low compactness progenitors.  In the following, we investigate how basic properties of the models (e.g., the accretion history or the neutrino luminosities), ultimately connected with the compactness, relate to differences in the explosion process and properties.  For a similar discussion in self-consistent 1D and 2D supernova simulations, see \citet{suwa14}. Again, we choose the 19.2 and 20.0 ZAMS mass progenitor runs with $t_{\rm rise}=150 \, {\rm ms}$ and $k_{\rm push}=3.0$, as representatives of the HC and LC samples, respectively.

In Figure~\ref{fig:compactness comparison}, we show the temporal evolution of several quantities of interest for both the 19.2~\msun and 20.0~\msun models, with and without PUSH. The evolution before $t_{\rm on}$ follows the well known early shock dynamics in CCSNe  (see, for example, \cite{Burrows1993}).  In both models, a few tens of milliseconds after core bounce, the expanding shock turns into an accretion front, and the mantle between the PNS surface and the shock reaches a quasi-stationary state.  In this accretion phase, $\dot{M}_{\rm shock}$ and $\dot{M}_{\rm PNS}$ are firmly related.  However, the two different density profiles already affect the evolution of the shock.  Since $\rho_{19.2}/\rho_{20.0} \gtrsim 1.2 $ outside the shock and up to a radius of $2 \times 10^{8}$~cm (while the infalling velocities of the unshocked matter are initially almost identical), $\dot{M}_{\rm shock}$ (and in turn also $\dot{M}_{\rm PNS}$) starts to differ between the two models around $t_{\rm pb} \approx 30$~ms.

The difference in the accretion rates has a series of immediate consequences. For the HC case,
(i) neutrino luminosities are larger (Figure~\ref{fig:compactness comparison}c); 
(ii) the shock is subject to a larger ram pressure (i.e., a larger momentum transfer provided by the collectively infalling mass flowing through the shock), and, as visible in the case without PUSH, shock stalling happens earlier and at a smaller radius (Figure~\ref{fig:compactness comparison}b);
(iii) the PNS mass grows faster.  Since the mass of the PNS at bounce is almost identical for the two models ($M_{\rm PNS} \approx 0.63$~\msun), the stronger gravitational potential implied by (iii) increases the differences in the accretion rates even further by augmenting the ratio of the radial velocities inside the gain region (larger by 12--15\% at $t \approx t_{\rm on}$ for the 19.2~\msun case).

For $t>t_{\rm on}$, the differences between the two runs amplify as a result of the PUSH action.  In the LC case, due to the lower accretion rate, a relatively small energy deposition by PUSH in the gain region (smaller than or comparable to the energy deposition by $\nu_e$ and $\bar{\nu}_e$ from IDSA, as visible in Figure~\ref{fig:energy derivaties}) is able to revive the shock expansion a few milliseconds after $t_{\rm on}$.  Later, the increasing ${\rm d}E_{\rm push}/{\rm d}t$ triggers an explosion in a few tens of milliseconds, even before $\mathcal{G}(t)$ reaches its maximum (Figure~\ref{fig:compactness comparison}b).
In the HC case, the energy deposition by neutrinos is more intense from the beginning due to the larger neutrino luminosities and harder neutrino spectra (Figures~\ref{fig:compactness comparison}c and \ref{fig:compactness comparison}d) and due to the higher density inside the gain region.  However, because of the larger accretion rate, the extra contribution provided by PUSH is initially only able to prevent the fast shock contraction observed in the model without PUSH. During this shock stalling phase, the accretion rate and the luminosity decrease, but only marginally and very similarly to the non-exploding case.  When PUSH reaches its maximum energy deposition rate ($t \approx t_{\rm on} + t_{\rm rise}$), the shock revives and the explosion sets in (Figure~\ref{fig:compactness comparison}b).

In Figure~\ref{fig:pressure ratio}, we plot the ratio of the ram pressure just above the shock front ($P_{\rm ram}(R_{\rm shock}^+)=\rho v^2$ calculated at $R_{\rm shock}^{+} = R_{\rm shock} + 1 \, \rm{km}$) to the thermal pressure just inside it ($P_{\rm th}(R_{\rm shock}^-)$ where $R_{\rm shock}^{-} = R_{\rm shock} - 1 \, \rm{km}$).  In the non-exploding runs (i.e., without PUSH), both these pressures decrease with time, but their ratio stays always well above unity. On the other hand, in runs with PUSH, the more efficient energy deposition by neutrinos reduces the decrease of the thermal pressure inside the shock. The corresponding drop in the pressure ratio below unity determines the onset of the explosion.

In both runs, once the explosion has been launched, the density in the gain region decreases and the PUSH energy deposition rate reduces accordingly.  The conversion from an accreting to an expanding shock front decouples $\dot{M}_{\rm shock}$ from $\dot{M}_{\rm PNS}$.  The latter drops steeply, together with the accretion neutrino luminosities (Figures~\ref{fig:compactness comparison}a and \ref{fig:compactness comparison}c), while $\dot{M}_{\rm shock}$ decreases first but then stabilizes around an almost constant (slightly decreasing) value.  In the case where the shock expansion velocity is much larger than the infalling matter velocity at $R_{\rm shock}$, $\dot{M}_{\rm shock}$ can be re-expressed as
\begin{equation}
 \dot{M}_{\rm shock} \approx 4 \pi R_{\rm shock}^2 \, \rho(R_{\rm shock}) \, v_{\rm shock}, 
\end{equation}
where $v_{\rm shock} = {\rm d} R_{\rm shock} / {\rm d} t \propto R_{\rm shock}^{\delta}$.  For $R>R_{\rm shock}$ we have in good approximation $\rho(R) \propto R^{-2}$, and thus
\begin{equation}
  \dot{M}_{\rm shock} \propto R_{\rm shock}^{\delta} .
\end{equation}
The stationary value of $\dot{M}_{\rm shock}$ implies that  $\delta \approx 0$.  Thus, after an initial exponential expansion, the shock velocity is almost constant during the first second after the explosion.

Despite the larger difficulties to trigger an explosion, the HC model explodes more energetically than the LC model. According to the analysis performed in Section~\ref{sec:eexpl_contr}, the difference in the explosion energy between the HC and the LC model depends ultimately on the different amount of energy deposited by neutrinos. Since the high compactness model requires a larger energy deposition to overcome the ram pressure and the gravitational potential, the total energy of the corresponding ejecta (and in turn the explosion energy) will be more substantially increased. 
In addition, after the explosion has been triggered, the larger neutrino luminosities and densities that characterize the HC model inject more energy in the expanding shock compared with the LC model.

\subsection{Fitting of SN~1987A}
\label{sec:method_sn1987a}
The ultimate goal of core-collapse supernova simulations is to reproduce the properties observed in real supernovae.  So far we have only focused on the dependence of dynamical features of the explosion (e.g., the explosion energy) on the parameter choices in the PUSH method. However, the ejected mass of radioactive nuclides (such as $^{56}$Ni) is an equally important property of the supernova explosion. Here, we describe how we calibrate the PUSH method by reproducing the explosion energy and mass of Ni ejecta of SN~1987A for a progenitor within the expected mass range for this supernova.

\subsubsection{Observational constraints from SN~1987A}
The analysis and the modeling of the observational properties of SN~1987A just after the luminosity peak have been the topics of a long series of works \citep[e.g.,][ and references therein]{Woosley1988,arnett89,Shigeyama1990,Kozma1998a,Kozma1998b,Blinnikov2000,Fransson.Kozma:2002,Utrobin2005,Seitenzahl2014}.  They provide observational estimates for the explosion energy, the progenitor mass, and the ejected masses of $^{56}$Ni, $^{57}$Ni, $^{58}$Ni, and $^{44}$Ti, all of which carry rather large uncertainties.
In Table \ref{tab:sn1987a}, the values used for the calibration of the PUSH method are summarized. 

The ZAMS progenitor mass is assumed to be between 18~\msun and 21~\msun, corresponding to typical values reported in the literature for the SN~1987A progenitor, see, e.g., \citet{Woosley1988,Shigeyama1990}. For the explosion energy we consider the estimate reported by \citet{Blinnikov2000}, $E_{\rm expl} = \left(1.1 \pm 0.3 \right) \times 10^{51}$~erg (for a detailed list of explosion energy estimates for SN~1987A, see for example Table~1 in \citet{handy14}). This value was obtained assuming $\sim$14.7~\msun of ejecta and an hydrogen-rich envelope of $\sim$10.3~\msun.  The uncertainties in the progenitor properties and in the SN distance were taken into account in the error bar. The employed values of the total ejecta and of the hydrogen-rich envelope are compatible (within a 15\% tolerance) with a significant fraction of our progenitor candidates, especially for $M_{\rm ZAMS} < 19.6$~\msun (see Table~\ref{tab:prog_compact}, where the total ejecta can be estimated subtracting $1.6$~\msun from the mass of the 
star at the onset of the collapse). Explosion models with larger ejected mass (i.e., less compatible with our candidate sample) tend to have larger explosion energies (see, for example, \citet{Utrobin2005}). Finally, we consider the element abundances for $^{56,57}$Ni and $^{44}$Ti provided by \citet{Seitenzahl2014}, which were obtained from a least squares fit of the decay chains to the bolometric lightcurve.  For $^{58}$Ni we use the value provided by \citet{Fransson.Kozma:2002}.

\begin{table}[h,t]
\caption{Observational properties of SN~1987A. \label{tab:sn1987a}}
\begin{center}
\begin{tabular}{lc}
  \tableline \tableline
  $E_{\rm expl}$ & $(1.1 \pm 0.3) \times 10^{51}$~erg \\
  $m_{\rm prog}$ & 18-21 \msun\\
  $m(^{56}{\rm Ni})$ & $(0.071 \pm 0.003)$~M$_{\odot}$ \\ 
  $m(^{57}{\rm Ni})$ & $(0.0041 \pm 0.0018)$~M$_{\odot}$ \\ 
  $m(^{58}{\rm Ni})$ & 0.006~M$_{\odot}$ \\ 
  $m(^{44}{\rm Ti})$ & $(0.55 \pm 0.17) \times 10^{-4}$~M$_{\odot}$ \\
  \tableline
\end{tabular}
\tablecomments{
The nucleosynthesis yields are taken from \cite{Seitenzahl2014} except for $^{58}$Ni which is taken from \cite{Fransson.Kozma:2002}. No error estimates were given for $^{58}$Ni. The explosion energy is adapted from \cite{Blinnikov2000}.  For the progenitor range we chose typical values found in the literature, see e.g.\ \cite{Shigeyama1990,Woosley1988}.}
\end{center}
\end{table}

\subsubsection{Fitting procedure}
\label{sec: fit}

\begin{figure*}[htp!] 
   \includegraphics[width=0.5\textwidth]{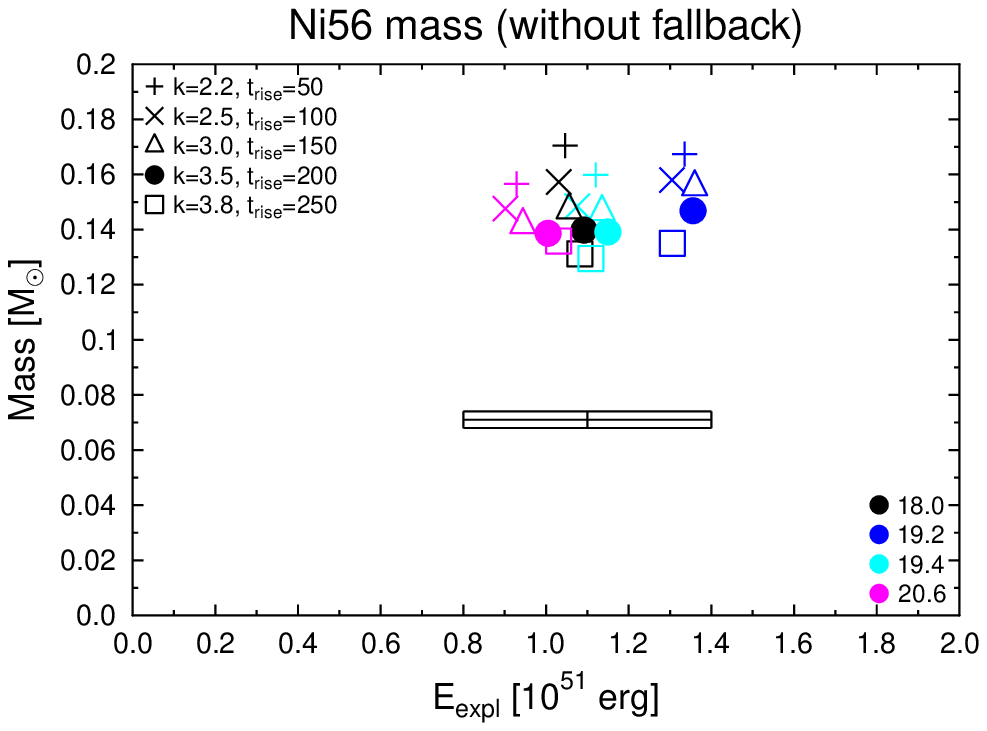} 
   \includegraphics[width=0.5\textwidth]{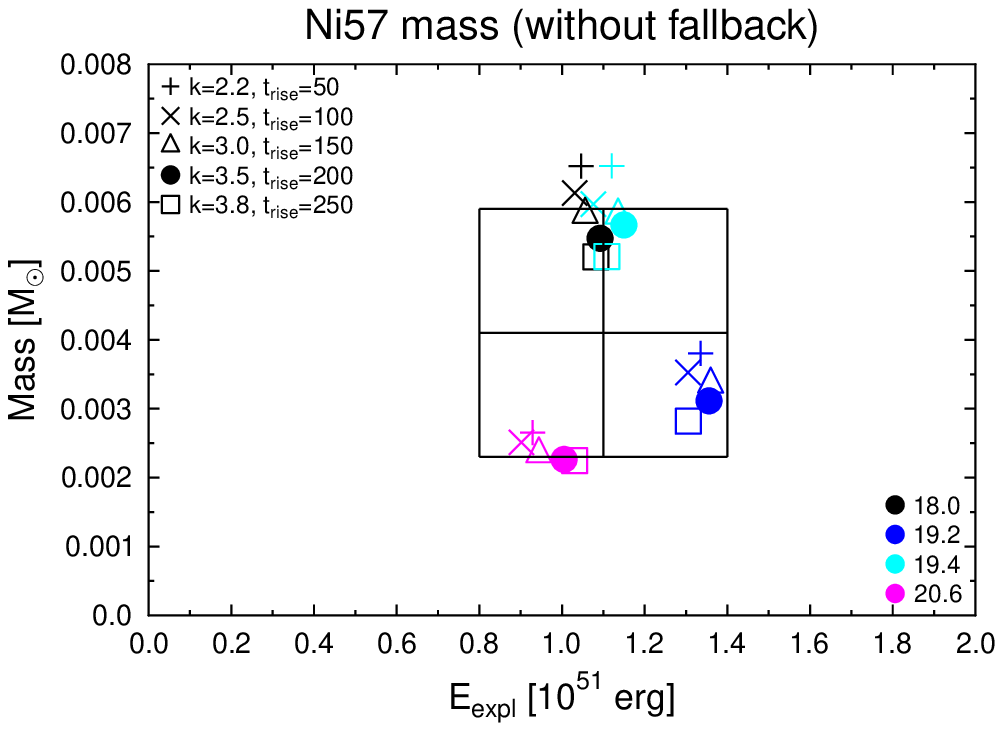} \\
   \includegraphics[width=0.5\textwidth]{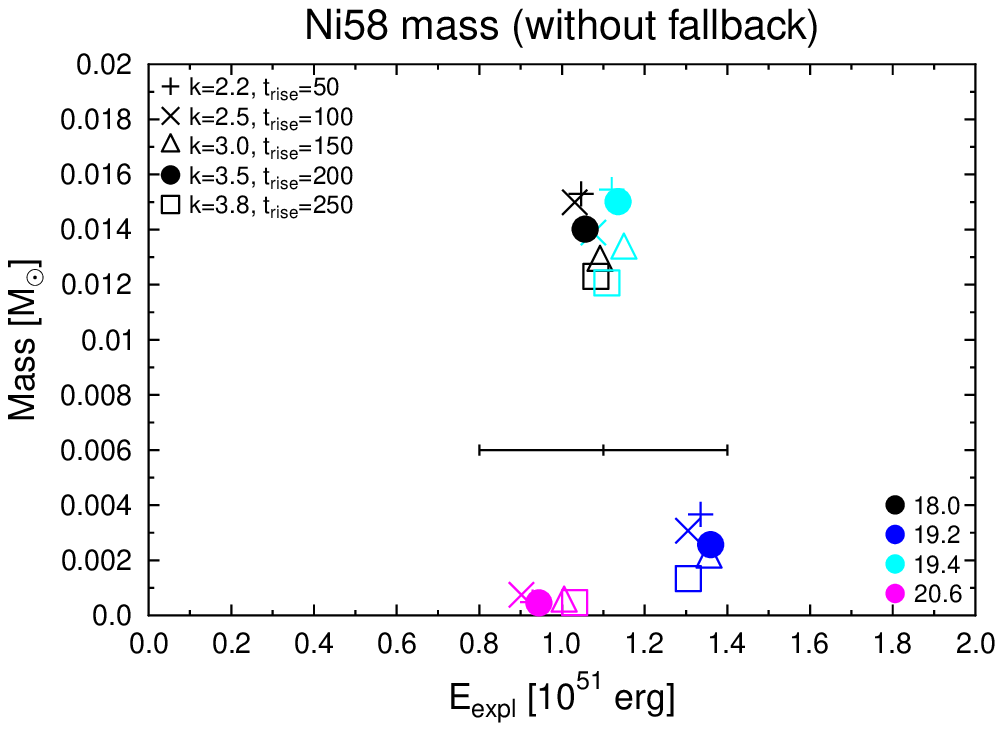} 
   \includegraphics[width=0.5\textwidth]{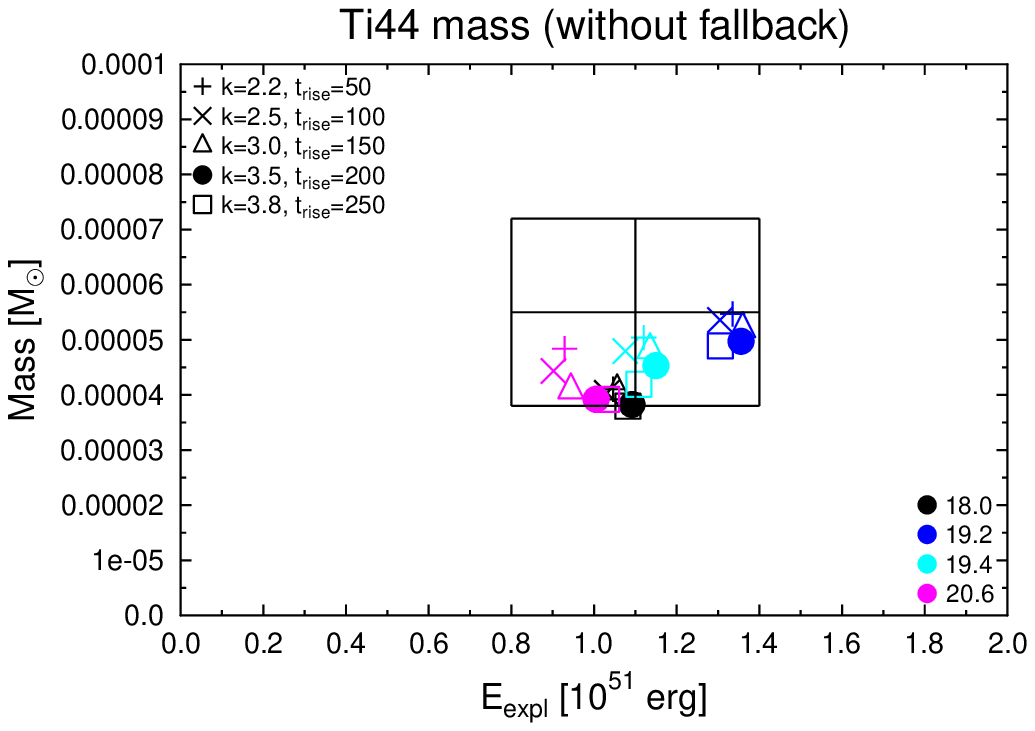} 
    \caption{Ejected mass of $^{56}$Ni (top left), $^{57}$Ni (top right), $^{58}$Ni (bottom left), and $^{44}$Ti (bottom right) and explosion energy for four representative HC progenitor models.  Five combinations of \kpush and \trise are shown, each with a different symbol. The error bar box represents the observational values from \cite{Seitenzahl2014} (for $^{56,57}$Ni and $^{44}$Ti) and from \citet{Fransson.Kozma:2002} (for $^{58}$Ni).  No error bars are reported for $^{58}$Ni.
\label{fig:calibration_summary} }
\end{figure*}

\begin{figure*}[htp!]  
   \includegraphics[width=0.5\textwidth]{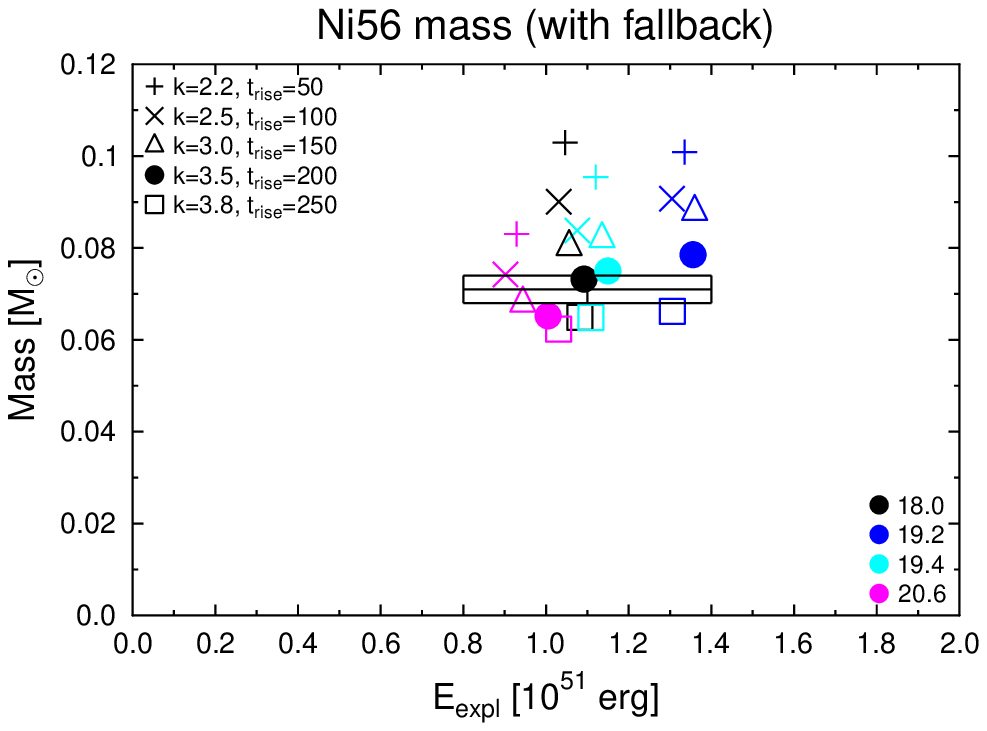} 
   \includegraphics[width=0.5\textwidth]{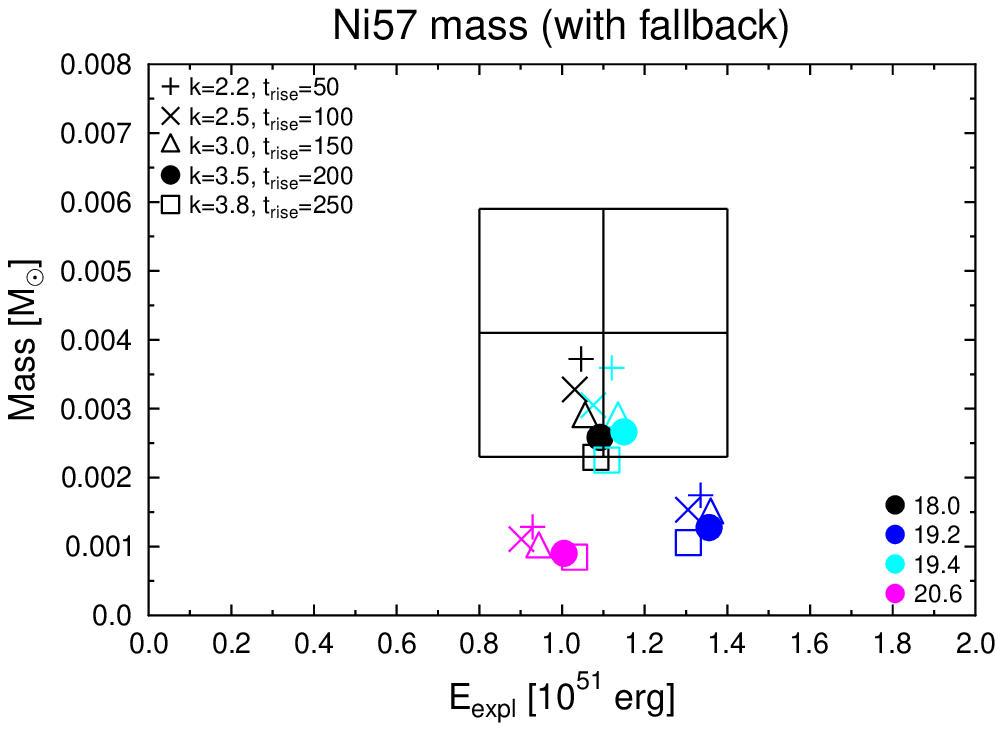} \\ 
   \includegraphics[width=0.5\textwidth]{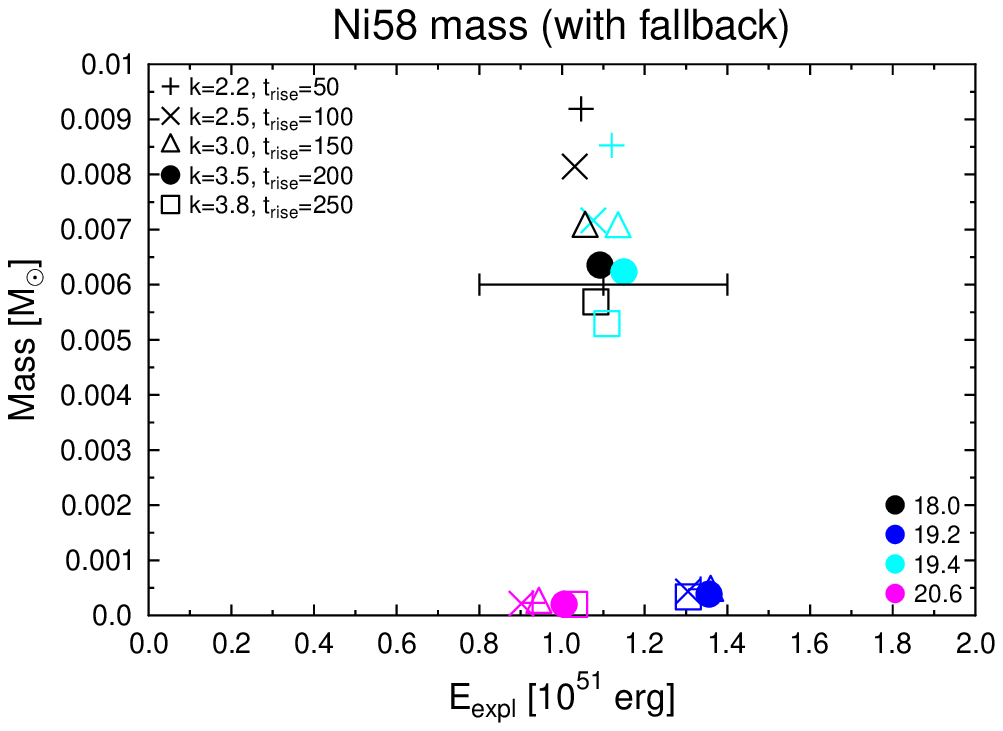}  
   \includegraphics[width=0.5\textwidth]{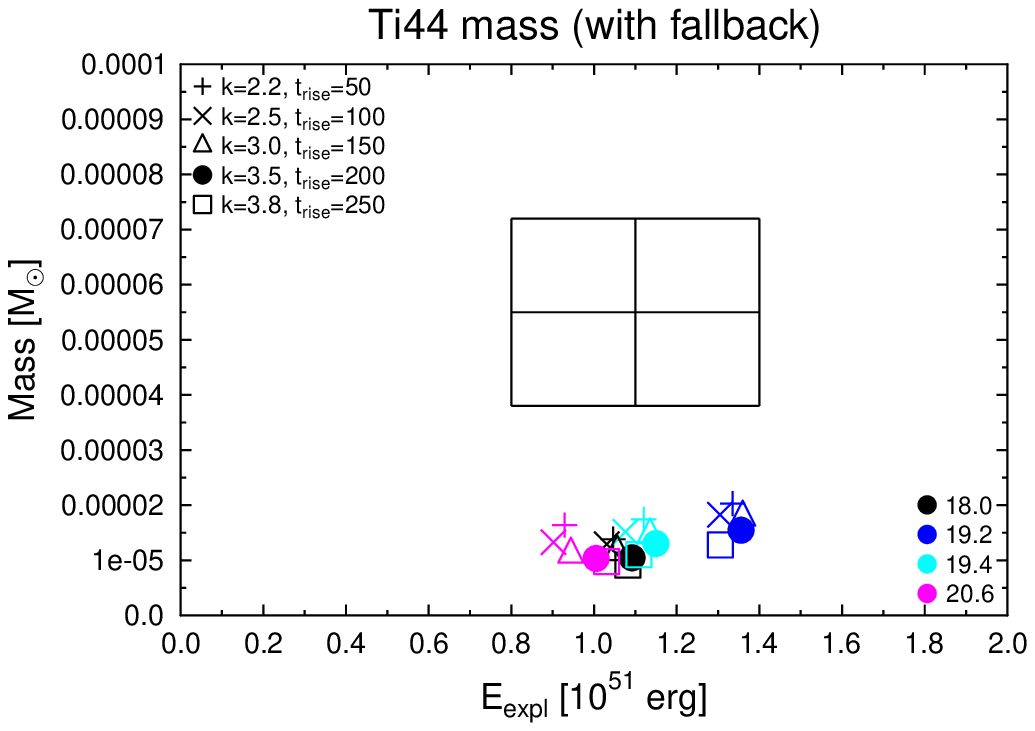}  
    \caption{Same as Figure~\ref{fig:calibration_summary}, but assuming 0.1~M$_{\odot}$ fallback. Note the different scale for $^{56}$Ni and $^{58}$Ni compared to Figure~\ref{fig:calibration_summary}.
\label{fig:calibration_summary_fb} }
\end{figure*}

We calibrate the PUSH method by finding a combination of progenitor mass, \kpush, and \trise which provides the best fit to the all observational quantities of SN~1987A mentioned above.  The weight given to each quantity is related to the uncertainty.  For example, due to the large uncertainty in the $^{44}$Ti mass, this does not provide a strong constraint on selecting the best fit.

Figure~\ref{fig:calibration_summary} shows the explosion energy and ejected mass of $^{56}$Ni, $^{57}$Ni, $^{58}$Ni, and $^{44}$Ti for different cases of \kpush and \trise and for four select HC progenitors used to calibrate the PUSH method.  We do not consider the LC progenitors, because of their generally lower explosion energies, see Figure~\ref{fig:kpush}. The different cases of \kpush and \trise span a wide range of explosion energies around 1~Bethe.  For all parameter combinations shown, at least one progenitor in the 18-21~\msun range fulfills the requirement of an explosion energy between 0.8~Bethe and 1.4~Bethe.  
There is a roughly linear correlation between the explosion energy and the synthesized $^{56}$Ni-mass. However, this correlation is not directly compatible with the observations, as the ejected $^{56}$Ni is systematically larger than expected (up to a factor of $\sim 2$ for models with an explosion energy around 1~Bethe). There is a weak trend that models with higher $t_{\rm rise}$ tend to give lower nickel masses for given explosion energy. Among the parameter combinations that produce robustly high explosion energies (i.e., $k_{\rm push} \geq 3$), $k_{\rm push}=3.5$ with the high value of $t_{\rm rise}$ of 200~ms gives the lowest $^{56}$Ni mass for similar explosion energies, but still much too high.

Our simulations can be reconciled with the observations by taking into account fallback from the initially unbound matter.  Since we do not model the explosion long enough to see the development of the reverse shock and the appearance of the related fallback when the shock reaches the hydrogen-rich envelope, we have to impose it, removing some matter from the innermost ejecta\footnote{Note that we did not modify the explosion energy due to the fallback. This is based on the expectation that at the late time when fallback forms, the explosion energy is approximately equally distributed among the total ejected mass, which is about two orders of magnitude higher than our fallback mass.}. With a value of $\sim 0.1$~\msun we can match both the expected explosion energy and $^{56}$Ni ejecta mass, see Figure~\ref{fig:calibration_summary_fb}. In this way we have fixed the final mass cut by observations. However, we point out that we are able to identify the amount of late-time fallback only because we also have the 
dynamical mass cut from our hydrodynamical simulations. This is not possible in other methods such as pistons or thermal bombs. Our value of $\sim 0.1$~\msun of fallback in SN~1987A will be further discussed and compared with other works in Section~\ref{sec_fallback}.

\begin{figure*}[htp!]
\begin{tabular}{cc}
 \includegraphics[width=0.35\textwidth,angle=-90]{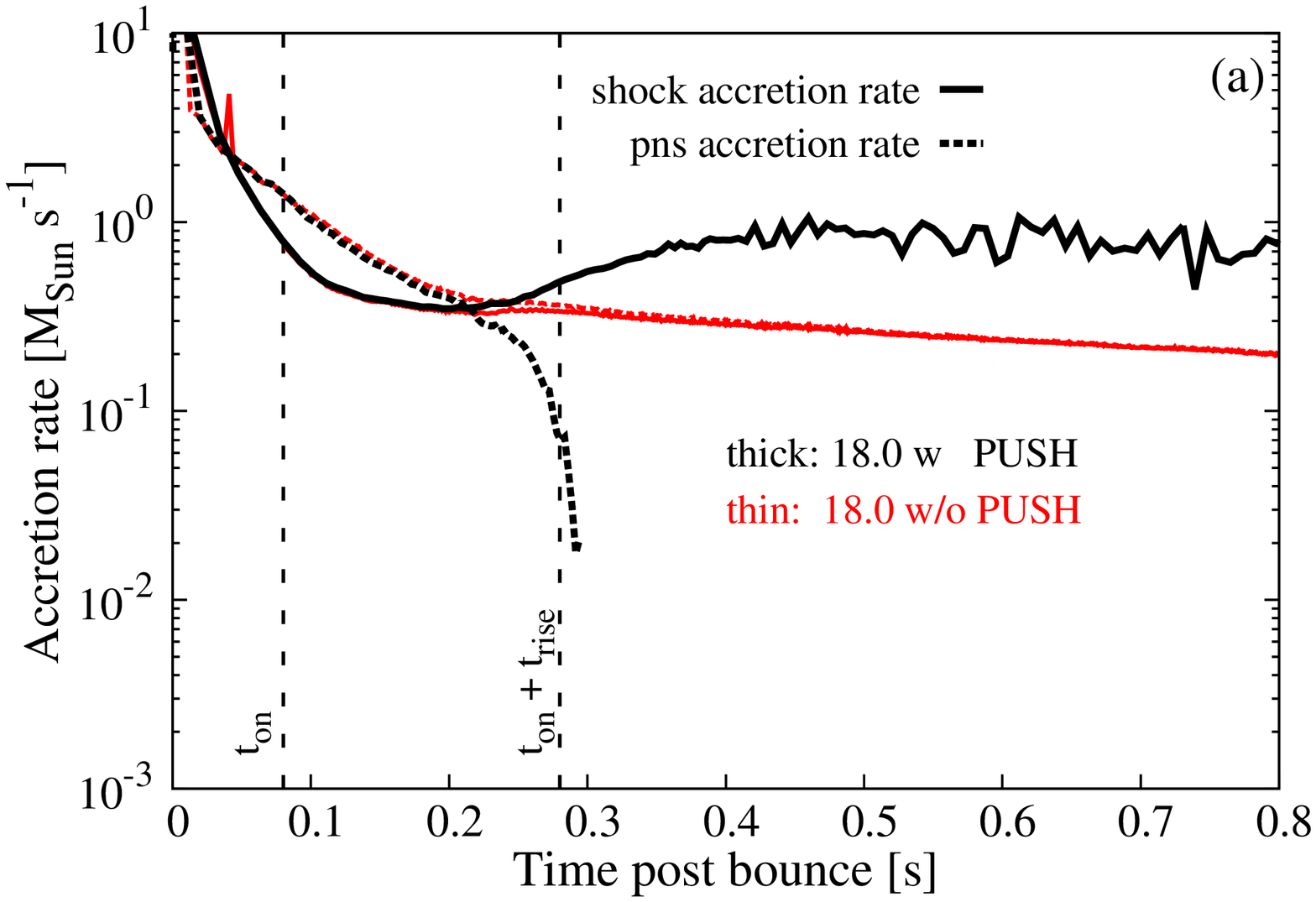} 
 \includegraphics[width=0.35\textwidth,angle=-90]{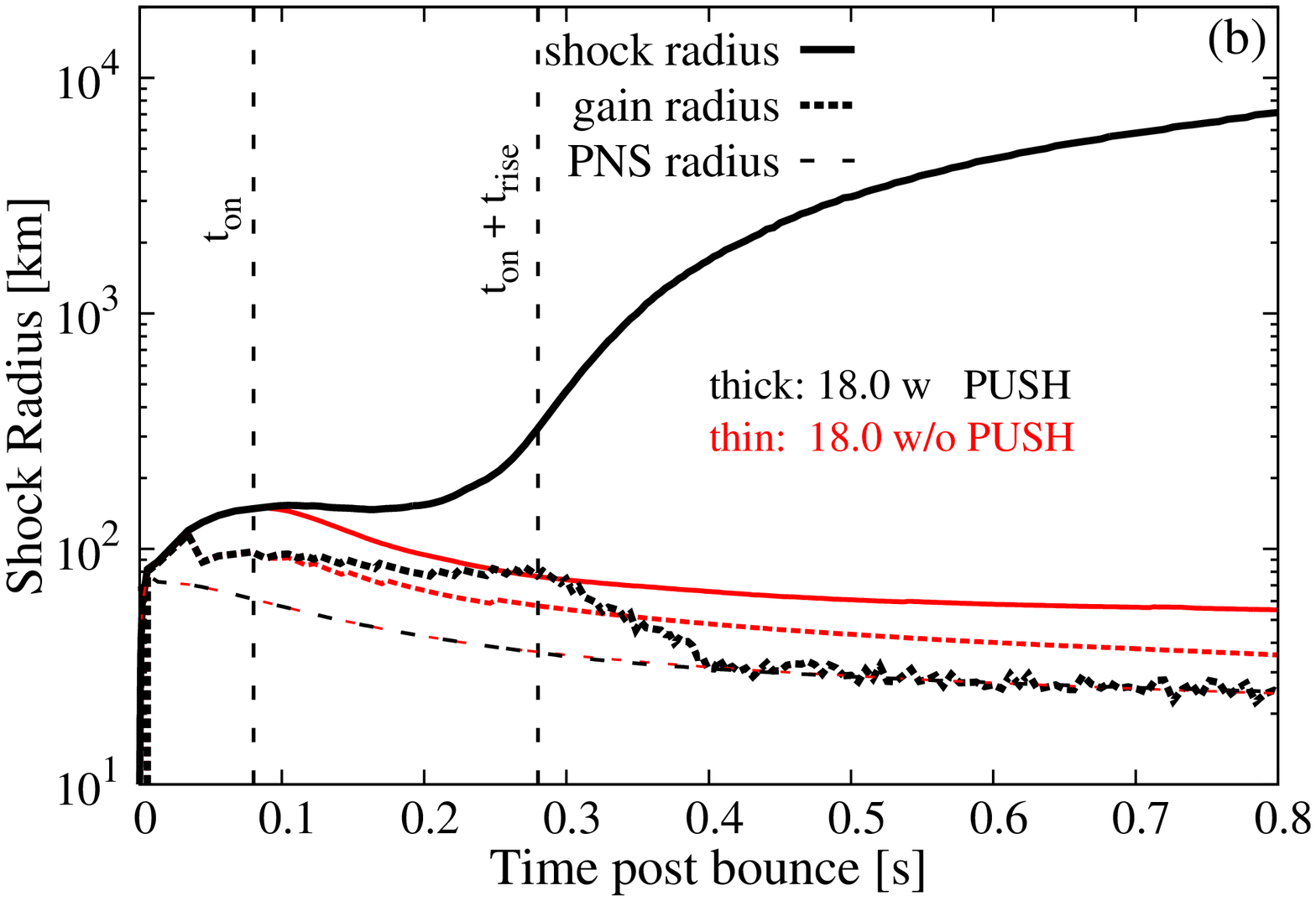}
 \\
 \includegraphics[width=0.35\textwidth,angle=-90]{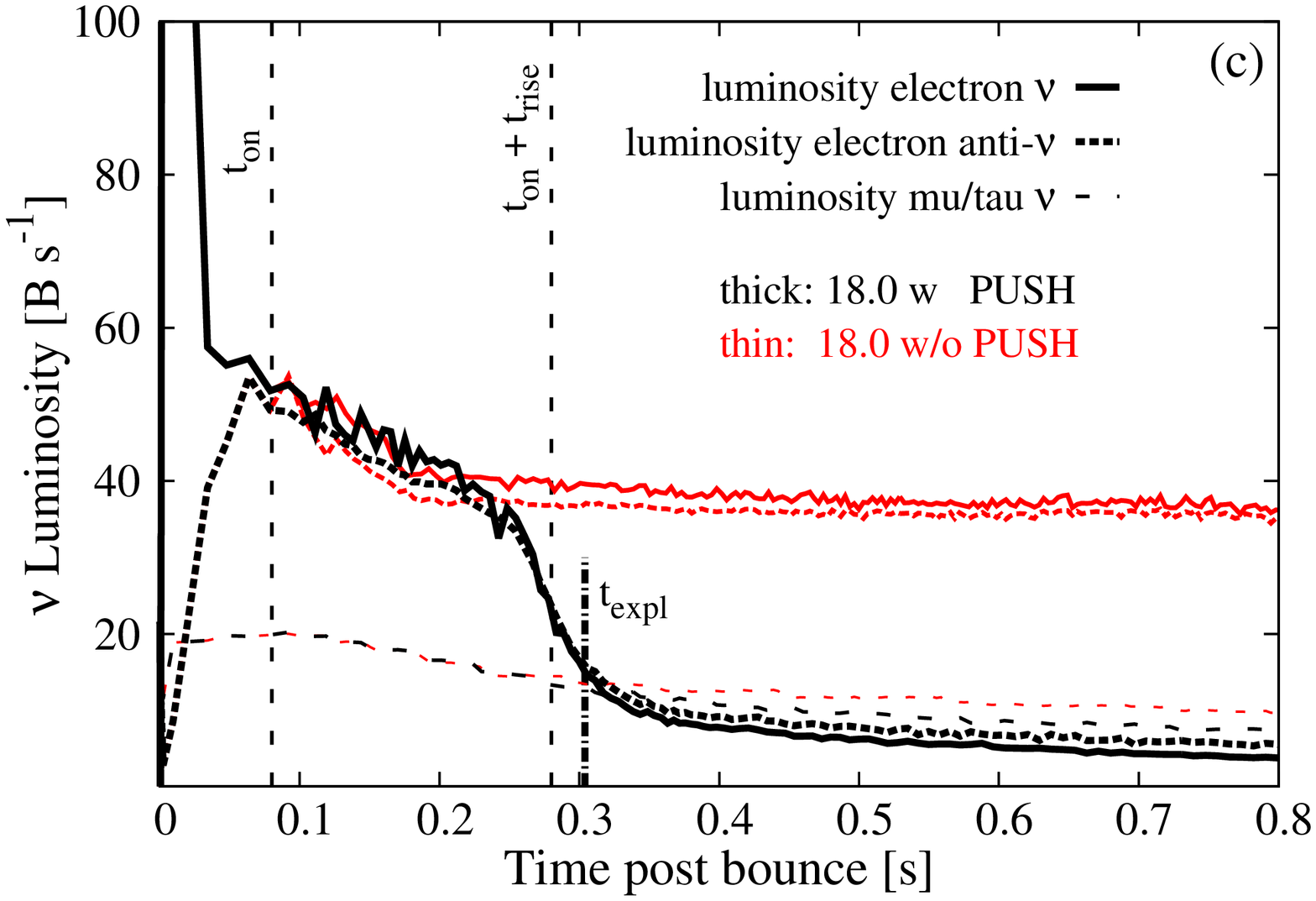}
 \includegraphics[width=0.35\textwidth,angle=-90]{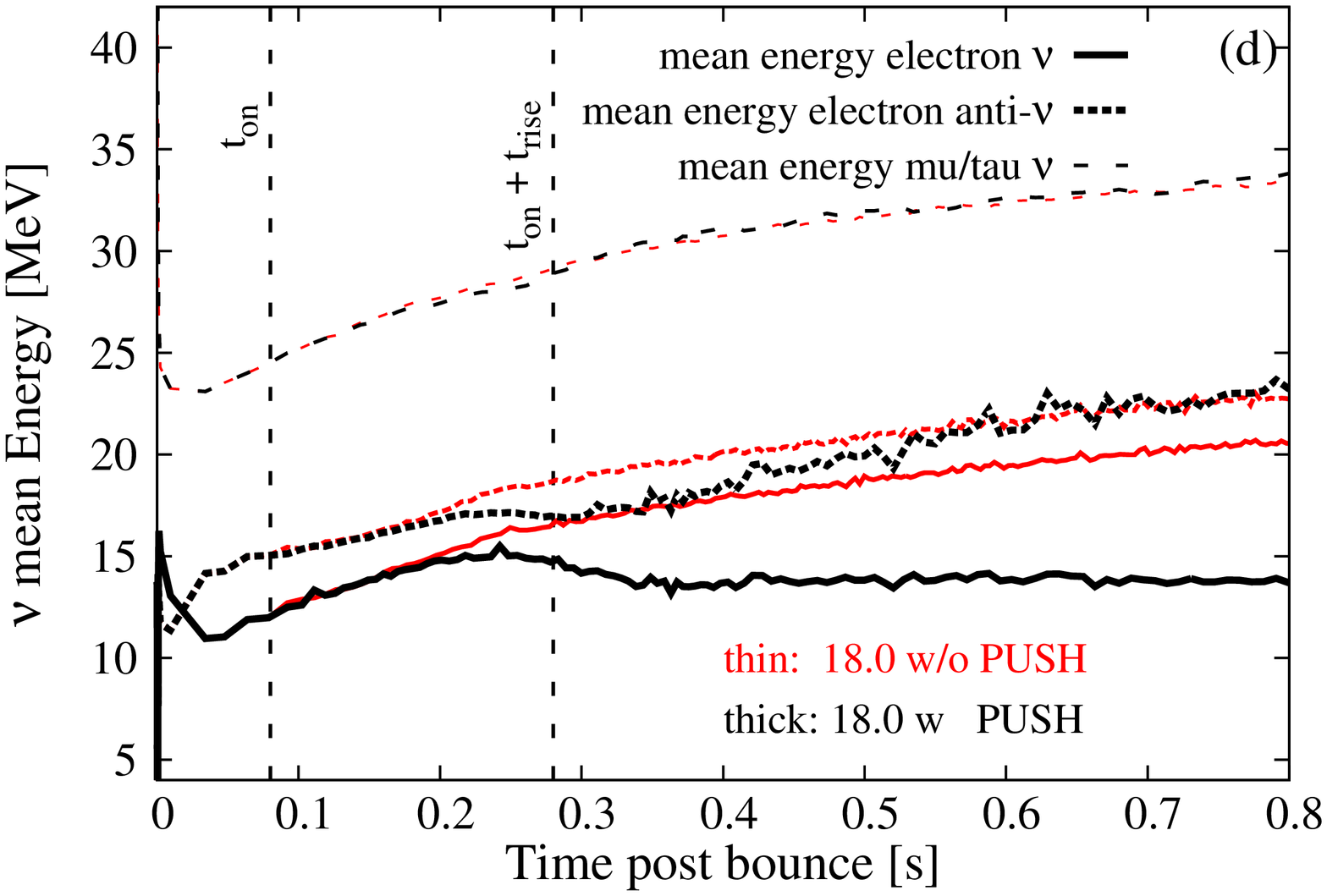} 
\end{tabular}  
\caption{Same as in Figure~\ref{fig:compactness comparison}, but for the SN~1987A best fit model: 18.0~\msun progenitor, with $t_{\rm on}=80$~ms, $t_{\rm rise}=200$~ms, and $k_{\rm push}=3.5$.
\label{fig:fitting model} }
\end{figure*}

The observed yield of $^{56}$Ni provides a strong constraint on which parameter combination would fit the data. From the observed yields of $^{57}$Ni and $^{58}$Ni, only the 18.0 and 19.4 progenitors remain viable candidates. Without fallback our predicted $^{44}$Ti yields are compatible with the observed yields (see Figure~\ref{fig:calibration_summary}).  However, if we include fallback (which is needed to explain the observed Ni yields), $^{44}$Ti becomes underproduced compared to the oberved value.  Since this behavior is true for all out models, we exclude the constraint given by $^{44}$Ti from our calibration procedure. From the considered parameter combinations, we obtained the best fit to SN~1987A for the 18.0~M$_{\odot}$ progenitor model with $k_{\rm push}=3.5$, $t_{\rm rise}=200$~ms, and a fallback of 0.1~M$_{\odot}$. These parameters are summarized in Table \ref{tab:bestfit}. In Figure~\ref{fig:fitting model}, we show the temporal evolution of the accretion rates, of the relevant radii, and of the 
neutrino luminosities and mean energies for our best fit model.
For comparison purposes, we present also the results obtained for the same model without PUSH. Note that in this non-exploding case the $\nu_e$ and $\bar{\nu}_e$ luminosities stay almost constant over several $ \sim 100$ ms after core bounce, despite the decreasing accretion rate. This is due to the relatively slow variation of $\dot{M}_{\rm PNS}$ (for example, compared with the variation obtained in the 19.2 \msun model, Figure~\ref{fig:compactness comparison}) and due to the simultaneous increase of the PNS gravitational potential, proportional to $M_{\rm PNS}/R_{\rm PNS}$ \citep[see, for example,][]{Fischer2009}.
A summary of the most important results of the simulations using this parameter set for the different progenitors in the 18-21~\msun window is given in Table~\ref{tab:values}. For the remnant mass and for the $^{56}$Ni yields of our best-fit model, we provide both the values obtained with and without assuming a fallback of $0.1$~\msun.

\begin{table}[h,t]
\caption{Parameter values for best fit to SN~1987A. \label{tab:bestfit}}
\begin{center}
  \begin{tabular}{cccc}
  \tableline \tableline
  \kpush & \trise & $t_{\rm on}$ & $t_{\rm off}$  \\
     ~(-) & (ms)  & (ms) & (s) \\
  \tableline
    3.5   & 200   & 80   & 1   \\
  \tableline
\end{tabular}
\tablecomments{We identified the 18.0~\msun model as the progenitor which fits best, whereas we had to impose a late-time fallback of 0.1~M$_{\odot}$. }
\end{center}
\end{table}

\begin{table}[ht] 
\begin{center}
\caption{Summary of simulations for $k_{\rm push}=3.5 $ and $t_{\rm rise}=200$~ms. \label{tab:values} }
\begin{tabular}{cccccc}
    \tableline \tableline
    ZAMS & $E_{\mathrm{expl}}$ & $t_{\mathrm{expl}}$ & M$^B_{\mathrm{remnant}}$ & M$^G_{\mathrm{remnant}}$ & M($^{56}$Ni ) \\
    (M$_{\odot}$) & (Bethe) & (s) & (M$_{\odot}$) & (M$_{\odot}$) & (M$_{\odot}$)  \\
    \tableline
    18.0  & 1.092  &  0.304  &  1.563  &  1.416  &  0.158  \\
    18.2  & 0.808  &  0.249  &  1.509  &  1.371  &  0.110  \\
    18.4  & 1.358  &  0.318  &  1.728  &  1.549  &  0.144  \\
    18.6  & 0.702  &  0.239  &  1.529  &  1.388  &  0.090  \\
    18.8  & 0.721  &  0.236  &  1.522  &  1.382  &  0.093  \\
    19.0  & 1.366  &  0.317  &  1.716  &  1.54   &  0.161  \\
    19.2  & 1.356  &  0.318  &  1.724  &  1.546  &  0.152  \\
    19.4  & 1.15   &  0.326  &  1.608  &  1.452  &  0.158  \\
    19.6  & 0.371  &  0.230  &  1.584  &  1.433  &  0.04   \\
    19.8  & 0.661  &  0.225  &  1.523  &  1.383  &  0.088  \\
    20.0  & 0.613  &  0.222  &  1.474  &  1.342  &  0.085  \\
    20.2  & 0.379  &  0.224  &  1.554  &  1.408  &  0.039  \\
    20.4  & 0.743  &  0.263  &  1.674  &  1.506  &  0.094  \\
    20.6  & 1.005  &  0.277  &  1.781  &  1.592  &  0.141  \\
    20.8  & 0.959  &  0.277  &  1.764  &  1.578  &  0.135  \\
    21.0  & 1.457  &  0.316  &  1.733  &  1.554  &  0.198  \\
    \tableline
    18.0 (fb) &  1.092  &  0.304  &  1.663  &  1.497  &  0.073   \\
    \tableline
\end{tabular}
\tablecomments{For the model 18.0 (fb), which is our best fit to SN~1987A, we have included 0.1~M$_{\odot}$ of fallback, determined from obervational constraints. See the text for more details. }
\end{center}
\end{table}

\subsection{Ni and Ti yields, progenitor dependence}  
\label{sec_yields}

\begin{figure*}[htp!]
   \includegraphics[width=0.5\textwidth]{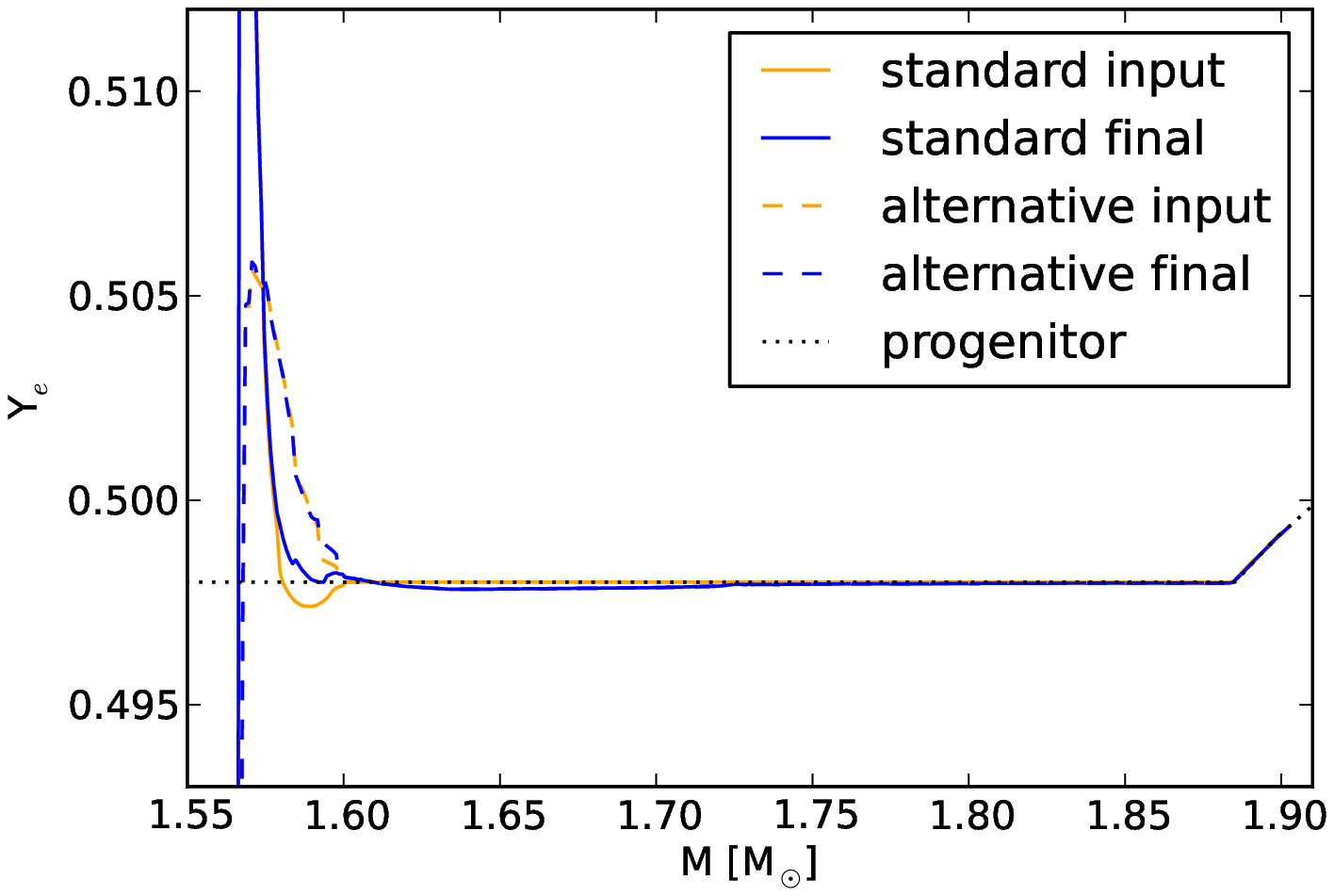}
   \includegraphics[width=0.5\textwidth]{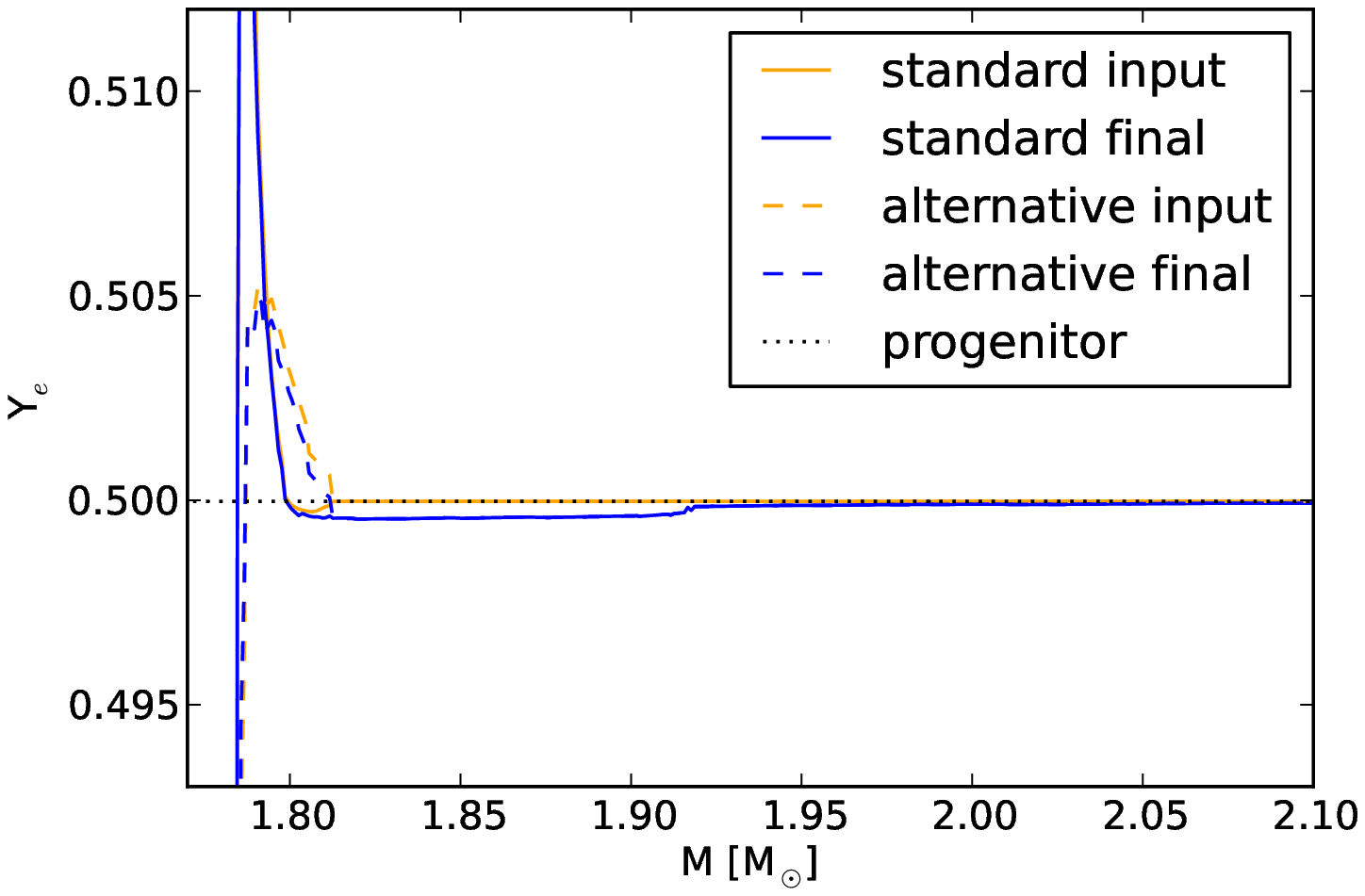}
   \includegraphics[width=0.5\textwidth]{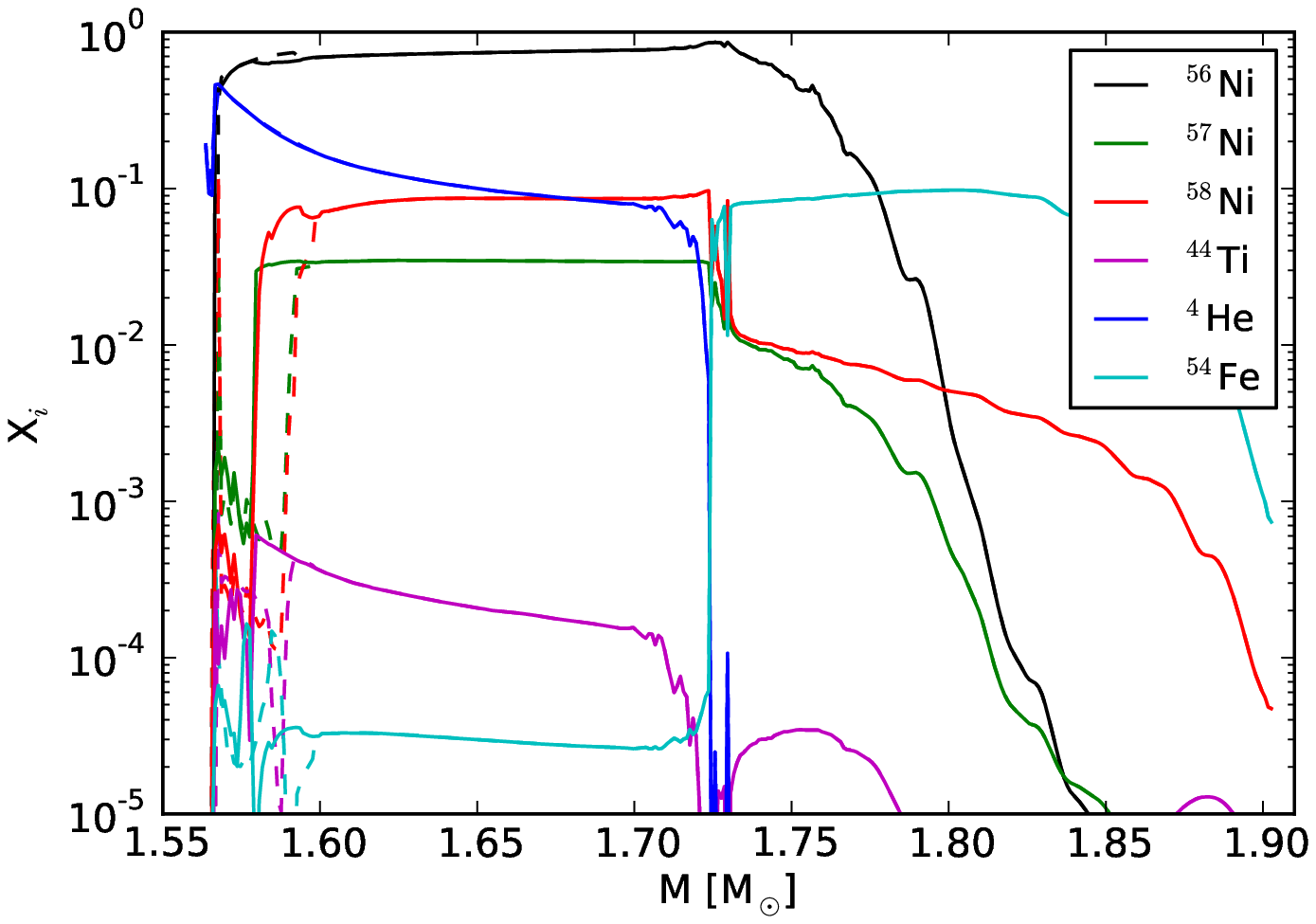} 
   \includegraphics[width=0.5\textwidth]{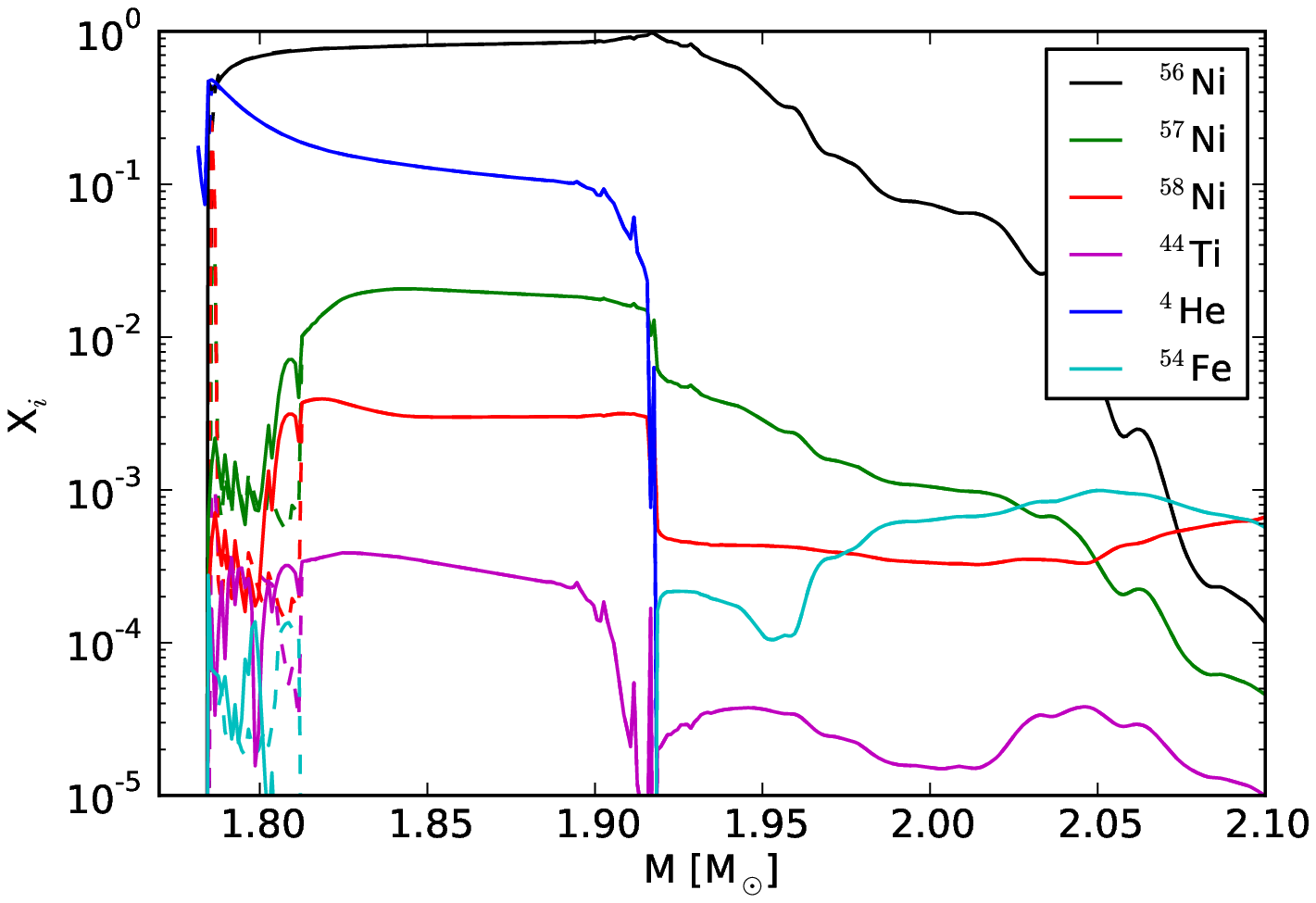} 
   \caption{Electron fraction profiles (top) and nuclear compositions at 100~s 
(bottom) above the mass cut for the 18.0~M$_{\odot}$ (left) and the 20.6~M$_{\odot}$ (right) progenitors with the parameters $k_{\rm push}=3.5$ and $t_{\rm rise}=200$~ms. The electron fraction is plotted for two different times in the network: the input values for the first timestep (``input'') and the value after post-processing (``final''). The dashed lines in all panels correspond to the alternative case, where $Y_e^{\rm hydro} (t=4.6~\rm s)$ is taken as the initial electron fraction in the network, whereas the solid lines represent the standard case (using $Y_e^{\rm hydro} (T=10~\rm GK)$).
   \label{fig:composition_ye_profiles}}
\end{figure*}

Figures~\ref{fig:calibration_summary} and \ref{fig:calibration_summary_fb} show that the composition of the ejecta is highly dependent on the progenitor model, especially for the amount of $^{57}$Ni and $^{58}$Ni ejected. From the four HC progenitors shown, two (18.0~M$_{\odot}$ and 19.4~M$_{\odot}$) produce a fairly high amount of those isotopes, while the other two (19.2~M$_{\odot}$ and 20.6~M$_{\odot}$) do not reach the amount observed in SN~1987A.  A thorough investigation of the composition profile of the ejecta reveals that $^{57}$Ni and $^{58}$Ni are mainly produced in the slightly neutron-rich layers ($Y_e < 0.5$), where the alpha-rich freeze-out leads to nuclei only one or two neutron units away from the $N=Z$ line. A comparison of the $Y_e$ and composition profiles for the 18.0~M$_{\odot}$ and the 20.6~M$_{\odot}$ progenitors is shown in Figure~\ref{fig:composition_ye_profiles}. For the 18.0~M$_{\odot}$ model, the cutoff mass is 1.56~M$_{\odot}$ and a large part of the silicon shell is ejected. In 
this shell, the initial matter composition is slightly neutron-rich (due to a small contribution from $^{56}$Fe) with $Y_e \simeq 0.498$ (dotted line in top left graph) and the conditions for the production of $^{57}$Ni and $^{58}$Ni are favorable. The increase in $Y_e$ around 1.9~M$_{\odot}$ marks the transition to the oxygen shell. The same transition for the 20.6~M$_{\odot}$ model happens around 1.74~M$_{\odot}$, i.e., inside the mass cut. Therefore, this model ejects less $^{57}$Ni and $^{58}$Ni (see also \citealt{thielemann90}). In all our models, $^{44}$Ti is produced within the innermost 0.15~M$_{\odot}$ of the ejecta (see Figure \ref{fig:composition_ye_profiles}). Since we assume 0.1~M$_{\odot}$ fallback onto the PNS, most of the synthesized $^{44}$Ti is not ejected in our simulations.

\section{Implications and Discussion}
\label{sec:discuss}

\subsection{Sensitivities of nucleosynthesis yields}  
\label{sec: ye dependence of nucleosynthesis}

While post-processing the ejecta trajectories for nucleosynthesis, $Y_e$ is evolved by the nuclear network independently of the hydrodynamical evolution. This leads to a discrepancy at later times between the electron fraction in the initial trajectory ($Y_e^{\rm hydro}$) and in the network ($Y_e^{\rm nuc}$). In order to estimate the possible error in our nucleosynthesis calculations arising from this discrepancy, we have performed reference calculations using $Y_e^{\rm hydro}(t=t_{\rm final})$ instead of $Y_e^{\rm hydro}(T=10\;{\rm GK})$ as a starting value for the network (see Section~\ref{sec:winnet}). The results are shown in Figure~\ref{fig:composition_ye_profiles} for two progenitors: 18.0~M$_{\odot}$ and 20.6~M$_{\odot}$. The label ``standard'' refers to the regular case which uses $Y_e^{\rm hydro}(T=10\;{\rm GK})$ as input. The calculation using $Y_e^{\rm hydro}(t=t_{\rm final})$ as input is labeled ``alternative'' and is represented by the dashed lines. The point in time at which the $Y_e$ profile 
is shown is indicated by the supplements ``input'' (before the first timestep) and ``final'' (at $t=100$~s). 
The corresponding nuclear compositions of the ejecta, each at the final calculation time of 100~s, are shown in the bottom panels. For the alternative 
$Y_e$ profile of the 18.0~M$_{\odot}$ progenitor (top left) the minimum around 1.59~M$_{\odot}$ disappears, leading to an increase in $^{56}$Ni in this region at the expense of $^{57}$Ni and $^{58}$Ni (bottom left). For the 20.6~M$_{\odot}$ progenitor the situation is similar, with only a very small region just above 1.8~M$_{\odot}$ showing significant differences. In general, we observe that the uncertainties in $Y_e$ in our calculations are only present up to 0.05~\msun above the mass cut. The resulting uncertainties in the composition of the ejecta are very small or even inexistent in the scenarios where we consider fallback. 

The radioactive isotope $^{44}$Ti can be detected in supernovae and supernova remnants. Several groups have used different techniques to estimate the $^{44}$Ti yield \citep{chugai97,Fransson.Kozma:2002,jerkstrand11,larsson11,grebenev12,grefenstette14,Seitenzahl2014}. The inferred values span a broad range, $(0.5 - 4) \times 10^{-4}$~\msun. Traditional supernova nucleosynthesis calculations \citep[e.g.][]{tnh96,ww95} typically predict too low $^{44}$Ti yields. Only very few models predict high $^{44}$Ti yields:  \citet{thielemann90} report $^{44}$Ti yields around $10^{-4}$ and above in the best fits of their artificial SN explosions to SN~1987A.  \citet{rauscher2002} argue that the yields of $^{56}$Ni and $^{44}$Ti are very sensitive to the ``final mass cut'' (as we have shown, too), which is often determined by fallback. Ejecta in a supernova may be subject to convective overturn. To account for this, we can assume homogeneous mixing in the inner layers up to the outer boundary of the silicon shell before 
cutting off the fallback material (see, for example, \cite{Umeda2002} and references therein). For our best-fit model, the ejected $^{44}$Ti mass increases to $2.70~\times~10^{-5}$~M$_{\odot}$, if this prescription is applied. Comparing to the previous yield of $1.04~\times~10^{-5}$~M$_{\odot}$, we observe that the effect of homogeneous mixing is considerable, but not sufficient to match the observational values. The ejected $^{56-58}$Ni masses also show a slight increase. However, there are also uncertainties in the nuclear physics connected to the production and destruction of $^{44}$Ti. The final amount of produced $^{44}$Ti depends mainly on two reactions: $^{40}$Ca($\alpha,\gamma$)$^{44}$Ti and $^{44}$Ti($\alpha,p$)$^{47}$V. Recent measurements of the $^{44}$Ti($\alpha,p$)$^{47}$V reaction rate within the Gamow window concluded that it may be considerably smaller than previous theoretical predictions \citep{margerin2014}. In this study, an upper limit cross section is reported that is a factor of 2.2 
smaller than the cross section we have used in our calculations (at a confidence level of $68\%$). Using this smaller cross section for the $^{44}$Ti($\alpha,p$)$^{47}$V reaction, our yield of ejected $^{44}$Ti for our best-fit model (18.0~M$_{\odot}$ progenitor, $k_{\rm push}=3.5$, $t_{\rm rise}=200$~ms) rises to $1.49~\times~10^{-5}$~M$_{\odot}$ with fallback and $5.65~\times~10^{-5}$~M$_{\odot}$ without fallback. This corresponds to a relative increase of $43\%$ with fallback and $48\%$ without fallback. If we include both the new cross section and homogeneous mixing, the amount of $^{44}$Ti in the ejecta is $3.99~\times~10^{-5}$~M$_{\odot}$ including fallback. This value, however, is still below the expected value derived from observations, but within the error box.

\subsection{Wind ejecta}  
In the analysis of the nucleosynthesis yields above, we have used a mass resolution of 0.001~\msun for the tracers. This is too coarse to resolve the ejecta of the late neutrino-driven wind. Note that in our best-fit approach, where no mixing is assumed, none of the neutrino-driven wind is ejected because it is part of the fallback. Nevertheless, in the following we report briefly on the properties of the wind obtained by our detailed neutrino-transport scheme.  For our best-fit model, the 18.0~M$_{\odot}$ progenitor, at $t_{\rm final}$ we find an electron fraction around 0.32, entropies up to 80~$k_{B}$ per baryon, and fast expansion velocities ($\sim 10^9$~cm/s). Similar conditions are also found for the other progenitors. They are not sufficient for a full r-process (see, for example, \citet{farouqi2010}). On the other hand, we have found that the entropy is still increasing and the electron fraction still decreasing in the further evolution. The high asymmetries are only obtained if we include the 
nucleon mean-field interaction potentials in the neutrino charged-current rates \citep{martinez12}.  However, they are much higher than found in other long-term simulations which also include these potentials \citep{roberts12,martinez12,martinez14}. This could be related to the missing neutrino-electron scattering in our neutrino transport, which is an important source of thermalization and down-scattering, especially for the high energy electron antineutrinos at late times, see \citet{fischer12}. More detailed comparisons are required to identify the origin of the found differences which will be addressed in a future study.

\subsection{Amount of fallback}
\label{sec_fallback}

To reconcile our models with the nucleosynthesis observables of SN~1987A we need to invoke 0.1~\msun of fallback (see Section \ref{sec: fit}). The variation in the amount of synthesized Ni isotopes between runs obtained with different PUSH parameters (Figure~\ref{fig:calibration_summary}) suggests that a smaller $t_{\rm rise}$ (and consequently smaller $k_{\rm push}$) could also be compatible with SN~1987A observables, if a larger fallback is assumed. On the one hand, assuming that $t_{\rm rise}$ ranges between 50 ms and 250 ms, fallback for the 18.0~\msun model compatible with observations is between 0.14~\msun (for $t_{\rm rise}=50$~ms) and 0.09~\msun (for $t_{\rm rise}=250$~ms). On the other hand, if the amount of fallback has been fixed, the observed yields (especially of $^{56}$Ni) reduce the uncertainty in $t_{\rm rise}$ to $\lesssim 50$~ms.

Our choice of 0.1~\msun is compatible with the fallback obtained by \cite{Ugliano.Janka.ea:2012} in exploding spherically symmetric models for progenitor stars in the same ZAMS mass window. Moreover, \citet{chevalier89} estimated a total fallback around 0.1~\msun for SN~1987A, which is supposed to be an unusually high value compared to ``normal'' type II supernovae. Recent multi-dimensional numerical simulations by \citet{bernal13,fraija14} confirmed this scenario and furthermore showed that such a hypercritical accretion can lead to a submergence of the magnetic field, giving a natural explanation why the neutron star (possibly) born in SN~1987A has not been found yet.

\subsection{Compact Remnant of SN~1987A}  

From the observational side, the compact remnant in SN~1987A is still obscure.  From the neutrino signal (see, e.g., \citet{arnett89,koshiba92} and \citet{vissani14} for a recent detailed analysis) one can conclude that a PNS star was formed and that it lasted at least for about 12~s. The mass cut in our calibration run is located at an enclosed baryon mass of  1.56~\msun without fallback. If we include the 0.1~\msun of late-time fallback required to fit the observed nickel yields and the explosion energy, we have a final baryonic mass of 1.66~\msun. For the employed HS(DD2) EOS this corresponds to a gravitational mass of a cold neutron star of 1.42~\msun (without fallback) or 1.50~\msun (with fallback). The CCSN simulations with artificial explosions of \citet{thielemann90}, where a final kinetic energy of 1~Bethe was obtained by hand and where the mass-cut was deduced from a $^{56}$Ni yield of $(0.07\pm 0.01)$~\msun, lead to a similar baryonic mass of $(1.6 \pm 0.045)$~\msun. These authors also wrote that 
uncertainties in the stellar models could increase this value to $1.7$~\msun which would also be fully compatible with our result.

The prediction of the neutron star mass has important consequences. From the observations of \citet{demorest2010} and \citet{antoniadis2013} it follows that the maximum gravitational mass of neutron stars has to be above two solar masses. The maximum mass of the HS(DD2) EOS is 2.42~\msun, corresponding to a baryonic mass of 2.92~\msun. If the compact remnant in SN~1987A was a black hole, and not a neutron star, it means that at least $\sim 0.5$~\msun of additional accreted mass were required, if we just take the two solar mass limit. If we use the maximum baryonic mass of HS(DD2) we even have to accrete $\sim$1.3~\msun of additional material. Obviously, if such a huge amount of material would be accreted onto the neutron star, our predictions for the explosion energy and the nucleosynthesis would not apply any more. 

Nevertheless, we have the impression that it would be difficult to fit the SN~1987A observables and obtain a black hole as the compact remnant at the same time. For spherical fallback, it is certainly excluded. The only possibility could be a highly anisotropic explosion and aspherical accretion, which we cannot address with our study. To show if such a scenario can be realized remains a task for future multi-dimensional studies. In the 2D simulations of \citet{yamamoto2013} the remnant mass is decreasing with the explosion energy and an explosion energy above 1~Bethe would result in neutron stars below $\sim 2$~\msun baryonic mass.  Note that \citet{kifonidis2006} already came to the same conclusion that the formation of a black hole in SN~1987A ``is quite unlikely'', based on 2D simulations with a 15~\msun progenitor. 

Another possibility was proposed by \citet{chan2009}. These authors argued that the time delay of $\sim 5$~s observed for the neutrino signal by the IMB detector could be related to a collapse to a quark star. Due to the proposed faster neutrino cooling of quark stars, this would give a natural explanation why it has not been observed until today. The end of our simulations is also around 5~s, thus we can make statements about the conditions at which the phase transition to quark matter took place in SN~1987A, if the scenario of  \citet{chan2009} was true. We have a central mass density of $4.56 \times 10^{14}$~g/cm$^3$ corresponding to $n_B^c=0.272$~fm$^{-3}$ or $n_B^c=1.83~n_B^0$, a temperature of 23.2~MeV, and an electron fraction of 0.24. Some simplified models for quark matter predict that the phase transition in symmetric matter is shifted to higher densities compared with supernova conditions \citep{fischer11}. Under that hypothesis, a phase transition around 2~$\rho_0$ and 20 MeV cannot be excluded.

A simpler explanation is given by the possibility that a pulsar in the SN~1987A remnant is simply not (yet) observable. \citet{oegelman2004,graves2005} showed that the non-detection of any compact remnant puts important limits on the magnetic field the NS can have (either unusually low or very high, in the realm of magnetars). Furthermore, for both cases (NS and BH) \citet{graves2005} put severe constraints on currently ongoing accretion scenarios, e.g., spherical accretion is almost ruled out. \citet{graves2005} conclude that ``it seems unlikely that the remnant of SN~1987A currently harbors a pulsar''. Our simulations would be in line with the option of a neutron star with a very low magnetic field or with a ``normal'' magnetic field which is still (partly) buried in the crust due to the late time fallback, similar to what is observed for neutron stars in binary systems. In this respect, recent high-resolution radio observations of the remnant indicate the presence of a compact source or a pulsar wind 
nebula \citep{zanardo13,Zanardo2014}. Future observations will be able to clarify the nature of this emission.

\subsection{Correlations}  

\begin{figure}[hbp!]  
   \includegraphics[width=0.5\textwidth]{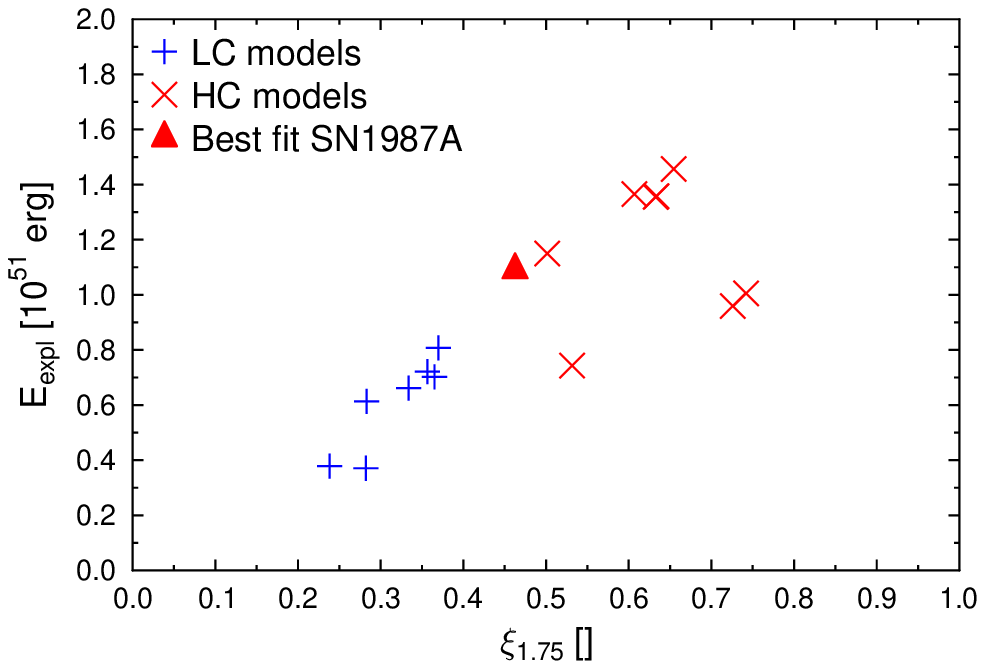}
   \includegraphics[width=0.5 \textwidth]{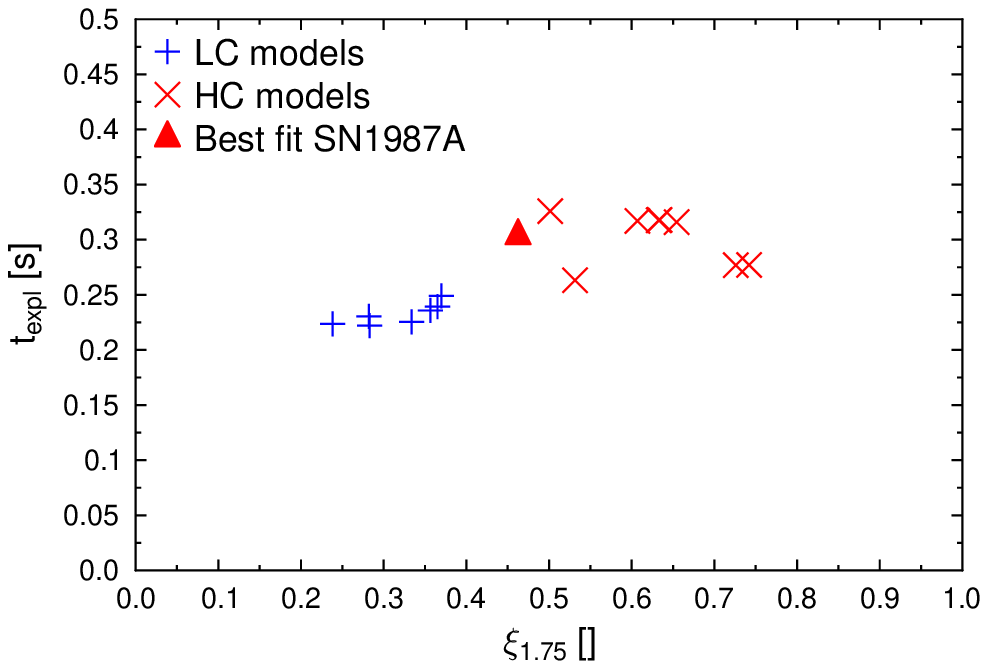}
   \includegraphics[width=0.5\textwidth]{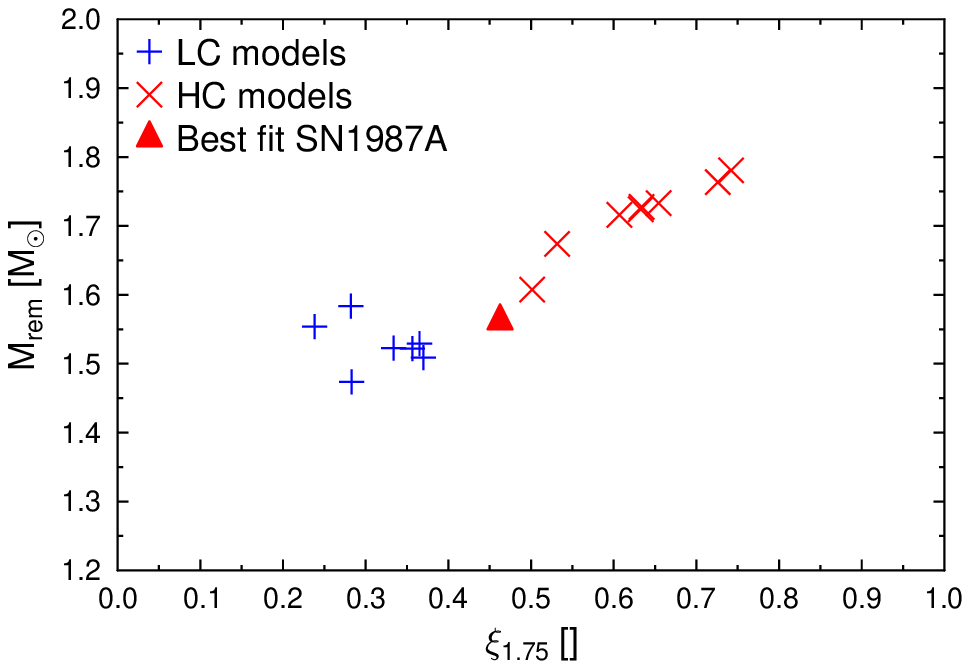}
   \caption{Explosion energies (top), explosion times (middle), and (baryonic) remnant mass (bottom) as function of compactness for the PUSH parameters of our best-fit model (\kpush $=3.3$ and $t_{\rm rise}=0.15$~s) for all progenitors in the 18-21~\msun window. HC models are denoted by a red cross, LC models by a blue plus. Our best-fit model for SN~1987A is highlighted by a filled triangle.
   \label{fig:correls_with-xi} }
\end{figure}

As a byproduct of exploring the 18-21~\msun window and the fitting procedure to SN~1987A we have found interesting correlations between different quantities, which we will discuss here. In Figure~\ref{fig:correls_with-xi}, we plot the explosion energy, the explosion time, and the (baryonic) remnant mass as function of the progenitor compactness. The results obtained with the calibrated runs indicate a general trend with progenitor compactness for $E_{\rm expl}$. The explosion time, $t_{\rm expl}$, is almost constant within each the LC and the HC group, while the difference between the two groups is related to the difference between how LC and HC models explode (discussed in Section~\ref{sec:hc and lc}). The remnant mass increases with compactness, as expected. Nevertheless, we notice significant deviations from the described trends: for $E_{\rm expl}$ and $t_{\rm expl}$ in the HC sample, for $M_{\rm rem}$ mainly in the LC sample.

\begin{figure}[htp!] 
   \includegraphics[width=0.5\textwidth]{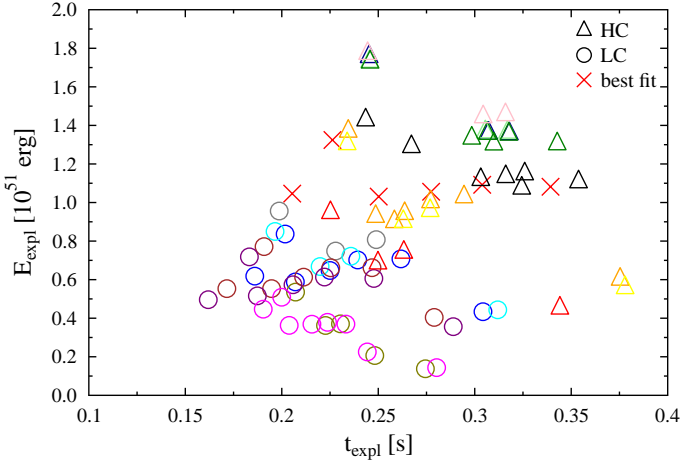} \\
   \caption{Explosion energy $E_{\rm expl}$ versus explosion time $t_{\rm expl}$ for all the progenitors in the 18-21~\msun range and for different combinations of \kpush and $t_{\rm rise}$, however only the exploding models are included. HC models are indicated by a triangle, LC models by a circle. The best fit model is indicated by a cross. The different colors distinguish different progenitors.
   \label{fig:correls_Eexpl-texpl} }
\end{figure}

Figure~\ref{fig:correls_Eexpl-texpl} shows explosion times and explosion energies for all the exploding runs in our sample. We can identify a correlation between $t_{\rm expl}$ and $E_{\rm expl}$ for a given progenitor: the larger $t_{\rm expl}$ the lower is $E_{\rm expl}$. This correlation is more pronounced for the HC models than for the LC models.  It means that the explosion in PUSH cannot set in too late, if the observed explosion energy should be achieved.

\subsection{Heating efficiency and residence time}

\begin{figure}[htp!]  
 \includegraphics[width=0.35\textwidth,angle=-90]{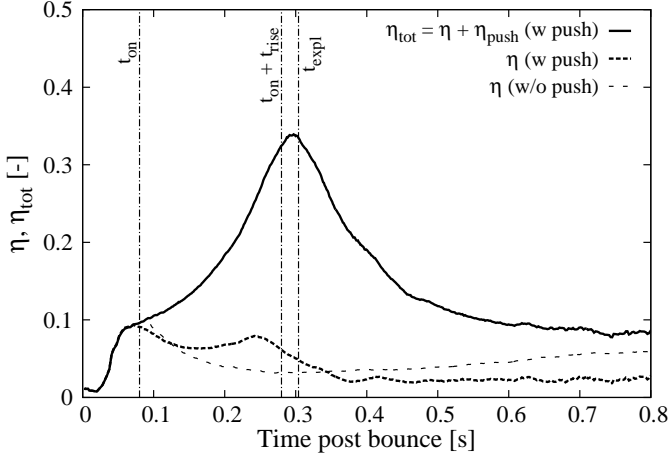}
 \caption{Neutrino heating efficiency for the SN~1987A best fit model: 18.0~\msun model with $t_{\rm on}=80$~ms, $t_{\rm rise}=200$~ms, and $k_{\rm push}=3.5$. The solid lines represent the total efficiency (i.e., due to $\nu_e$ and $\bar{\nu}_e$ absorption and due to PUSH), the short-thick dashed lines the efficiency only due to $\nu_e$ and $\bar{\nu}_e$ absorption. For comparison, the heating efficiency of the corresponding non-exploding model ($k_{\rm push} = 0$) is also presented (long-thin dashed lines). 
\label{fig:eta plot} }
\end{figure}

In the context of CCSNe, the heating efficiency $\eta$ is often defined as the ratio between the volume-integrated, net energy deposition inside the gain region and the sum of the $\nu_e$ and $\bar{\nu}_e$ luminosities at infinity:
\begin{equation}
 \eta = \frac{\int_{V_{\rm gain}} \rho \, \dot{e}_{\nu_e,\bar{\nu}_e}{\rm d}V}{L_{\nu_{e}} + L_{\bar{\nu}_{e}}},
\end{equation}
see, e.g., \citet{Murphy2008,Marek2009,Mueller.Janka.ea:2012,Couch2014,Suwa2013}.
In non-exploding, spherically symmetric simulations, $\eta$ usually rises within a few tens of milliseconds after core bounce and reaches its maximum around $\eta \sim 0.1$ at $t \approx 100 \, \rm{ms}$, when the shock approaches its maximum radial extension. As soon as the shock starts to recede and the volume of the gain region decreases, $\eta$ diminishes quickly to a few percents (see, for example, the long-dashed lines in Figure~\ref{fig:eta plot}).

In multi-dimensional simulations, where the shock contraction is delayed or even not happening, energy deposition is expected to be slightly more efficient ($\eta \sim $ 0.10 -- 0.15 at maximum) and to decrease more slowly, within a few hundreds of milliseconds after bounce or at the onset of an explosion \citep[see, for example,][]{Murphy2008,Mueller.Janka.ea:2012,Couch2013a,Couch2014}. These differences arise not only because the gain region does not contract, but also because neutrino-driven convection efficiently mixes low and high entropy matter between the neutrino cooling and the heating regions below the shock front.  Furthermore, convective motion and SASIs are expected to increase significantly the residence time of fluid particles inside the gain region during which they are subject to intense neutrino heating \citep[see, e.g., ][]{Murphy2008,handy14}. Since the increase of the particle internal energy is given by the time integral of the energy absorption rate over the residence time, this 
translates to a larger energy variation \citep{handy14}.

In spherically symmetric models, the imposed radial motion does not allow the increase of the residence time. This constraint limits the energy gain of a mass element traveling through the gain region. In models exploded using the light-bulb approximation, a large enough internal energy variation is provided by increasing the neutrino luminosity above a critical value, which depends on the mass accretion rate and on the dimensionality of the model
\citep[e.g.,][]{Burrows1993,Yamasaki2005,iwakami08,Murphy2008,iwakami09,Nordhaus2010,Hanke2012,Couch2013a,Dolence2013,handy14,suwa14}.
Since in our model the neutrino luminosities are univocally defined by the cooling of the PNS and by the accretion rate history, we increase the energy gain by acting on the neutrino heating efficiency. This effect can be made visible by defining a heating efficiency that takes the PUSH contribution into account, $\eta_{\rm tot}$:
\begin{equation}
 \eta_{\rm tot} = \eta + \eta_{\rm push} = \frac{\int_{V_{\rm gain}} \rho \, \left( \dot{e}_{\nu_e,\bar{\nu}_e} + \dot{Q}^{+}_{\rm push} \right){\rm d}V} {L_{\nu_{e}} + L_{\bar{\nu}_{e}}}.
\end{equation}

In Figure~\ref{fig:eta plot}, we plot $\eta_{\rm tot}$ as a function of time for our SN~1987A calibration model, with PUSH ($k_{\rm push} = 3.5$) and without it ($k_{\rm push} = 0$). We first notice that the heating efficiency provided by $\nu_{e}$ and $\bar{\nu}_{e}$ can differ between exploding (short-thick dashed lines) and non-exploding models (long-thin dashed lines). In the case of the exploding model, PUSH provides an increasing contribution to $\eta_{\rm tot}$. It continues to increase steeply up to $t \approx t_{\rm on} + t_{\rm rise}$, but also later, up to $t \approx t_{\rm expl}$, due to the shock expansion preceding the explosion. Thus, the increasing heating efficiency in our spherically symmetric models can be interpreted as an effective way to include average residence times longer than the advection timescale.

\begin{figure}[htp!]
   \includegraphics[width=0.35\textwidth,angle=-90]{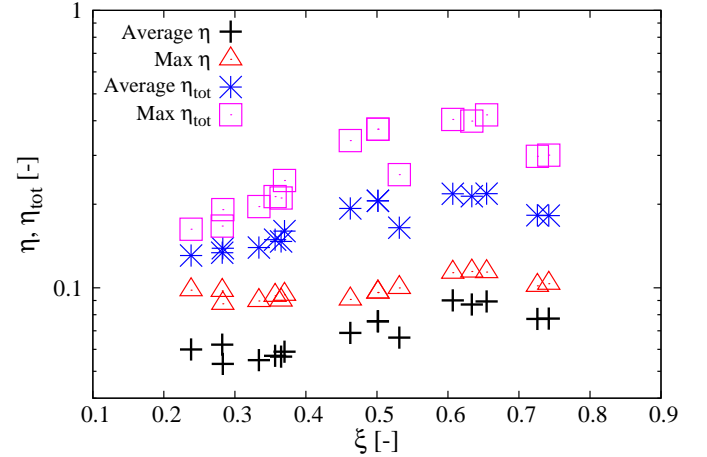}
   \caption{Average and maximum heating efficiencies, calculated between $t = t_{\rm on}$ and $t = t_{\rm expl}$ for the runs obtained with the fitted parameters, Table~\ref{tab:bestfit}, and plotted as a function of the progenitor compactness $\xi_{1.75}$. The black crosses and the red triangles refer to the average and the maximum efficiency due to $\nu_e$ and $\bar{\nu}_e$ ($\eta$), while the blue stars and the magenta squares to the average and the maximum total efficiency ($\eta_{\rm tot}$), including also the PUSH contribution.
   \label{fig:av_eta}}
\end{figure}

In Figure~\ref{fig:av_eta}, we collect the average and the maximum heating efficiencies, for all the models obtained with the set of parameters resulting from the fit procedure (Table~\ref{tab:bestfit}). Both the average and the maximum values are computed within the interval $ t_{\rm on} \le t \le t_{\rm expl}$.  We plot them as a function of the compactness and we distinguish between $\eta$ and $\eta_{\rm tot}$. The maximum of $\eta$ is usually realized at $t \approx t_{\rm on}$, while the maximum of $\eta_{\rm tot}$ is reached around $t \approx t_{\rm expl}$ (see also Figure~\ref{fig:eta plot}). Since the explosion sets in later for HC models, when $t_{\rm expl} \gtrsim t_{\rm on} + t_{\rm rise}$, the PUSH factor $\mathcal{G}$ has time to rise to \kpush for these models. This increases not only the maximum but also the average $\eta_{\rm tot}$ compared with the LC cases. We notice that all four quantities show a correlation with $\xi_{1.75}$, but much weaker in the case of $\eta$ than in the case of $\eta_
{\rm tot}$. Moreover, in the HC region, we recognise deviations from monotonic behaviors which reproduce the irregularities already observed in the explosion properties.

\subsection{Alternative measures of the explosion energy}

In the following, we discuss alternative measures of the explosion energy used in the literature for reasons of comparison. We investigate their behaviors at early simulation times and their general rate of convergence. The diagnostic energy $E^+(t)$, see e.g.~\citet{Bruenn.Mezzacappa.ea:2013}, is given by the integral of the specific explosion energy $e_{\rm expl}$ over regions where it is positive (again, excluding the PNS core, see Section~\ref{sec:def_eexpl}). The quantity $E^+(t)$ is often used in multi-dimensional simulations as an estimate of the explosion energy at early simulation times, see e.g.~\citet{buras2006b,suwa2010,janka2012,Couch2014,takiwaki2014}. 

The overburden $E_{\rm ov}(t)$, see \cite{Bruenn.Mezzacappa.ea:2013}, is given by the integral of the specific explosion energy of the still gravitationally bound regions between the expanding shock front and the surface of the progenitor star. If we define $E_{\rm ov}^+(t)$ as the sum of the overburden and of the diagnostic energy, we recover a measure of the explosion energy equivalent to the one defined in Equation~(\ref{eq:expl energy t}): 
\begin{equation}
  E_{\rm expl}(t)\equiv E_{\rm ov}^+(t)=E^+(t) + E_{\rm ov}(t).
\end{equation}
For long enough simulation times, all matter above the mass-cut should get positive specific explosion energies, and thus the overburden should approach zero and the diagnostic energy should become equal to the explosion energy $E_{\rm expl}(t)$. 

Finally, an upper limit for the explosion energy is obtained by also taking into account the ``residual recombination energy'' $E_{\rm rec}(t)$ \citep{Bruenn.Mezzacappa.ea:2013}:
\begin{equation}
  E_{\rm ov,r}^+(t)=E_{\rm ov}^+(t) + E_{\rm rec}(t) \, ,
\end{equation}
where $E_{\rm rec}(t)$ is the energy that would be released if all neutron-proton pairs and all \textsuperscript{4}He recombined to \textsuperscript{56}Ni in the regions of positive specific explosion energy. We call it \textit{residual} recombination energy to make clear that this is energy which is \textit{not} liberated in our simulations, in contrast to the energy of the recombination processes which we identified in Section~\ref{sec:eexpl_contr}.

\label{sec:eexpl_evo}
\begin{figure}[htp!]
   \includegraphics[angle=-90,width=0.5\textwidth]{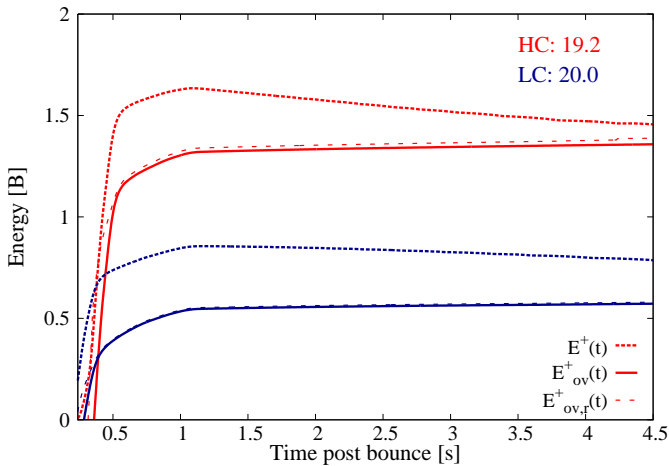} 
   \caption{Temporal evolution of the diagnostic energy $E^+$, the explosion energy $E^+_{\rm ov}$, and the upper limit of the explosion energy also including the recombination energy $E_{\rm rec}$ for a HC (19.2~\msun progenitor) and a LC case (20.0~\msun progenitor), for PUSH parameters reported in Table~\ref{tab:HC vs LC}.
   \label{fig:diagexplosion_energies}}
\end{figure}

In Figure \ref{fig:diagexplosion_energies}, we investigate the behavior of the diagnostic energy $E^+(t)$, and we compare it with our estimate of the explosion energy $E_{\rm expl}(t)\equiv E_{\rm ov}^+(t)$ and with its upper limit represented by $E^+_{\rm ov,r}(t)$. We want to emphasize that these quantites are obtained from mass integrals above the time-dependent mass-cut, in contrast to most of the energies investigated in Section~\ref{sec:eexpl_contr}, where a fixed mass domain was considered.

While $E_{\rm ov}^+(t)$ and $E^+_{\rm ov,r}(t)$ have already saturated to a constant value at $t\approx1.5$~s, even at $t\approx4.6$~s the diagnostic energy has not yet converged. $E_{\rm ov}^+(t)$ and $E^+_{\rm ov,r}(t)$ approach their asymptotic values from below, and any late time increase ($t \gtrsim 1.5$~s) is due to the energy carried by the neutrino-driven wind ejected from the PNS surface. On the other hand, $E^+(t)$ reaches its maximum around $\la 1$~s after $t_{\rm expl}$, when the neutrino absorption and the nuclear recombination have released most of their energy in the expanding shock wave, and then it decreases towards $E_{\rm ov}^+(t)$, since matter with negative total specific energy is accreted at the shock. The difference between $E_{\rm ov}^+(t)$ and $E^+(t)$ is mainly represented by the gravitational binding energy of the stellar layers above the shock front. Thus, the rate of convergence of the diagnostic energy depends on the amount of gravitational binding energy contained in the outer 
envelope of the star and on the relative speed at which the shock propagates inside it. Since the gravitational binding energy of the outer layers is similar between the two explored models, the different rate of convergence depends mostly on the different expansion velocity of the shock wave, which is larger for more energetic HC models.

\citet{yamamoto2013} found for a 15~\msun progenitor that the diagnostic energy saturates and thus reaches the asymptotic explosion energy already between 1 and 2~s post-bounce. This difference to what we find is related to the different progenitors used and, in particular, to the different binding energy of the outer envelopes, which is expected to be much smaller for a $15$~\msun progenitor than for a $\sim 20$~\msun progenitor \citep[see, for example, Figure 5 of][]{burrows13}. Nevertheless, we conclude that the diagnostic energy is in general (i.e., without further considerations) not suited to give an accurate estimate of the explosion energy at early times.

\subsection{Comparison with other works}

A similar fitting to SN~1987A energetics has been done for multi-dimensional simulations (2D and 3D) using a light-bulb scheme for the neutrinos by \citet{handy14}. As initial conditions they used a post-collapse model based on the 15~\msun blue supergiant progenitor model of \citet{Woosley1988}. Even if they did not provide the corresponding compactness, the values of the accretion rate ($\sim 0.2-0.3$~\msun~s$^{-1}$) and of the electron neutrino luminosity ($\sim 1.8-3.5 \times 10^{51}$~erg~s$^{-1}$) at the onset of the explosion  are more compatible with our LC models. In their fitting, only the diagnostic explosion energy $E^+$ was used at a time of $t_{\rm pb} = 1.5$~s when it is expected to have saturated to $E_{\rm expl}$ (cf. \cite{yamamoto2013}). But no estimates for the nucleosynthesis yields were given. The time when the shock reaches 500 km (which corresponds for us to $t_{\rm expl}$) is significantly lower in their models (90-140 ms after bounce), mainly due to the different extension and 
evolution of the shock during the first 100 ms after core bounce. A more detailed quantitative comparison (albeit limited by the different dimensionality of the two models) requires to use a more similar progenitor. However, the advection timescale and the mass in the gain region are larger than the corresponding values we have obtained in all our models, as expected from the larger average residence time resulting from multi-dimensional hydrodynamical effects.

\citet{Ugliano.Janka.ea:2012} also calibrated their spherically symmetric exploding models with the observational constraints from SN~1987A, and used progenitor models identical to the ones we have adopted \citep{Woosley.Heger:2002}. They also found that the remnant mass and the properties of the explosion exhibit a large variability inside the narrow 18-21~\msun ZAMS mass window (they even found some non-exploding models). However, they did not find any clear trend with progenitor compactness (for example, their calibration model is represented by the 19.8~\msun ZAMS mass progenitor which belongs to the LC sample). The explosion timescales for models in the 18-21~\msun ZAMS mass interval are much longer in their case ($t_{\rm expl} \sim 0.3 - 1$~s), while their range for the explosion energy (0.6~--~1.6~Bethe) is relatively compatible with ours (0.4~--~1.6~Bethe). Clearly, all these differences are related to the numerous diversities between the two models.

A possible relation between explosion properites and progenitor compactness has been first pointed out by \citet{OConnor2010}, who searched for a minimum enhanced neutrino energy deposition in spherically symmetric models. Similarly to us, they found that more compact progenitors require larger heating efficiency to explode. However, they do not investigate the explosion energy of their models. Moreover, they consider it to be unlikely that a model which requires $\eta \gtrsim 0.23$ ($\xi_{2.5} \gtrsim 0.45$) will explode in nature. In our analysis, we have interpreted a large neutrino heating efficiency in spherically symmetric models as an effective way to take into account longer residence time inside the gain region. We have pointed out that HC models, characterized by larger $\eta_{\rm tot}$, are required to obtain the observed properties of SN~1987A. However, these models still have $\xi_{2.5} < 0.45$ and our average heating efficiency are below the critical value of \citet{OConnor2010}.

A clear correlation between explosion properties and progenitor compactness has been recently discussed by \citet{Nakamura2014}. They performed systematic 2D calculations of exploding CCSNe for a large variety of progenitors, using the IDSA to model $\nu_e$ and $\bar{\nu}_e$ transport. Due to computational limitations and due to the usage of only a NSE EOS, their simulations were limited to $\sim 1$~s after core bounce. Thus, they could not ensure the convergence of the diagnostic energy and could not directly compare their results with CCSN observables. However, they found trends with compactness similar to the ones we have found in our reduce sample. 

Other authors have also compared the predicted explosion energy and Ni yield from their models to the observational constraints. For example, \citet{yamamoto2013}, using the neutrino light-bulb method to trigger explosions in spherical symmetry, found a similar trend between explosion energies and nickel masses as we found (see Table~\ref{tab:values}). They also compared to a thermal bomb model with similar explosion energies and mass cut, and found that the neutrino heating mechanism leads to systematically larger $^{56}$Ni yields. They related it to higher peak temperatures, which appear because a greater thermal energy is required to unbind the accreting envelope. They also concluded that the neutrino-driven mechanism is more similar to piston-driven models by comparing with \citet{young2007}. The problem of overproducing $^{56}$Ni is lessened in the 2D simulations of \citet{yamamoto2013} because of slightly lower peak temperatures and the occurrence of fallback.

The conclusions drawn in Section~\ref{sec:eexpl_contr} about the contributions of nuclear reactions to the explosion energy are somewhat opposite to what can be found in other works in the literature. For example, \citet{yamamoto2013} state that the contribution of the nuclear reactions to the explosion energy is comparable to or greater than that of neutrino heating. Furthermore, they identify the recombinations of nucleons into nuclei in NSE as the most important nuclear reactions. However, they also point out that this ``recombination energy eventually originates from neutrino heating''. We think that this aspect is crucial for understanding the global energetics. Indeed, if we had started the analysis presented in Figure~\ref{fig:explosion energy contributions} not at bounce but at $t_{\rm expl}$ we would also have identified a strong contribution from the nuclear reactions, given roughly by the difference between $-(E_{\rm mass}-E_{\rm mass,0})$ at $t_{\rm expl}$ (which is close to the minimum) and the 
final value. However, as is clear from the Figure, roughly the same amount of energy was actually taken from the thermal energy before $t_{\rm expl}$. The dominant net contribution to the explosion energy originates from neutrino heating, as is evident from Figure~\ref{fig:total energy contributions} and as we haved discussed in detail in Section~\ref{sec:eexpl_contr}.

\section{Summary and Conclusions}
\label{sec:concl}
 
The investigation of the explosion mechanism of CCSNe as well as accurate explorations
of all the aspects related with it, is a long lasting, but still fascinating problem.
Sophisticated multi-dimensional hydrodynamical simulations, possibly including detailed neutrino 
transport, microphysical EOS, magnetic fields and aspherical properites
of the progenitor structure, are ultimately required to address this problem.
The high computational costs of such models and the uncertanties in several necessary ingredients 
still motivate the usage of effective spherically symmetric models to perform 
extended progenitor studies.

In this work we have presented a new method, PUSH, for artificially triggering parametrized core-collapse supernova explosions of massive stars in spherical symmetry.
The method provides a robust and computationally affordable framework to study important aspects of core-collapse supernovae that require modeling of the explosion for several seconds after its onset for extended sets of progenitors. 
For example, the effects of the shock passage through the star, the neutron star mass distribution, the determination of the explosion energy, or explosive supernova nucleosynthesis. Here, we have focused on the exploration of basic explosion properties and on the calibration of PUSH by reproducing observables of SN~1987A. We considered progenitors in the ZAMS mass range of 18~--~21~\msun which corresponds to typical values for the progenitor mass of SN~1987A \citep{Shigeyama1990}.

Unlike traditional methods (such as thermal bombs, pistons, or neutrino light-bulbs), our method does not require any external source of energy to trigger the explosion nor a modification of the charged-current neutrino reactions. Instead, the PUSH method taps a fraction of the energy from muon- and tau-neutrinos which are emitted by the PNS. This additional energy is deposited inside the gain region for a limited time after core bounce.
The introduction of a local heating term that is only active where electron-neutrinos are heating and where neutrino-driven convection can occur is inspired by qualitative properties of multi-dimensional CCSN simulations. We have two major free parameters, $t_{\rm rise}$, describing the temporal evolution of PUSH, and $k_{\rm push}$, controlling the strength. They are determined by comparing the outcome of our simulations with observations.

Our setup allows us to model the entire relevant domain, including the PNS and the ejecta.
In particular, 
(i) the thermodynamic properties of matter both in NSE and non-NSE conditions are treated accurately; 
(ii) the neutrino luminosities are directly related to the PNS evolution and to the mass accretion history;
(iii) the evolution of the electron fraction is followed by a radiative transport scheme for electron flavor neutrinos, which is important for the nucleosynthesis calculations.

We have studied the evolution of the explosion energy and how it is generated. The energy deposition by neutrinos is the main cause of the increase of the total energy of the ejecta and, thus, the main source of the explosion energy. The net nuclear binding energy released by the ejecta during the whole supernova (including both the initial endothermic photodissociation and the final exothermic explosive burning) is positive, but much smaller than the energy provided by neutrinos. Furthermore, we obtain an approximate convergence of the explosion energy typically only after 1 to 2 seconds and only if the full progenitor structure is taken into account. Vice-versa, we find that the so-called ``diagnostic energy'' is, in general, not suited to give an accurate estimate of the explosion energy at early times. 

Our broad parameter exploration has revealed a distinction between high compactness ($\xi_{1.75}>0.45$) and low compactness ($\xi_{1.75}<0.45$) progenitor models for the ZAMS mass range of 18~--~21~\msun. The LC models tend to explode earlier, with lower explosion energy, and with a lower remnant mass. When the HC models explode, they tend to explode later, more energetically, and producing more massive remnants. This is due to different accretion histories of the LC and HC models. The HC models have larger accretion rates, which produce larger neutrino luminosities, (marginally) harder neutrino spectra, and a stronger ram pressure at the shock. In order to overcome this pressure a more intense neutrino energy deposition is required behind the shock.
And, once the explosion has been launched, a more intense energy deposition inside the expanding shock is observed.
Thus, HC models require more time to explode but the resulting explosions are more energetic.

The fitting of the PUSH parameters to observations of SN~1987A has lead to several interesting conclusions. The requirement of an explosion energy around 1 Bethe has restricted our progenitor search to HC models. At the same time, our parameter space exploration has shown that a constraint on the explosion energy is equivalent to a tight correlation between the two most relevant PUSH parameters, $t_{\rm rise}$ and $k_{\rm push}$: if a certain explosion energy has to be achived, a longer timescale for PUSH to reach its maximum efficiency ($t_{\rm rise}$) has to be compensated by a larger PUSH strength ($k_{\rm push}$). This degeneracy can be broken by including nucleosynthesis yields in the calibration of the free parameters.

We find an overproduction of $^{56}{\rm Ni}$ for runs with an explosion energy around and above 1~Bethe. This problem is observed for all the tested parameter choices and progenitors that provide a sufficiently high explosion energy. Thus, fallback is necessary in our models to reproduce the observed nucleosynthesis yields. This fallback could be associated with the formation of a reverse shock when the forward shock reaches the hydrogen shell. The relatively large amount of fallback that we use (0.1~\msun) is consistent with observational constraints from SN~1987A and with explicit calculations of the fallback for exploding models of $\sim 20$~\msun ZAMS mass progenitors \citep{chevalier89,Ugliano.Janka.ea:2012}. 

The production of ${~}^{57-58}{\rm Ni}$ is sensitive to the electron fraction of the innermost ejecta. A final mass cut initially located inside the silicon shell can provide slightly neutron rich ejecta, corresponding to the conditions required to fit the ${~}^{57-58}{\rm Ni}$ yields of SN~1987A. We found that this is only possible for the 18.0 \msun and 19.4 \msun ZAMS mass progenitors, whereas for the other HC models, characterized by larger $\xi_{1.75}$, the mass cut is located inside the oxygen shell. The 18.0 \msun and 19.4 \msun ZAMS mass progenitors can explain the energetics and all nickel yields if fallback is included.
For $^{44}$Ti, in contrast, we find that it is underproduced. However, we have shown that uncertainties in the relevant nuclear reaction rates, together with mixing of the ejecta, can help reducing this discrepancy.

Our work has demonstrated that the progenitor structure and composition are of great importance for the nucleosynthesis yields. Recently, it has been pointed out that asphericities in the progenitor structure could aid the multi-dimensional neutrino-driven supernova mechanism \citep{Couch2013b,Mueller.Janka:2015,Couch.ea:2015}. For our work, the compositional changes induced by multi-dimensional effects in the progenitor evolution \citep{Arnett2015} would be of particular interest and could be the subject of future work. However, at present, databases with large sets of progenitors are only available for calculations that were done in spherical symmetry.

Finally, we have identified a progenitor (18.0~\msun ZAMS mass, compactness $\xi_{1.75} = 0.463$ at collapse) that fits the observables of SN~1987A for a suitable choice of the PUSH parameters ($t_{\rm on}=80$~ms, $t_{\rm rise}=200$~ms, and $k_{\rm push}=3.5$) and assuming 0.1~\msun of fallback. The associated explosion energy is $E_{\rm expl}=1.092$ Bethe, while $M({~}^{56}{\rm Ni})=0.073$ \msun. The formation of a BH in SN~1987A seems to be unlikely, since it would require a much larger fallback compared with our analysis and/or an extremely asymmetric explosion. Instead, we predict that in SN~1987A a neutron star with a baryonic mass of 1.66~\msun was born, corresponding to a gravitational mass of 1.50~\msun for a cold neutron star with our choice of the EOS. This will hopefully be probed by observations soon \citep{Zanardo2014}.

For our best model of SN~1987A the explosion happens on a timescale of a few hundereds of milliseconds after core bounce. This timescale is consistent with the overall picture of a neutrino-driven supernova, and broadly compatible with the first results obtained in exploding, self-consistent, multi-dimensional simulations.

From exploring the progenitor range of 18~--~21~\msun ZAMS mass we found indications of a correlation between explosion properties and the compactness of the progenitor model. However, a more complete analysis will require the exploration of a larger set of progenitors with the PUSH method. This will be the topic of a forthcoming work. An extended study considering all possible progenitors for core-collapse supernovae in the mass range of 8~--~100~\msun will be relevant for several open questions in nuclear astrophysics, for example for the comparison of predicted to observed explosion energies, neutron-star remnant masses, and ejected $^{56}$Ni (see, e.g., \citet{Bruenn2014}) or for the prediction of complete nucleosynthesis yields of all elements which is a crucial input to galactic chemical evolution. A full progenitor study could also give more insight into the extent to which the compactness parameter affects the supernova energetics and nucleosynthesis.

\acknowledgments
The authors thank Marcella Ugliano for useful discussions and Tobias Fischer for useful comments to the manuscript. 
The work of A.P. is supported by the Helmholtz-University Investigator grant No. VH-NG-825.
C.F.\ acknowledges support from the Department of Energy through an Early Career Award (DOE grant no.\  SC0010263) and through the Topical Collaboration in Nuclear Science ``Neutrinos and Nucleosynthesis in Hot and Dense Matter'' (DOE grant no.\ DE-SC0004786).
M.H., K.E., and M.E. acknowledge support from the Swiss National Science
Foundation (SNSF). Partial support comes from ``NewCompStar'', COST Action MP1304.
The research leading to these results has received funding from the European Research Council under the European Union's Seventh Framework Programme (FP7/2007-2013) / ERC grant agreement n° 321263 - FISH. M.H. and F.K.T. are also grateful for participating in the ENSAR/THEXO project.

\bibliographystyle{apj}
\bibliography{references_push}

\begin{thebibliography}{}
\expandafter\ifx\csname natexlab\endcsname\relax\def\natexlab#1{#1}\fi

\bibitem[{{Antoniadis} {et~al.}(2013){Antoniadis}, {Freire}, {Wex}, {Tauris},
  {Lynch}, {van Kerkwijk}, {Kramer}, {Bassa}, {Dhillon}, {Driebe}, {Hessels},
  {Kaspi}, {Kondratiev}, {Langer}, {Marsh}, {McLaughlin}, {Pennucci}, {Ransom},
  {Stairs}, {van Leeuwen}, {Verbiest}, \& {Whelan}}]{antoniadis2013}
{Antoniadis}, J., {Freire}, P.~C.~C., {Wex}, N., {et~al.} 2013, Science, 340,
  448

\bibitem[{{Arnett} {et~al.}(1989){Arnett}, {Bahcall}, {Kirshner}, \&
  {Woosley}}]{arnett89}
{Arnett}, W.~D., {Bahcall}, J.~N., {Kirshner}, R.~P., \& {Woosley}, S.~E. 1989,
  \araa, 27, 629

\bibitem[{{Arnett} {et~al.}(2015){Arnett}, {Meakin}, {Viallet}, {Campbell},
  {Lattanzio}, \& {Mo{\'c}ak}}]{Arnett2015}
{Arnett}, W.~D., {Meakin}, C., {Viallet}, M., {et~al.} 2015, ArXiv e-prints,
  arXiv:1503.00342

\bibitem[{{Audi} {et~al.}(2003){Audi}, {Wapstra}, \& {Thibault}}]{audi2003}
{Audi}, G., {Wapstra}, A.~H., \& {Thibault}, C. 2003, {Nucl. Phys. A}, 729, 337

\bibitem[{{Bartl} {et~al.}(2014){Bartl}, {Pethick}, \& {Schwenk}}]{Bartl14}
{Bartl}, A., {Pethick}, C.~J., \& {Schwenk}, A. 2014, Physical Review Letters,
  113, 081101

\bibitem[{{Bernal} {et~al.}(2013){Bernal}, {Page}, \& {Lee}}]{bernal13}
{Bernal}, C.~G., {Page}, D., \& {Lee}, W.~H. 2013, \apj, 770, 106

\bibitem[{{Bethe} \& {Wilson}(1985)}]{bethe_85}
{Bethe}, H.~A., \& {Wilson}, J.~R. 1985, \apj, 295, 14

\bibitem[{{Blinnikov} {et~al.}(2000){Blinnikov}, {Lundqvist}, {Bartunov},
  {Nomoto}, \& {Iwamoto}}]{Blinnikov2000}
{Blinnikov}, S., {Lundqvist}, P., {Bartunov}, O., {Nomoto}, K., \& {Iwamoto},
  K. 2000, \apj, 532, 1132

\bibitem[{{Blondin} \& {Mezzacappa}(2006)}]{Blondin2006}
{Blondin}, J.~M., \& {Mezzacappa}, A. 2006, \apj, 642, 401

\bibitem[{{Blondin} {et~al.}(2003){Blondin}, {Mezzacappa}, \&
  {DeMarino}}]{Blondin2003}
{Blondin}, J.~M., {Mezzacappa}, A., \& {DeMarino}, C. 2003, \apj, 584, 971

\bibitem[{{Bruenn}(1985)}]{Bruenn85}
{Bruenn}, S.~W. 1985, \apjs, 58, 771

\bibitem[{{Bruenn} {et~al.}(2013){Bruenn}, {Mezzacappa}, {Hix}, {Lentz},
  {Bronson Messer}, {Lingerfelt}, {Blondin}, {Endeve}, {Marronetti}, \&
  {Yakunin}}]{Bruenn.Mezzacappa.ea:2013}
{Bruenn}, S.~W., {Mezzacappa}, A., {Hix}, W.~R., {et~al.} 2013, \apjl, 767, L6

\bibitem[{{Bruenn} {et~al.}(2014){Bruenn}, {Lentz}, {Hix}, {Mezzacappa},
  {Harris}, {Bronson Messer}, {Endeve}, {Blondin}, {Chertkow}, {Lingerfelt},
  {Marronetti}, \& {Yakunin}}]{Bruenn2014}
{Bruenn}, S.~W., {Lentz}, E.~J., {Hix}, W.~R., {et~al.} 2014, ArXiv e-prints,
  arXiv:1409.5779

\bibitem[{{Buras} {et~al.}(2006){Buras}, {Janka}, {Rampp}, \&
  {Kifonidis}}]{buras2006b}
{Buras}, R., {Janka}, H.-T., {Rampp}, M., \& {Kifonidis}, K. 2006, \aap, 457,
  281

\bibitem[{{Burrows}(2013)}]{burrows13}
{Burrows}, A. 2013, Reviews of Modern Physics, 85, 245

\bibitem[{{Burrows} \& {Goshy}(1993)}]{Burrows1993}
{Burrows}, A., \& {Goshy}, J. 1993, \apjl, 416, L75

\bibitem[{{Chan} {et~al.}(2009){Chan}, {Cheng}, {Harko}, {Lau}, {Lin}, {Suen},
  \& {Tian}}]{chan2009}
{Chan}, T.~C., {Cheng}, K.~S., {Harko}, T., {et~al.} 2009, \apj, 695, 732

\bibitem[{{Chevalier}(1989)}]{chevalier89}
{Chevalier}, R.~A. 1989, \apj, 346, 847

\bibitem[{{Chieffi} \& {Limongi}(2013)}]{Chieffi2013}
{Chieffi}, A., \& {Limongi}, M. 2013, \apj, 764, 21

\bibitem[{{Chugai} {et~al.}(1997){Chugai}, {Chevalier}, {Kirshner}, \&
  {Challis}}]{chugai97}
{Chugai}, N.~N., {Chevalier}, R.~A., {Kirshner}, R.~P., \& {Challis}, P.~M.
  1997, \apj, 483, 925

\bibitem[{{Couch}(2013)}]{Couch2013a}
{Couch}, S.~M. 2013, \apj, 775, 35

\bibitem[{{Couch} {et~al.}(2015){Couch}, {Chatzopoulos}, {Arnett}, \&
  {Timmes}}]{Couch.ea:2015}
{Couch}, S.~M., {Chatzopoulos}, E., {Arnett}, W.~D., \& {Timmes}, F.~X. 2015,
  ArXiv e-prints, arXiv:1503.02199

\bibitem[{{Couch} \& {O'Connor}(2014)}]{Couch2014}
{Couch}, S.~M., \& {O'Connor}, E.~P. 2014, \apj, 785, 123

\bibitem[{{Couch} \& {Ott}(2013)}]{Couch2013b}
{Couch}, S.~M., \& {Ott}, C.~D. 2013, \apjl, 778, L7

\bibitem[{{Demorest} {et~al.}(2010){Demorest}, {Pennucci}, {Ransom}, {Roberts},
  \& {Hessels}}]{demorest2010}
{Demorest}, P.~B., {Pennucci}, T., {Ransom}, S.~M., {Roberts}, M.~S.~E., \&
  {Hessels}, J.~W.~T. 2010, Nature, 467, 1081

\bibitem[{{Dolence} {et~al.}(2013){Dolence}, {Burrows}, {Murphy}, \&
  {Nordhaus}}]{Dolence2013}
{Dolence}, J.~C., {Burrows}, A., {Murphy}, J.~W., \& {Nordhaus}, J. 2013, \apj,
  765, 110

\bibitem[{{Ertl} {et~al.}(2015){Ertl}, {Janka}, {Woosley}, {Sukhbold}, \&
  {Ugliano}}]{ertl2015}
{Ertl}, T., {Janka}, H.-T., {Woosley}, S.~E., {Sukhbold}, T., \& {Ugliano}, M.
  2015, ArXiv e-prints, arXiv:1503.07522

\bibitem[{{Farouqi} {et~al.}(2010){Farouqi}, {Kratz}, {Pfeiffer}, {Rauscher},
  {Thielemann}, \& {Truran}}]{farouqi2010}
{Farouqi}, K., {Kratz}, K.-L., {Pfeiffer}, B., {et~al.} 2010, \apj, 712, 1359

\bibitem[{{Fern{\'a}ndez}(2010)}]{Fernandez2010}
{Fern{\'a}ndez}, R. 2010, \apj, 725, 1563

\bibitem[{{Fischer} {et~al.}(2014){Fischer}, {Hempel}, {Sagert}, {Suwa}, \&
  {Schaffner-Bielich}}]{fischer14}
{Fischer}, T., {Hempel}, M., {Sagert}, I., {Suwa}, Y., \& {Schaffner-Bielich},
  J. 2014, European Physical Journal A, 50, 46

\bibitem[{{Fischer} {et~al.}(2012){Fischer}, {Mart{\'{\i}}nez-Pinedo},
  {Hempel}, \& {Liebend{\"o}rfer}}]{fischer12}
{Fischer}, T., {Mart{\'{\i}}nez-Pinedo}, G., {Hempel}, M., \&
  {Liebend{\"o}rfer}, M. 2012, \prd, 85, 083003

\bibitem[{{Fischer} {et~al.}(2009){Fischer}, {Whitehouse}, {Mezzacappa},
  {Thielemann}, \& {Liebend{\"o}rfer}}]{Fischer2009}
{Fischer}, T., {Whitehouse}, S.~C., {Mezzacappa}, A., {Thielemann}, F.-K., \&
  {Liebend{\"o}rfer}, M. 2009, \aap, 499, 1

\bibitem[{{Fischer} {et~al.}(2010){Fischer}, {Whitehouse}, {Mezzacappa},
  {Thielemann}, \& {Liebend{\"o}rfer}}]{fischer10}
---. 2010, \aap, 517, A80

\bibitem[{{Fischer} {et~al.}(2011){Fischer}, {Sagert}, {Pagliara}, {Hempel},
  {Schaffner-Bielich}, {Rauscher}, {Thielemann}, {K{\"a}ppeli},
  {Mart{\'{\i}}nez-Pinedo}, \& {Liebend{\"o}rfer}}]{fischer11}
{Fischer}, T., {Sagert}, I., {Pagliara}, G., {et~al.} 2011, \apjs, 194, 39

\bibitem[{{Foglizzo} {et~al.}(2006){Foglizzo}, {Scheck}, \&
  {Janka}}]{Foglizzo2006}
{Foglizzo}, T., {Scheck}, L., \& {Janka}, H.-T. 2006, \apj, 652, 1436

\bibitem[{{Fraija} {et~al.}(2014){Fraija}, {Bernal}, \&
  {Hidalgo-Gam{\'e}z}}]{fraija14}
{Fraija}, N., {Bernal}, C.~G., \& {Hidalgo-Gam{\'e}z}, A.~M. 2014, \mnras, 442,
  239

\bibitem[{{Fransson} \& {Kozma}(2002)}]{Fransson.Kozma:2002}
{Fransson}, C., \& {Kozma}, C. 2002, New Astronomy Review, 46, 487

\bibitem[{{Fr{\"o}hlich} {et~al.}(2006){Fr{\"o}hlich}, {Hauser},
  {Liebend{\"o}rfer}, {Mart{\'{\i}}nez-Pinedo}, {Thielemann}, {Bravo},
  {Zinner}, {Hix}, {Langanke}, {Mezzacappa}, \& {Nomoto}}]{cf06a}
{Fr{\"o}hlich}, C., {Hauser}, P., {Liebend{\"o}rfer}, M., {et~al.} 2006, \apj,
  637, 415

\bibitem[{{Graves} {et~al.}(2005){Graves}, {Challis}, {Chevalier}, {Crotts},
  {Filippenko}, {Fransson}, {Garnavich}, {Kirshner}, {Li}, {Lundqvist},
  {McCray}, {Panagia}, {Phillips}, {Pun}, {Schmidt}, {Sonneborn}, {Suntzeff},
  {Wang}, \& {Wheeler}}]{graves2005}
{Graves}, G.~J.~M., {Challis}, P.~M., {Chevalier}, R.~A., {et~al.} 2005, \apj,
  629, 944

\bibitem[{{Grebenev} {et~al.}(2012){Grebenev}, {Lutovinov}, {Tsygankov}, \&
  {Winkler}}]{grebenev12}
{Grebenev}, S.~A., {Lutovinov}, A.~A., {Tsygankov}, S.~S., \& {Winkler}, C.
  2012, Nature, 490, 373

\bibitem[{{Grefenstette} {et~al.}(2014){Grefenstette}, {Harrison}, {Boggs},
  {Reynolds}, {Fryer}, {Madsen}, {Wik}, {Zoglauer}, {Ellinger}, {Alexander},
  {An}, {Barret}, {Christensen}, {Craig}, {Forster}, {Giommi}, {Hailey},
  {Hornstrup}, {Kaspi}, {Kitaguchi}, {Koglin}, {Mao}, {Miyasaka}, {Mori},
  {Perri}, {Pivovaroff}, {Puccetti}, {Rana}, {Stern}, {Westergaard}, \&
  {Zhang}}]{grefenstette14}
{Grefenstette}, B.~W., {Harrison}, F.~A., {Boggs}, S.~E., {et~al.} 2014, \nat,
  506, 339

\bibitem[{{Guilet} \& {Foglizzo}(2012)}]{Guilet2012}
{Guilet}, J., \& {Foglizzo}, T. 2012, \mnras, 421, 546

\bibitem[{{Handy} {et~al.}(2014){Handy}, {Plewa}, \& {Odrzywo{\l}ek}}]{handy14}
{Handy}, T., {Plewa}, T., \& {Odrzywo{\l}ek}, A. 2014, \apj, 783, 125

\bibitem[{{Hanke} {et~al.}(2012){Hanke}, {Marek}, {M{\"u}ller}, \&
  {Janka}}]{Hanke2012}
{Hanke}, F., {Marek}, A., {M{\"u}ller}, B., \& {Janka}, H.-T. 2012, \apj, 755,
  138

\bibitem[{{Hanke} {et~al.}(2013){Hanke}, {M{\"u}ller}, {Wongwathanarat},
  {Marek}, \& {Janka}}]{Hanke2013}
{Hanke}, F., {M{\"u}ller}, B., {Wongwathanarat}, A., {Marek}, A., \& {Janka},
  H.-T. 2013, \apj, 770, 66

\bibitem[{{Hannestad} \& {Raffelt}(1998)}]{Hannestad98}
{Hannestad}, S., \& {Raffelt}, G. 1998, \apj, 507, 339

\bibitem[{{Hempel}(2014)}]{hempel14}
{Hempel}, M. 2014, arXiv:1410.6337

\bibitem[{{Hempel} \& {Schaffner-Bielich}(2010)}]{Hempel.SchaffnerBielich:2010}
{Hempel}, M., \& {Schaffner-Bielich}, J. 2010, Nuclear Physics A, 837, 210

\bibitem[{{Herant} {et~al.}(1994){Herant}, {Benz}, {Hix}, {Fryer}, \&
  {Colgate}}]{Herant94}
{Herant}, M., {Benz}, W., {Hix}, W.~R., {Fryer}, C.~L., \& {Colgate}, S.~A.
  1994, \apj, 435, 339

\bibitem[{{H{\"u}depohl} {et~al.}(2010){H{\"u}depohl}, {M{\"u}ller}, {Janka},
  {Marek}, \& {Raffelt}}]{huedepohl10}
{H{\"u}depohl}, L., {M{\"u}ller}, B., {Janka}, H.-T., {Marek}, A., \&
  {Raffelt}, G.~G. 2010, Physical Review Letters, 104, 251101

\bibitem[{{Iwakami} {et~al.}(2008){Iwakami}, {Kotake}, {Ohnishi}, {Yamada}, \&
  {Sawada}}]{iwakami08}
{Iwakami}, W., {Kotake}, K., {Ohnishi}, N., {Yamada}, S., \& {Sawada}, K. 2008,
  \apj, 678, 1207

\bibitem[{{Iwakami} {et~al.}(2009){Iwakami}, {Kotake}, {Ohnishi}, {Yamada}, \&
  {Sawada}}]{iwakami09}
---. 2009, \apj, 700, 232

\bibitem[{{Janka} \& {Mueller}(1996)}]{janka96}
{Janka}, H., \& {Mueller}, E. 1996, \aap, 306, 167

\bibitem[{{Janka}(2012)}]{janka12}
{Janka}, H.-T. 2012, Annual Review of Nuclear and Particle Science, 62, 407

\bibitem[{{Janka} {et~al.}(2012){Janka}, {Hanke}, {H{\"u}depohl}, {Marek},
  {M{\"u}ller}, \& {Obergaulinger}}]{janka12b}
{Janka}, H.-T., {Hanke}, F., {H{\"u}depohl}, L., {et~al.} 2012, Progress of
  Theoretical and Experimental Physics, 2012, 010000

\bibitem[{{Jerkstrand} {et~al.}(2011){Jerkstrand}, {Fransson}, \&
  {Kozma}}]{jerkstrand11}
{Jerkstrand}, A., {Fransson}, C., \& {Kozma}, C. 2011, \aap, 530, A45

\bibitem[{{Kifonidis} {et~al.}(2006){Kifonidis}, {Plewa}, {Scheck}, {Janka}, \&
  {M{\"u}ller}}]{kifonidis2006}
{Kifonidis}, K., {Plewa}, T., {Scheck}, L., {Janka}, H.-T., \& {M{\"u}ller}, E.
  2006, \aap, 453, 661

\bibitem[{{Korobkin} {et~al.}(2012){Korobkin}, {Rosswog}, {Arcones}, \&
  {Winteler}}]{korobkin2012}
{Korobkin}, O., {Rosswog}, S., {Arcones}, A., \& {Winteler}, C. 2012, \mnras,
  426, 1940

\bibitem[{{Koshiba}(1992)}]{koshiba92}
{Koshiba}, M. 1992, \physrep, 220, 229

\bibitem[{{Kozma} \& {Fransson}(1998{\natexlab{a}})}]{Kozma1998a}
{Kozma}, C., \& {Fransson}, C. 1998{\natexlab{a}}, \apj, 496, 946

\bibitem[{{Kozma} \& {Fransson}(1998{\natexlab{b}})}]{Kozma1998b}
---. 1998{\natexlab{b}}, \apj, 497, 431

\bibitem[{{Larsson} {et~al.}(2011){Larsson}, {Fransson}, {{\"O}stlin},
  {Gr{\"o}ningsson}, {Jerkstrand}, {Kozma}, {Sollerman}, {Challis}, {Kirshner},
  {Chevalier}, {Heng}, {McCray}, {Suntzeff}, {Bouchet}, {Crotts}, {Danziger},
  {Dwek}, {France}, {Garnavich}, {Lawrence}, {Leibundgut}, {Lundqvist},
  {Panagia}, {Pun}, {Smith}, {Sonneborn}, {Wang}, \& {Wheeler}}]{larsson11}
{Larsson}, J., {Fransson}, C., {{\"O}stlin}, G., {et~al.} 2011, \nat, 474, 484

\bibitem[{{Lattimer} \& {Swesty}(1991)}]{lattimer91}
{Lattimer}, J.~M., \& {Swesty}, D.~F. 1991, {Nucl. Phys. A}, 535, 331

\bibitem[{{Liebend{\"o}rfer}(2005)}]{Liebendoerfer.delept:2005}
{Liebend{\"o}rfer}, M. 2005, \apj, 633, 1042

\bibitem[{{Liebend{\"o}rfer} {et~al.}(2004){Liebend{\"o}rfer}, {Messer},
  {Mezzacappa}, {Bruenn}, {Cardall}, \& {Thielemann}}]{Liebendoerfer2004}
{Liebend{\"o}rfer}, M., {Messer}, O.~E.~B., {Mezzacappa}, A., {et~al.} 2004,
  \apjs, 150, 263

\bibitem[{{Liebend{\"o}rfer} {et~al.}(2001){Liebend{\"o}rfer}, {Mezzacappa}, \&
  {Thielemann}}]{Liebendoerfer.Agile}
{Liebend{\"o}rfer}, M., {Mezzacappa}, A., \& {Thielemann}, F.-K. 2001, \prd,
  63, 104003

\bibitem[{{Liebend{\"o}rfer} {et~al.}(2009){Liebend{\"o}rfer}, {Whitehouse}, \&
  {Fischer}}]{Liebendoerfer.IDSA:2009}
{Liebend{\"o}rfer}, M., {Whitehouse}, S.~C., \& {Fischer}, T. 2009, \apj, 698,
  1174

\bibitem[{{Limongi} \& {Chieffi}(2006)}]{Limongi2006}
{Limongi}, M., \& {Chieffi}, A. 2006, \apj, 647, 483

\bibitem[{{Marek} \& {Janka}(2009)}]{Marek2009}
{Marek}, A., \& {Janka}, H.-T. 2009, \apj, 694, 664

\bibitem[{{Margerin} {et~al.}(2014){Margerin}, {Murphy}, {Davinson},
  {Dressler}, {Fallis}, {Kankainen}, {Laird}, {Lotay}, {Mountford}, {Murphy},
  {Seiffert}, {Schumann}, {Stowasser}, {Stora}, {Wang}, \&
  {Woods}}]{margerin2014}
{Margerin}, V., {Murphy}, A.~S.~J., {Davinson}, T., {et~al.} 2014, Physics
  Letters B, 731, 358

\bibitem[{{Mart{\'{\i}}nez-Pinedo} {et~al.}(2014){Mart{\'{\i}}nez-Pinedo},
  {Fischer}, \& {Huther}}]{martinez14}
{Mart{\'{\i}}nez-Pinedo}, G., {Fischer}, T., \& {Huther}, L. 2014, Journal of
  Physics G Nuclear Physics, 41, 044008

\bibitem[{{Mart{\'{\i}}nez-Pinedo} {et~al.}(2012){Mart{\'{\i}}nez-Pinedo},
  {Fischer}, {Lohs}, \& {Huther}}]{martinez12}
{Mart{\'{\i}}nez-Pinedo}, G., {Fischer}, T., {Lohs}, A., \& {Huther}, L. 2012,
  Physical Review Letters, 109, 251104

\bibitem[{{Melson} {et~al.}(2015){Melson}, {Janka}, \& {Marek}}]{Melson2015}
{Melson}, T., {Janka}, H.-T., \& {Marek}, A. 2015, \apjl, 801, L24

\bibitem[{{Mezzacappa} \& {Bruenn}(1993{\natexlab{a}})}]{Mezzacappa1993b}
{Mezzacappa}, A., \& {Bruenn}, S.~W. 1993{\natexlab{a}}, \apj, 405, 669

\bibitem[{{Mezzacappa} \& {Bruenn}(1993{\natexlab{b}})}]{Mezzacappa1993c}
---. 1993{\natexlab{b}}, \apj, 410, 740

\bibitem[{{Mezzacappa} \& {Bruenn}(1993{\natexlab{c}})}]{Mezzacappa1993a}
---. 1993{\natexlab{c}}, \apj, 405, 637

\bibitem[{{M{\"o}ller} {et~al.}(1995{\natexlab{a}}){M{\"o}ller}, {Nix},
  {Myers}, \& {Swiatecki}}]{moller95}
{M{\"o}ller}, P., {Nix}, J.~R., {Myers}, W.~D., \& {Swiatecki}, W.~J.
  1995{\natexlab{a}}, Atomic Data and Nuclear Data Tables, 59, 185

\bibitem[{{M{\"o}ller} {et~al.}(1995{\natexlab{b}}){M{\"o}ller}, {Nix},
  {Myers}, \& {Swiatecki}}]{Moller.ea:1995}
---. 1995{\natexlab{b}}, Atomic Data and Nuclear Data Tables, 59, 185

\bibitem[{{M{\"u}ller} \& {Janka}(2015)}]{Mueller.Janka:2015}
{M{\"u}ller}, B., \& {Janka}, H.-T. 2015, \mnras, 448, 2141

\bibitem[{{M{\"u}ller} {et~al.}(2012{\natexlab{a}}){M{\"u}ller}, {Janka}, \&
  {Heger}}]{Mueller.Janka.ea:2012}
{M{\"u}ller}, B., {Janka}, H.-T., \& {Heger}, A. 2012{\natexlab{a}}, \apj, 761,
  72

\bibitem[{{M{\"u}ller} {et~al.}(2012{\natexlab{b}}){M{\"u}ller}, {Janka}, \&
  {Marek}}]{janka2012}
{M{\"u}ller}, B., {Janka}, H.-T., \& {Marek}, A. 2012{\natexlab{b}}, \apj, 756,
  84

\bibitem[{{Murphy} \& {Burrows}(2008)}]{Murphy2008}
{Murphy}, J.~W., \& {Burrows}, A. 2008, \apj, 688, 1159

\bibitem[{{Nakamura} {et~al.}(2014){Nakamura}, {Takiwaki}, {Kuroda}, \&
  {Kotake}}]{Nakamura2014}
{Nakamura}, K., {Takiwaki}, T., {Kuroda}, T., \& {Kotake}, K. 2014, ArXiv
  e-prints, arXiv:1406.2415

\bibitem[{{Nordhaus} {et~al.}(2010){Nordhaus}, {Burrows}, {Almgren}, \&
  {Bell}}]{Nordhaus2010}
{Nordhaus}, J., {Burrows}, A., {Almgren}, A., \& {Bell}, J. 2010, \apj, 720,
  694

\bibitem[{{O'Connor} \& {Ott}(2010)}]{OConnor2010}
{O'Connor}, E., \& {Ott}, C.~D. 2010, Classical and Quantum Gravity, 27, 114103

\bibitem[{{O'Connor} \& {Ott}(2011)}]{OConnor.Ott:2011}
---. 2011, \apj, 730, 70

\bibitem[{{O'Connor} \& {Ott}(2013)}]{OConnor.Ott:2013}
---. 2013, \apj, 762, 126

\bibitem[{{{\"O}gelman} \& {Alpar}(2004)}]{oegelman2004}
{{\"O}gelman}, H., \& {Alpar}, M.~A. 2004, \apjl, 603, L33

\bibitem[{{Perego} {et~al.}(2014){Perego}, {Rosswog}, {Cabez{\'o}n},
  {Korobkin}, {K{\"a}ppeli}, {Arcones}, \& {Liebend{\"o}rfer}}]{Perego2014}
{Perego}, A., {Rosswog}, S., {Cabez{\'o}n}, R.~M., {et~al.} 2014, \mnras, 443,
  3134

\bibitem[{{Plewa} \& {M{\"u}ller}(1999)}]{plewa98}
{Plewa}, T., \& {M{\"u}ller}, E. 1999, \aap, 342, 179

\bibitem[{{Podsiadlowski} {et~al.}(2007){Podsiadlowski}, {Morris}, \&
  {Ivanova}}]{PP.sn1987a:2007}
{Podsiadlowski}, P., {Morris}, T.~S., \& {Ivanova}, N. 2007, in American
  Institute of Physics Conference Series, Vol. 937, Supernova 1987A: 20 Years
  After: Supernovae and Gamma-Ray Bursters, ed. S.~{Immler}, K.~{Weiler}, \&
  R.~{McCray}, 125--133

\bibitem[{{Rauscher} {et~al.}(2002){Rauscher}, {Heger}, {Hoffman}, \&
  {Woosley}}]{rauscher2002}
{Rauscher}, T., {Heger}, A., {Hoffman}, R.~D., \& {Woosley}, S.~E. 2002, \apj,
  576, 323

\bibitem[{{Rauscher} \& {Thielemann}(2000)}]{Rauscher.FKT:2000}
{Rauscher}, T., \& {Thielemann}, F.-K. 2000, Atomic Data and Nuclear Data
  Tables, 75, 1

\bibitem[{{Reddy} {et~al.}(1998){Reddy}, {Prakash}, \& {Lattimer}}]{reddy98}
{Reddy}, S., {Prakash}, M., \& {Lattimer}, J.~M. 1998, \prd, 58, 013009

\bibitem[{{Roberts} {et~al.}(2012){Roberts}, {Reddy}, \& {Shen}}]{roberts12}
{Roberts}, L.~F., {Reddy}, S., \& {Shen}, G. 2012, \prc, 86, 065803

\bibitem[{{Scheck} {et~al.}(2008){Scheck}, {Janka}, {Foglizzo}, \&
  {Kifonidis}}]{Scheck2008}
{Scheck}, L., {Janka}, H.-T., {Foglizzo}, T., \& {Kifonidis}, K. 2008, \aap,
  477, 931

\bibitem[{{Scheck} {et~al.}(2006){Scheck}, {Kifonidis}, {Janka}, \&
  {M{\"u}ller}}]{Scheck2006}
{Scheck}, L., {Kifonidis}, K., {Janka}, H.-T., \& {M{\"u}ller}, E. 2006, \aap,
  457, 963

\bibitem[{{Seitenzahl} {et~al.}(2014){Seitenzahl}, {Timmes}, \&
  {Magkotsios}}]{Seitenzahl2014}
{Seitenzahl}, I.~R., {Timmes}, F.~X., \& {Magkotsios}, G. 2014, \apj, 792, 10

\bibitem[{{Shen} {et~al.}(1998){Shen}, {Toki}, {Oyamatsu}, \&
  {Sumiyoshi}}]{shen98}
{Shen}, H., {Toki}, H., {Oyamatsu}, K., \& {Sumiyoshi}, K. 1998, {Nucl. Phys.
  A}, 637, 435

\bibitem[{{Shigeyama} \& {Nomoto}(1990)}]{Shigeyama1990}
{Shigeyama}, T., \& {Nomoto}, K. 1990, \apj, 360, 242

\bibitem[{{Sukhbold} \& {Woosley}(2014)}]{sukhbold.woosley:2014}
{Sukhbold}, T., \& {Woosley}, S.~E. 2014, \apj, 783, 10

\bibitem[{{Suwa} {et~al.}(2010){Suwa}, {Kotake}, {Takiwaki}, {Whitehouse},
  {Liebend{\"o}rfer}, \& {Sato}}]{suwa2010}
{Suwa}, Y., {Kotake}, K., {Takiwaki}, T., {et~al.} 2010, \pasj, 62, L49

\bibitem[{{Suwa} {et~al.}(2013){Suwa}, {Takiwaki}, {Kotake}, {Fischer},
  {Liebend{\"o}rfer}, \& {Sato}}]{Suwa2013}
{Suwa}, Y., {Takiwaki}, T., {Kotake}, K., {et~al.} 2013, \apj, 764, 99

\bibitem[{{Suwa} {et~al.}(2014){Suwa}, {Yamada}, {Takiwaki}, \&
  {Kotake}}]{suwa14}
{Suwa}, Y., {Yamada}, S., {Takiwaki}, T., \& {Kotake}, K. 2014, ArXiv e-prints,
  arXiv:1406.6414

\bibitem[{{Takiwaki} {et~al.}(2014){Takiwaki}, {Kotake}, \&
  {Suwa}}]{takiwaki2014}
{Takiwaki}, T., {Kotake}, K., \& {Suwa}, Y. 2014, \apj, 786, 83

\bibitem[{{Thielemann} {et~al.}(1990){Thielemann}, {Hashimoto}, \&
  {Nomoto}}]{thielemann90}
{Thielemann}, F.-K., {Hashimoto}, M.-A., \& {Nomoto}, K. 1990, \apj, 349, 222

\bibitem[{{Thielemann} {et~al.}(1996){Thielemann}, {Nomoto}, \&
  {Hashimoto}}]{tnh96}
{Thielemann}, F.-K., {Nomoto}, K., \& {Hashimoto}, M.-A. 1996, \apj, 460, 408

\bibitem[{{Timmes} \& {Swesty}(2000)}]{Timmes.Swesty:2000}
{Timmes}, F.~X., \& {Swesty}, F.~D. 2000, \apjs, 126, 501

\bibitem[{{Typel} {et~al.}(2010){Typel}, {R{\"o}pke}, {Kl{\"a}hn}, {Blaschke},
  \& {Wolter}}]{typel10}
{Typel}, S., {R{\"o}pke}, G., {Kl{\"a}hn}, T., {Blaschke}, D., \& {Wolter},
  H.~H. 2010, \prc, 81, 015803

\bibitem[{{Ugliano} {et~al.}(2012){Ugliano}, {Janka}, {Marek}, \&
  {Arcones}}]{Ugliano.Janka.ea:2012}
{Ugliano}, M., {Janka}, H.-T., {Marek}, A., \& {Arcones}, A. 2012, \apj, 757,
  69

\bibitem[{{Umeda} \& {Nomoto}(2002)}]{Umeda2002}
{Umeda}, H., \& {Nomoto}, K. 2002, \apj, 565, 385

\bibitem[{{Umeda} \& {Nomoto}(2008)}]{Umeda2008}
---. 2008, \apj, 673, 1014

\bibitem[{{Utrobin} \& {Chugai}(2005)}]{Utrobin2005}
{Utrobin}, V.~P., \& {Chugai}, N.~N. 2005, \aap, 441, 271

\bibitem[{{Vissani}(2015)}]{vissani14}
{Vissani}, F. 2015, Journal of Physics G Nuclear Physics, 42, 013001

\bibitem[{{Winteler} {et~al.}(2012){Winteler}, {K{\"a}ppeli}, {Perego},
  {Arcones}, {Vasset}, {Nishimura}, {Liebend{\"o}rfer}, \&
  {Thielemann}}]{Winteler.ea:2012}
{Winteler}, C., {K{\"a}ppeli}, R., {Perego}, A., {et~al.} 2012, \apjl, 750, L22

\bibitem[{{Woosley}(1988)}]{Woosley1988}
{Woosley}, S.~E. 1988, \apj, 330, 218

\bibitem[{{Woosley} {et~al.}(2002){Woosley}, {Heger}, \&
  {Weaver}}]{Woosley.Heger:2002}
{Woosley}, S.~E., {Heger}, A., \& {Weaver}, T.~A. 2002, Reviews of Modern
  Physics, 74, 1015

\bibitem[{{Woosley} \& {Weaver}(1995)}]{ww95}
{Woosley}, S.~E., \& {Weaver}, T.~A. 1995, \apjs, 101, 181

\bibitem[{{Yamamoto} {et~al.}(2013){Yamamoto}, {Fujimoto}, {Nagakura}, \&
  {Yamada}}]{yamamoto2013}
{Yamamoto}, Y., {Fujimoto}, S.-i., {Nagakura}, H., \& {Yamada}, S. 2013, \apj,
  771, 27

\bibitem[{{Yamasaki} \& {Yamada}(2005)}]{Yamasaki2005}
{Yamasaki}, T., \& {Yamada}, S. 2005, \apj, 623, 1000

\bibitem[{{Young} \& {Fryer}(2007)}]{young2007}
{Young}, P.~A., \& {Fryer}, C.~L. 2007, \apj, 664, 1033

\bibitem[{{Zanardo} {et~al.}(2013){Zanardo}, {Staveley-Smith}, {Ng},
  {Gaensler}, {Potter}, {Manchester}, \& {Tzioumis}}]{zanardo13}
{Zanardo}, G., {Staveley-Smith}, L., {Ng}, C.-Y., {et~al.} 2013, \apj, 767, 98

\bibitem[{{Zanardo} {et~al.}(2014){Zanardo}, {Staveley-Smith}, {Indebetouw},
  {Chevalier}, {Matsuura}, {Gaensler}, {Barlow}, {Fransson}, {Manchester},
  {Baes}, {Kamenetzky}, {Laki{\'c}evi{\'c}}, {Lundqvist}, {Marcaide},
  {Mart{\'{\i}}-Vidal}, {Meixner}, {Ng}, {Park}, {Sonneborn}, {Spyromilio}, \&
  {van Loon}}]{Zanardo2014}
{Zanardo}, G., {Staveley-Smith}, L., {Indebetouw}, R., {et~al.} 2014, \apj,
  796, 82

\end{thebibliography}

\end{document}